\documentclass[a4paper,11pt]{article}

\usepackage{jcappub}
\usepackage[T1]{fontenc}
\usepackage{lmodern}
\usepackage{mathtools,bm}
\usepackage{microtype}
\usepackage{booktabs}
\usepackage{enumitem}
\usepackage[nameinlink,noabbrev]{cleveref}

\newcommand{\dd}{\mathrm{d}}
\newcommand{\ii}{\mathrm{i}}
\newcommand{\ee}{\mathrm{e}}

\newcommand{\cD}{\mathcal D}

\newcommand{\avg}[1]{\overline{#1}}
\newcommand{\vev}[1]{\left\langle #1\right\rangle}

\title{\centering Disorder at the Cosmological Collider:\\ \textit{Quenched Random Fields}}

\author[a]{Matheus~Curado~Ferreira,}

\affiliation[a]{Brazilian Center for Research in Physics (CBPF), \\ Dr. Xavier Sigaud st. 150, zip 22290-180, Rio de Janeiro, RJ, Brazil}

\emailAdd{matheuscurado@cbpf.br}

\abstract{Cosmological-collider signals are usually interpreted as clean probes of the spectrum of massive particles present during inflation.~This interpretation implicitly assumes that heavy fields evolve in a clean, deterministic background. Yet random sources can shape the dynamics of cosmological fields, as illustrated by stochastic inflation, where short-wavelength quantum fluctuations act as effective noise for long-wavelength modes. In this work, we study a massive spectator field excited by a spatially quenched random source, which may arise, for instance, from a frozen realization of a light field or environmental sector whose relaxation time is long compared with the inflationary timescale. We determine which features of the collider signal remain reliable probes of the heavy field.~After Gaussian disorder averaging, the random source modifies only the statistical component of the propagator, leaving the spectral function and the heavy-field poles unchanged. The particle spectrum is therefore preserved, and the logarithmic frequency of the collider oscillations remains fixed by the heavy-field mass. What changes is how the massive modes are excited and imprinted on cosmological correlators: the temporal profile of the source controls the amplitude, phase, and envelope of the oscillations. For a source with power-law time dependence, the problem at fixed disorder realization is exactly solvable in terms of Lommel functions, allowing us to separate the forced response from the homogeneous massive contribution. A decaying source suppresses the forced response, making the nonanalytic clock signal parametrically dominant in the squeezed limit, whereas spatial white noise can interfere destructively with the standard collapsed-trispectrum signal. These results show that mass measurements from cosmological colliders are robust against this class of quenched disorder, while inferences based on signal amplitude, including the interpretation of a null detection, can depend sensitively on the surrounding random environment.
}

\begin{document}
\maketitle

\clearpage

\section{Introduction}

Cosmological correlators provide a spectroscopic probe of massive degrees of freedom present during inflation. In particular, the exchange of a field with mass of order the Hubble scale produces nonanalytic momentum dependence governed by its de Sitter scaling dimensions, leading to the characteristic oscillatory signals of the cosmological collider~\cite{Arkani-Hamed:2015bza,
Arkani-Hamed:2018kmz,
Pinol_2023,
Qin:2022pmb,
Qin:2025xct,
Aoki:2024uyi,
Jiang:2025mlm,
Jazayeri:2023xcj,
Jazayeri:2023kji,
Jazayeri:2022kjy,
DuasoPueyo:2023kyh,
BaumannGreenSUSY,
Baumann_2018,
Lee_2016,
Wu_2019,
Saito_2018,
Lu_2020,
Lu_2021,
Sou_2021,
Hook_2020,
Hook_20201,
Liu_2020,
Wang_2020,
Fan_2021,
Kumar_2020,
Hubisz:2024xnj,
Cui_2022,
Chen:2018trispectrum,
ChenWangXianyuSM,
Chen_2012,
Chen:2009zp,
Kehagias_2017,
Gong_2013,
NoumiEFTQSFI,
Guimaraes:2020vfd,
Iacconi:2024hmg,
Falciano:2013uaa,
Mukhanov:1990me,
Maldacena:2002vr,
Martin2014,
Planck:2018jri,
BaumannSpinningBootstrap,
PimentelWangBoostless,
Aoki:2026vbc}. These signatures are most clearly visible in the squeezed limit of the bispectrum and the collapsed limit of the trispectrum, where their oscillatory structure carries information about the mass and spin of the exchanged field. The observable collider signal, however, is sensitive not only to the intrinsic particle spectrum, but also to the way in which massive fields are excited during inflation. Changes in the initial state, particle-production mechanism, dispersion relation,
or time dependence of the inflationary background can modify the amplitude and
momentum dependence with which heavy degrees of freedom appear in primordial
correlators~\cite{Yin:2023jlv,
ReeceWangXianyu:2022,
BodasScalarChemicalPotential,
Chua:2018dqh,
ChenEbadiKumar:2022,
Aoki:2023dynamical,
Ferreira:2026ghostspectators,
Ferreira:2026rsl,
Wang:2026lff,
Sundrum:2026ghf}. Cosmological-collider signals can therefore retain information about the excitation history of the exchanged field in addition to its spectral data. Here we explore this interplay in the presence of random fields.

Stochasticity is already familiar in inflationary dynamics~\cite{Starobinsky:1986fx,
Starobinsky:1994bd,
Martin:2011ib,
Vennin:2015hra,
Grain:2017dqa,
Pinol:2020kvw,
Gow:2023bke}. In stochastic inflation, short-wavelength quantum fluctuations can be described as an effective random source for the long-wavelength sector, motivating a statistical description of the resulting field dynamics. Replica methods have also been applied in this context~\cite{Kuhnel:2008ue,Kuhnel:2008ry}. The stochasticity considered here is of a different type. Rather than a dynamical noise continually generated during the evolution, we study a spatially quenched random field whose realization is fixed in comoving space, while its temporal profile is deterministic.~Quenched disorder has a long history in quantum field theory, where Gaussian disorder can be represented through the replica method as interactions among copies of the theory~\cite{Fujita:2008zv,Aharony:2018mjm,Svaiter:2016lha}. Random-source scalar theories are especially tractable because the field remains exactly solvable for each realization of the source~\cite{Piazza:2024jiz,Arias:2011pw}. Related ideas have also appeared in inflation: Ref.~\cite{Green:2014disorder} developed a formalism for quenched disorder in the effective field theory of inflation, with random parameters depending on time but not on position, while spatially dependent disorder was identified as a qualitatively different possibility. General replica generating functionals for random quantum field theories, including in-in observables, were subsequently developed in Ref.~\cite{Jain:2015random}, while Schwinger--Keldysh methods provide a natural language for stochastic effects in real-time cosmology~\cite{Salcedo_2025,Salcedo:2024smn}; in particular, Ref.~\cite{Colas:2025opencc} showed that the local and nonlocal sectors of heavy-field exchange admit a transparent interpretation in the Keldysh basis.

Motivated by these developments, in this work we consider a massive spectator field coupled linearly to a spatially quenched random source during inflation. We take the source to have the form
\begin{equation}
h(\bm x,\eta)=(-H\eta)^p h(\bm x),
\label{eq:introProfile}
\end{equation}
where $h(\bm x)$ is quenched in comoving space and $p$ controls a deterministic time dependence. The case $p=0$ corresponds to a persistent random field, while $p>0$ makes the forcing decay toward the future boundary. For every fixed realization the source remains deterministic, so the setup differs from a dynamical bath: the randomness resides in the spatial ensemble, while its temporal envelope is fixed.

The main effect of the disorder can be stated directly at the level of the propagator. After Gaussian disorder averaging, the random source contributes to the statistical sector while leaving the spectral function and the heavy-field poles unchanged. The spectrum of the exchanged particle is therefore preserved, but the coefficients with which its massive modes appear in cosmological correlators are modified. For principal-series fields, the logarithmic clock frequency remains fixed by the mass, while the disorder changes the amplitude, phase, and soft weighting of the corresponding oscillatory signal. In this sense, the particle spectrum and the history of its excitation leave distinct imprints on the collider correlator.

The fixed-realization problem is exactly solvable for the power-law family in Eq.~\eqref{eq:introProfile}. General sourced Klein--Gordon equations in de Sitter space have been studied at the level of fundamental solutions and Cauchy problems~\cite{Yagdjian:2009dS}, while forced scalar and fluid equations in cosmological backgrounds are known to reduce to inhomogeneous Bessel equations involving Lommel functions~\cite{Choi:1998moduli,Matarrese:2002IGM}. The special-function structure itself is therefore classical: the Lommel function solves the inhomogeneous Bessel equation and admits a standard representation in terms of ${}_1F_2$~\cite{Zullo:2024lommel}. In the present setup it appears as the exact retarded response to a spatially quenched source in Lorentzian de Sitter space.

For each realization, the massive field can be written as the sum of the usual free Bunch--Davies operator and a deterministic retarded response. For the profile in Eq.~\eqref{eq:introProfile}, the latter reduces to a Lommel function $S_{p-5/2,\nu}$. Disorder averaging then generates an additive contribution to the statistical propagator, with no corresponding modification of the spectral function. In the Schwinger--Keldysh language, this contribution arises from a $qq$ kernel that is rank one in time, while in replica space it lies entirely in the singlet sector. Since the same c-number contribution appears in every $\pm$ contour propagator, its contribution to the cosmological-collider exchange factorizes.

The temporal profile plays an important role in separating the forced and massive parts of the response. For the persistent realization,
\begin{equation}
h(\bm x,\eta)=h(\bm x),
\end{equation}
the inhomogeneous equation produces, away from the resonant massless limit, a static late-time contribution
\begin{equation}
R_k(\eta)\longrightarrow \frac{1}{m^2},
\qquad \eta\to0,
\end{equation}
in addition to the homogeneous massive tails $(-k\eta)^{3/2\pm\nu}$. These two sectors behave differently in cosmological correlators. The static contribution is integrable for the $\ell_1=\ell_2=0$ exchange seed relevant to the disorder trispectrum, whereas it generates a local late-time logarithm for the $\ell=-2$ vertex entering the bispectrum. This endpoint behavior originates from a source that remains active until arbitrarily late times and is distinct from both folded and cosmological-collider singularities.

A decaying source provides a controlled way of separating this local forced contribution from the massive response. More generally, we may write
\begin{equation}
h(\bm x,\eta)=h(\bm x)g(\eta),
\qquad
g(\eta)\xrightarrow[\eta\to0]{}0,
\end{equation}
with the power-law family
\begin{equation}
g(\eta)\sim(-H\eta)^p,
\qquad p>0.
\end{equation}
The forced response then decays toward the future boundary, while the massive modes excited during the disordered stage remain. The late-time solution takes the form
\begin{equation}
R_{k,p}(\eta)
\sim
\frac{(-H\eta)^p}
{H^2\left[M^2+p(p-3)\right]}
+B_+(p)(-k\eta)^{3/2+\nu}
+B_-(p)(-k\eta)^{3/2-\nu}
+\cdots,
\quad
M^2\equiv\frac{m^2}{H^2}.
\end{equation}
For a principal-series field, $\nu=i\mu$, the last two terms retain the usual logarithmic clock frequency $\mu$. The temporal profile instead modifies the coefficients $B_\pm(p)$ and therefore the relative weight and phase of the massive branches.

This distinction is inherited by the cosmological-collider correlators. A decaying source removes the future-boundary obstruction of the persistent bispectrum contribution and progressively suppresses the analytic forced sector. For sufficiently fast decay, the nonanalytic massive contribution becomes the dominant soft signal. Spatial white noise provides a particularly simple realization in which the disorder correction can also interfere with the ordinary collapsed trispectrum collider contribution. These effects arise without shifting the heavy-field poles and therefore modify the observable realization of the collider signal rather than its underlying mass spectrum.

A Mellin--Barnes representation provides a complementary description of the same separation. The analytic forced response and the homogeneous massive completion arise from distinct pole structures, while the heavy-field exponents remain fixed. This representation also makes the exact factorization of the disorder contribution transparent and provides a convenient route to the folded and soft limits of the resulting seed functions.

To our knowledge, the combination of a spatially quenched random source, an exactly solvable time-dependent response in de Sitter space, and its imprint on cosmological-collider correlators has not been studied previously. The construction provides a simple setting in which an external environment modifies the amplitude, phase, and soft behavior of a massive collider signal while preserving its characteristic mass-dependent frequency. In the following, we formulate the problem entirely in Lorentzian de Sitter space, solve the power-law family for arbitrary $p$, derive the disorder correction in the Schwinger--Keldysh formalism, and construct the corresponding exact exchange seed. We then study its folded and soft limits and the resulting bispectrum and trispectrum phenomenology.

\section{Random fields during inflation}
\label{sec:physical}

We take the spatial pattern of the disorder to be quenched while allowing its overall amplitude to carry a deterministic power-law weight,
\begin{equation}
  h(\bm x,\eta)=(-H\eta)^p h(\bm x).
  \label{eq:timeWeightedDisorder}
\end{equation}
The persistent random field is the special case $p=0$.  For real $p>0$ the forcing vanishes toward the future boundary.  The power-law family is useful because it remains exactly solvable and provides a Mellin basis for more general deterministic time profiles.  This differs from the quenched disorder studied in the effective field theory of inflation in Ref.~\cite{Green:2014disorder}, where the stochastic parameters are functions of time and are independent of position.  Here the stochastic variable is always the spatial field $h(\bm x)$; the factor $(-H\eta)^p$ is deterministic.

During inflation, the slow-roll parameters are parametrically small, so that, at leading order in the slow-roll, the inflationary background can be approximated by \(dS_4\) \cite{Maldacena:2002vr,Chen:2009zp}. Then, we take
\begin{equation}
  S[\sigma;h]
  =\frac12\int\dd\eta\,\dd^3x\,
  a^2\left[(\sigma')^2-(\bm\nabla\sigma)^2-a^2m^2\sigma^2\right]
  +\int\dd\eta\,\dd^3x\,a^4(-H\eta)^p h(\bm x)\sigma(\eta,\bm x),
  \label{eq:physicalaction}
\end{equation}
with
\begin{equation}
  a(\eta)=-\frac{1}{H\eta}.
\end{equation}
The disorder ensemble is specified by
\begin{equation}
  \avg{h(\bm k)}=0,
  \qquad
  \avg{h(\bm k)h(\bm k')}
  =(2\pi)^3\delta^{(3)}(\bm k+\bm k')f(k).
  \label{eq:hkvariance}
\end{equation}
With the Fourier convention used here, $f(k)$ has mass dimension three.  It is therefore convenient to parameterize spatial white disorder as
\begin{equation}
  f(k)=f_0 H^3,\qquad [f_0]=0,
  \label{eq:whiteNoiseNormalization}
\end{equation}
whereas a momentum-dependent $f(k)$ allows the random environment itself to carry a nontrivial spatial scale dependence.  The white-noise choice is understood as an idealized fixed-momentum benchmark: in position space it corresponds to a delta-correlated random distribution, whereas a microscopic realization would generically possess a finite correlation length, represented here by a smooth ultraviolet-regulated $f(k)$.  No coincident-point disorder observable is required below.  We keep $f(k)$ general in the derivation of the disorder-averaged propagators and specialize to \cref{eq:whiteNoiseNormalization} only when discussing the cosmological-collider seed and its phenomenology.

For a fixed realization of the random field, the Fourier modes obey
\begin{equation}
  \sigma_{\bm k}''
  -\frac{2}{\eta}\sigma_{\bm k}'
  +\left(k^2+\frac{m^2}{H^2\eta^2}\right)\sigma_{\bm k}
  =\frac{(-H\eta)^p h_{\bm k}}{H^2\eta^2}.
  \label{eq:EOMsource}
\end{equation}
The corresponding homogeneous Bunch--Davies mode is
\begin{equation}
  u_k(\eta)
  =\frac{\sqrt\pi}{2}
  H\,\ee^{\ii\pi(2\nu+1)/4}
  (-\eta)^{3/2}H_\nu^{(1)}(-k\eta),
  \qquad
  \nu^2=\frac94-\frac{m^2}{H^2}.
  \label{eq:BDmode}
\end{equation}
For a principal-series field, $\nu=\ii\mu$ with $\mu=\sqrt{m^2/H^2-9/4}$.

The retarded Green function is
\begin{equation}
  G_R(k;\eta,\eta')
  =\ii\,\theta(\eta-\eta')
  \left[
  u_k(\eta)u_k^*(\eta')
  -u_k^*(\eta)u_k(\eta')
  \right],
  \label{eq:GR}
\end{equation}
with
\begin{equation}
  \left[
  \partial_\eta^2-\frac{2}{\eta}\partial_\eta
  +k^2+\frac{m^2}{H^2\eta^2}
  \right]G_R(k;\eta,\eta')
  =\frac{\delta(\eta-\eta')}{a^2(\eta')}.
  \label{eq:GRnormalization}
\end{equation}
Since the source enters linearly, the exact field for each realization separates into the free quantum field and a deterministic response,
\begin{equation}
  \sigma_{\bm k}(\eta)
  =\sigma^{(0)}_{\bm k}(\eta)
  +h_{\bm k}R_{k,p}(\eta),
  \qquad
  R_{k,p}(\eta)=
  \int_{-\infty(1-\ii\epsilon)}^{\eta}
  \dd\eta'\,a^4(\eta')(-H\eta')^pG_R(k;\eta,\eta').
  \label{eq:Rdef}
\end{equation}
We impose Bunch--Davies initial conditions for the massive field and describe the response to the external source through the retarded Green function. As usual in the in--in formalism, the integration contour in the asymptotic past is slightly deformed, so that the oscillatory time integrals are well defined. For a time-independent source, \(p=0\), this prescription has the standard early-time behavior. In particular, for \(p>0\) the source grows toward the far past, and the resulting oscillatory integral is understood through the same \(i\epsilon\) prescription, or equivalently by analytic continuation from the region where the integral converges. The power-law profile should therefore be viewed as an analytically tractable component of the time-dependent source rather than as a physical source extending unchanged to arbitrarily early times.

\subsection{Exact response in terms of Lommel functions}
\label{subsec:lommel}

Substituting $\sigma_{\bm k}=\sigma^{(0)}_{\bm k}+h_{\bm k}R_{k,p}$ into \cref{eq:EOMsource} gives
\begin{equation}
  R_{k,p}''-\frac{2}{\eta}R_{k,p}'
  +\left(k^2+\frac{m^2}{H^2\eta^2}\right)R_{k,p}
  =\frac{(-H\eta)^p}{H^2\eta^2}.
  \label{eq:Reqeta}
\end{equation}
Introduce
\begin{equation}
  x\equiv-k\eta>0,
  \qquad
  M^2\equiv\frac{m^2}{H^2},
  \qquad
  \Delta_p\equiv M^2+p(p-3)
  =\left(p-\frac32\right)^2-\nu^2.
  \label{eq:xdef}
\end{equation}
The response equation becomes
\begin{equation}
  \frac{\dd^2R_{k,p}}{\dd x^2}
  -\frac{2}{x}\frac{\dd R_{k,p}}{\dd x}
  +\left(1+\frac{M^2}{x^2}\right)R_{k,p}
  =\left(\frac{H}{k}\right)^p\frac{x^{p-2}}{H^2}.
  \label{eq:Reqx}
\end{equation}
Writing
\begin{equation}
  R_{k,p}(x)
  =\left(\frac{H}{k}\right)^p\frac{x^{3/2}}{H^2}F_{p,\nu}(x)
  \label{eq:RtoF}
\end{equation}
gives
\begin{equation}
  x^2F_{p,\nu}''+xF_{p,\nu}'+(x^2-\nu^2)F_{p,\nu}
  =x^{p-3/2}.
  \label{eq:LommelEqMultiplied}
\end{equation}
Equivalently,
\begin{equation}
  F_{p,\nu}''+\frac1xF_{p,\nu}'
  +\left(1-\frac{\nu^2}{x^2}\right)F_{p,\nu}
  =x^{p-7/2}.
  \label{eq:LommelEq}
\end{equation}
This is the Lommel equation with
\begin{equation}
  \rho=p-\frac52.
\end{equation}
The general solution is
\begin{equation}
  F_{p,\nu}(x)
  =s_{p-5/2,\nu}(x)+c_1J_\nu(x)+c_2Y_\nu(x).
  \label{eq:LommelGeneral}
\end{equation}
The retarded prescription fixes the homogeneous admixture.  The second Lommel function obeys
\begin{equation}
  S_{\rho,\nu}(x)\sim x^{\rho-1}\left[1+\mathcal O(x^{-2})\right],
  \qquad x\to\infty,
  \label{eq:LommelLargeX}
\end{equation}
so that
\begin{equation}
  R_{k,p}(x)
  \sim
  \left(\frac{H}{k}\right)^p\frac{x^{p-2}}{H^2},
  \qquad x\to\infty,
\end{equation}
matching the adiabatic response to the weighted source.  The exact retarded solution is therefore
\begin{equation}
  R_{k,p}(\eta)
  =\left(\frac{H}{k}\right)^p
  \frac{(-k\eta)^{3/2}}{H^2}
  S_{p-\frac52,\nu}(-k\eta).
  \label{eq:Rlommel}
\end{equation}
Setting $p=0$ immediately reproduces the persistent response $x^{3/2}S_{-5/2,\nu}/H^2$.  A direct Mellin--Barnes derivation for arbitrary $p$ is given in Appendix~\ref{app:MBresponse}.

For later use define
\begin{align}
  \mathcal C_{p,\nu}
  \equiv{}&2^{p-7/2}
  \Gamma\left(\frac p2-\frac34+\frac\nu2\right)
  \Gamma\left(\frac p2-\frac34-\frac\nu2\right),
  \\
  \theta_{p,\nu}
  \equiv{}&\frac\pi2\left(p-\frac52-\nu\right).
  \label{eq:Ctheta}
\end{align}
The Lommel connection formula gives
\begin{equation}
  S_{p-\frac52,\nu}(x)
  =s_{p-\frac52,\nu}(x)
  +\mathcal C_{p,\nu}
  \left[
  \sin\theta_{p,\nu}\,J_\nu(x)
  -\cos\theta_{p,\nu}\,Y_\nu(x)
  \right].
  \label{eq:SminusS}
\end{equation}
The particular solution is
\begin{equation}
  s_{p-\frac52,\nu}(x)
  =\frac{x^{p-3/2}}{\Delta_p}
  {}_1F_2\left(
  1;
  \frac p2+\frac14-\frac\nu2,
  \frac p2+\frac14+\frac\nu2;
  -\frac{x^2}{4}
  \right).
  \label{eq:smallLommelHyper}
\end{equation}
Hence
\begin{align}
  R_{k,p}(x)
  ={}&\left(\frac{H}{k}\right)^p
  \frac{x^p}{H^2\Delta_p}
  {}_1F_2\left(
  1;
  \frac p2+\frac14-\frac\nu2,
  \frac p2+\frac14+\frac\nu2;
  -\frac{x^2}{4}
  \right)
  \nonumber\\
  &+\left(\frac{H}{k}\right)^p
  \frac{\mathcal C_{p,\nu}}{H^2}x^{3/2}
  \left[
  \sin\theta_{p,\nu}\,J_\nu(x)
  -\cos\theta_{p,\nu}\,Y_\nu(x)
  \right].
  \label{eq:Rexplicit}
\end{align}
For a fixed realization,
\begin{equation}
  \sigma_{\bm k}(\eta)
  =a_{\bm k}u_k(\eta)
  +a_{-\bm k}^\dagger u_k^*(\eta)
  +h_{\bm k}R_{k,p}(\eta).
  \label{eq:fullDisorderedField}
\end{equation}

\subsection{Late-time response and massive oscillations}
\label{subsec:lateLommel}

For generic parameters, the exact response has the late-time expansion
\begin{equation}
  \begin{aligned}
  R_{k,p}(x)
  =\left(\frac{H}{k}\right)^p\Bigg\{&
  \frac{x^p}{H^2\Delta_p}
  \left[
  1-\frac{x^2}{M^2+(p+2)(p-1)}+\mathcal O(x^4)
  \right]
  \\
  &+\mathcal A_{+,p}(\nu)x^{3/2+\nu}
  +\mathcal A_{-,p}(\nu)x^{3/2-\nu}
  +\cdots\Bigg\},
  \end{aligned}
  \label{eq:Rlate}
\end{equation}
where
\begin{align}
  \mathcal A_{+,p}(\nu)
  ={}&\frac{\mathcal C_{p,\nu}}{H^2}
  \frac{2^{-\nu}}{\Gamma(1+\nu)}
  \left[
  \sin\theta_{p,\nu}-\cos\theta_{p,\nu}\cot(\pi\nu)
  \right],
  \label{eq:Aplus}\\
  \mathcal A_{-,p}(\nu)
  ={}&\frac{\mathcal C_{p,\nu}}{H^2}
  \frac{2^{\nu}}{\Gamma(1-\nu)}
  \cos\theta_{p,\nu}\,\csc(\pi\nu).
  \label{eq:Aminus}
\end{align}
The first line is the analytic forced response and the last two terms are the homogeneous massive tails.  In terms of conformal time the leading forced term is
\begin{equation}
  R_{k,p}^{\rm forced}(\eta)
  \sim
  \frac{(-H\eta)^p}{H^2\Delta_p}.
  \label{eq:RweightedLate}
\end{equation}
Thus $p=0$ gives $R_{k,0}\to1/m^2$, whereas every real $p>0$ removes the static offset at the future boundary.

For the principal series, $\nu=\ii\mu$, one has
\begin{equation}
  \Delta_p=\left(p-\frac32\right)^2+\mu^2>0,
\end{equation}
and reality implies $\mathcal A_{-,p}(\ii\mu)=\mathcal A_{+,p}(\ii\mu)^*$.~The temporal weight changes the amplitude and phase of the clock through $\mathcal C_{p,\nu}$ and $\theta_{p,\nu}$, but not its frequency $\mu$ or its de Sitter exponents $3/2\pm\ii\mu$.

The generic expansion should be analytically continued when a forced power collides with a homogeneous Frobenius exponent.  The resonance condition is
\begin{equation}
  p+2n=\frac32\pm\nu,
  \qquad n=0,1,2,\ldots .
  \label{eq:generalResonance}
\end{equation}
At such values the separated coefficients develop compensating poles and the degenerate pair is replaced by a logarithm, while the complete Lommel solution remains finite.  The familiar persistent examples are recovered by setting $p=0$.  In particular, $m=0$ ($\nu=3/2$) produces a collision between the constant forced term and the slow homogeneous branch, while $m^2=2H^2$ ($\nu=1/2$) produces a collision between the $x^2$ forced term and the fast homogeneous branch.  Their explicit $p=0$ limits are
\begin{align}
  R_{k,0}^{(m=0)}(x)
  ={}&\frac{1}{3H^2}
  \Bigg\{
  1-
  \bigl(\cos x+x\sin x\bigr)\operatorname{Ci}(x)\nonumber
  \\[-1mm]
  &\hspace{19mm}
  +\bigl(x\cos x-\sin x\bigr)
  \left[\operatorname{Si}(x)-\frac{\pi}{2}\right]
  \Bigg\},
  \label{eq:RmasslessExact}\\
  R_{k,0}^{(m=0)}(x)
  ={}&\frac{1-\gamma_{\rm E}-\log x}{3H^2}
  +\mathcal O(x^2\log x),
  \label{eq:RmasslessLate}
\end{align}
and
\begin{align}
  R_{k,0}^{(m^2=2H^2)}(x)
  ={}&\frac{1}{2H^2}
  \left\{
  1-x\sin x\,\operatorname{Ci}(x)
  +x\cos x\left[\operatorname{Si}(x)-\frac{\pi}{2}\right]
  \right\},
  \label{eq:RconformalExact}\\
  R_{k,0}^{(m^2=2H^2)}(x)
  ={}&\frac{1}{2H^2}
  -\frac{\pi}{4H^2}x
  +\frac{x^2}{2H^2}\left(1-\gamma_{\rm E}-\log x\right)
  +\mathcal O(x^3).
  \label{eq:RconformalLate}
\end{align}
All the above cases can contribute to a different signal in the cosmological collider template.~The endpoint pathology corresponds to \(p=0\); however, the trispectrum seed remains completely finite.

\subsection{Persistent limit and finite-time weights}
\label{subsec:persistentTransient}

The persistent model is now simply the point $p=0$ of the exact family.  Equation~\eqref{eq:RweightedLate} then gives
\begin{equation}
  R_{k,0}(0)=\frac1{m^2},
  \label{eq:persistentStatic}
\end{equation}
and no homogeneous boundary-condition choice can cancel this constant for a principal-series field because the homogeneous modes vanish as $(-k\eta)^{3/2\pm\ii\mu}$.

For $p>0$ the forced response vanishes at late times while the homogeneous massive tails remain.  This is enough to improve the $\ell=-2$ physical Schwinger--Keldysh endpoint.  Near $\eta=0$,
\begin{equation}
  \int^{0}\dd\eta\,(-\eta)^{-2}
  \left(e^{\ii K\eta}-e^{-\ii K\eta}\right)R_{s,p}(\eta)
  \sim
  \int^{0}\dd\eta\,(-\eta)^{p-1},
  \label{eq:ellminus2weightedScaling}
\end{equation}
which is integrable for every $p>0$ in the principal series.  The special value $p=0$ gives the logarithm $\int\dd\eta/(-\eta)$.

The pure power law is an exactly solvable prototype rather than the only possible history.  If a deterministic envelope admits a Mellin representation
\begin{equation}
  g(y)=\int_{\mathcal C_p}\frac{\dd p}{2\pi\ii}\,\widetilde g(p)y^p,
  \qquad y=-H\eta,
  \label{eq:gMellinSuperposition}
\end{equation}
then linearity gives the response as the superposition
\begin{equation}
  R_{k,g}(\eta)
  =\int_{\mathcal C_p}\frac{\dd p}{2\pi\ii}\,
  \widetilde g(p)R_{k,p}(\eta).
  \label{eq:RMellinSuperposition}
\end{equation}
A profile that switches off completely at a finite time subsequently evolves purely homogeneously,
\begin{equation}
  R_{k,g}(\eta)
  =B_+(k)(-k\eta)^{3/2+\nu}
  +B_-(k)(-k\eta)^{3/2-\nu}+\cdots,
  \label{eq:Rmemory}
\end{equation}
so the collider oscillation can be viewed as the memory left after the forcing has disappeared.  The power-law family provides the analytic building blocks for this more general construction.

\subsection{Disorder-averaged correlators}
\label{subsec:disordercorr}

For a fixed realization,
\begin{equation}
  \vev{\sigma_{\bm k}(\eta)}_h=h_{\bm k}R_{k,p}(\eta).
\end{equation}
The ensemble-averaged one-point function vanishes because $\avg{h}=0$, whereas fluctuations of the realization-dependent one-point function survive in the two-point function.  Using \cref{eq:hkvariance},
\begin{align}
  \avg{\vev{\sigma_{\bm k}(\eta_1)\sigma_{\bm k'}(\eta_2)}_h}
  &=(2\pi)^3\delta^{(3)}(\bm k+\bm k')
  \left[
  G_0(k;\eta_1,\eta_2)
  +f(k)R_{k,p}(\eta_1)R_{k,p}(\eta_2)
  \right].
\end{align}
Hence
\begin{equation}
  G_{\rm dis}(k;\eta_1,\eta_2)
  =G_0(k;\eta_1,\eta_2)
  +f(k)R_{k,p}(\eta_1)R_{k,p}(\eta_2).
  \label{eq:Gdis}
\end{equation}
The second term is the variance of the disorder-induced expectation value \footnote{Since $\langle h(\mathbf{k})h(\mathbf{k}')\rangle=(2\pi)^3\delta^{(3)}(\mathbf{k}+\mathbf{k}')f(k)$ implies $[f]=3$, the white-noise parameterization $f(k)=f_0H^3$ isolates the dimensionless disorder strength $f_0$.}.  It is disconnected with respect to quantum fluctuations at fixed realization, where $\langle\sigma\rangle_h=hR$, but it is correlated in the combined quantum-plus-disorder ensemble because the same quenched realization appears at both insertions.  This distinction is precisely what allows $fR_pR_p$ to act as an exchange-like statistical contribution after the disorder average.  Since the random source is additive and Gaussian, it modifies the statistical two-point function but does not shift the poles of the retarded propagator at this linear level.

At equal time,
\begin{equation}
  P_\sigma^{\rm dis}(k,\eta)
  =P_\sigma^{(0)}(k,\eta)+f(k)R_{k,p}(\eta)^2,
  \label{eq:powerdis}
\end{equation}
where $P_\sigma^{(0)}=|u_k|^2$.  Using \cref{eq:Rlommel},
\begin{equation}
  P_\sigma^{\rm dis}(k,\eta)
  =|u_k(\eta)|^2
  +\frac{f(k)}{H^4}
  \left(\frac{H}{k}\right)^{2p}
  (-k\eta)^3
  S_{p-\frac52,\nu}(-k\eta)^2.
  \label{eq:powerLommel}
\end{equation}
At late times the forced contribution behaves as
\begin{equation}
  P_\sigma^{\rm dis}-P_\sigma^{(0)}
  \supset
  \frac{f(k)(-H\eta)^{2p}}{H^4\Delta_p^2},
\end{equation}
which reduces to $f(k)/m^4$ at $p=0$ and vanishes for $p>0$.  The nonanalytic terms retain the massive oscillatory response of \cref{eq:Rlate}.

\section{Direct Schwinger--Keldysh derivation in dS}
\label{sec:SK}

The result in \cref{eq:Gdis} can be derived directly in the real-time in-in formalism.  This derivation is useful for two reasons.  First, it makes explicit which part of the Schwinger--Keldysh propagator is modified by the random field.  Second, it clarifies the relation between the usual contour basis, labelled by $+$ and $-$, and the Keldysh basis, in which causal response and statistical fluctuations are separated.  We will move between these two bases below; they are related by a linear change of variables and contain exactly the same information.  Replica constructions for in-in observables in quantum field theories with random potentials were discussed in Ref.~\cite{Jain:2015random}.  More recently, Schwinger--Keldysh methods have also been applied to wave propagation in static random backgrounds in Ref.~\cite{Amoruso:2026gw}. Related recent applications have also used the Schwinger--Keldysh formalism to treat prescribed classical stochastic backgrounds in de Sitter space~\cite{VicenteGarcia-Consuegra:2026vaf}. Our purpose here is narrower: to exploit the exact solvability of a linearly forced massive field in de Sitter and derive its disorder-averaged propagators explicitly.

For a fixed realization $h(\bm x)$ with deterministic weight $(-H\eta)^p$, the in-in generating functional contains a forward and backward copy of the field,
\begin{equation}
  \begin{aligned}
  Z_h[J_+,J_-]
  ={}&\int\cD\sigma_+\cD\sigma_-\,
  \exp\Bigg\{
  \ii S_0[\sigma_+]-\ii S_0[\sigma_-]
  \\
  &\hspace{14mm}
  +\ii\int\dd\eta\,\dd^3x\,a^4(-H\eta)^p h(\sigma_+-\sigma_-)
  +\ii\int\dd\eta\,\dd^3x\,(J_+\sigma_+-J_-\sigma_-)
  \Bigg\}.
  \end{aligned}
  \label{eq:SKfixedh}
\end{equation}
The labels $+$ and $-$ refer to the forward and backward branches of the closed-time contour.  They are not replica indices.  The same realization of $h$ appears on both branches because the bra and ket describe the same physical universe.  In particular, unitarity implies $Z_h[J,J]=1$ for each fixed realization when the two sources are identified.

The stochastic amplitude is quenched in comoving coordinates; its only time dependence is the deterministic factor $(-H\eta)^p$.  Its coupling to the contour fields therefore has the characteristic difference form
\begin{equation}
  \int\dd\eta\,a^4(\eta)(-H\eta)^p h(\bm x)
  \left[\sigma_+(\eta,\bm x)-\sigma_-(\eta,\bm x)\right].
  \label{eq:SKdifferencecoupling}
\end{equation}
This observation is the starting point for the Keldysh rotation.

\subsection{From the contour basis to the Keldysh basis}
\label{subsec:Keldyshrotation}

The usual Schwinger--Keldysh basis is the contour basis $(\sigma_+,\sigma_-)$.  It is the natural basis for diagrammatic in-in perturbation theory because every interaction vertex is assigned to one of the two branches.  For the present problem, however, it is more convenient to reorganize the same degrees of freedom into the average and difference fields
\begin{equation}
  \boxed{
  \sigma_c=\frac12(\sigma_++\sigma_-),
  \qquad
  \sigma_q=\sigma_+-\sigma_- .}
  \label{eq:Keldyshrotation}
\end{equation}
The inverse transformation is
\begin{equation}
  \boxed{
  \sigma_+=\sigma_c+\frac12\sigma_q,
  \qquad
  \sigma_-=\sigma_c-\frac12\sigma_q .}
  \label{eq:Keldyshinverse}
\end{equation}
Thus no approximation is involved: the contour and Keldysh bases are simply two linear bases for the same doubled field space.  In matrix notation,
\begin{equation}
  \begin{pmatrix}\sigma_c\\ \sigma_q\end{pmatrix}
  =
  \begin{pmatrix}
  \frac12 & \frac12\\
  1 & -1
  \end{pmatrix}
  \begin{pmatrix}\sigma_+\\ \sigma_-\end{pmatrix}.
  \label{eq:Keldyshmatrixrotation}
\end{equation}
It is useful to rotate the sources at the same time,
\begin{equation}
  J_c=\frac12(J_++J_-),
  \qquad
  J_q=J_+-J_- ,
\end{equation}
for which
\begin{equation}
  J_+\sigma_+-J_-\sigma_-
  =J_q\sigma_c+J_c\sigma_q.
  \label{eq:sourcerotation}
\end{equation}
The $c/q$ basis is particularly useful because it separates response from fluctuations; see, for example, Ref.~\cite{Kamenev:2009jj} for a pedagogical discussion of the Keldysh formalism.

For the quadratic spectator action, integration by parts gives, up to the usual boundary and initial-state terms,
\begin{equation}
  S_0[\sigma_+]-S_0[\sigma_-]
  =-
  \int\dd\eta\,\dd^3k\,
  a^2(\eta)\,
  \sigma_q(-\bm k,\eta)\,
  \cD_k\sigma_c(\bm k,\eta),
  \label{eq:cleanKeldyshaction}
\end{equation}
where
\begin{equation}
  \cD_k
  \equiv
  \partial_\eta^2-\frac{2}{\eta}\partial_\eta
  +k^2+\frac{m^2}{H^2\eta^2}.
  \label{eq:DoperatorSK}
\end{equation}
Equation~\eqref{eq:SKdifferencecoupling} becomes simply
\begin{equation}
  S_h^{\rm SK}
  =\int\dd\eta\,\dd^3x\,a^4(\eta)(-H\eta)^p h(\bm x)\sigma_q(\eta,\bm x).
  \label{eq:hqcoupling}
\end{equation}
Thus the random field couples only to the Keldysh difference field $\sigma_q$.  This is why the Keldysh basis is the natural basis in which to perform the disorder average.

\subsection{Disorder average and the Keldysh noise kernel}
\label{subsec:noisekernel}

We first suppress replicas, since the Gaussian average can be performed directly for normalized in-in observables.  In momentum space, the disorder is characterized by
\begin{equation}
  \avg{h_{\bm k}h_{\bm k'}}
  =(2\pi)^3\delta^{(3)}(\bm k+\bm k')f(k).
\end{equation}
Averaging the exponential of \cref{eq:hqcoupling} gives
\begin{align}
  &\avg{
  \exp\left[
  \ii\int\dd\eta\int_{\bm k}
  a^4(\eta)(-H\eta)^p h_{\bm k}\sigma_q(\eta,-\bm k)
  \right]}
  \nonumber\\
  &\qquad =
  \exp\left[
  -\frac12\int_{\bm k}\int\dd\eta\,\dd\eta'\,
  \sigma_q(\eta,\bm k)
  N_{\rm dis}^{(p)}(k;\eta,\eta')
  \sigma_q(\eta',-\bm k)
  \right],
  \label{eq:noisekernelinfluence}
\end{align}
where
\begin{equation}
  \boxed{
  N_{\rm dis}^{(p)}(k;\eta,\eta')
  =f(k)a^4(\eta)a^4(\eta')(-H\eta)^p(-H\eta')^p.}
  \label{eq:noisekernel}
\end{equation}
The notation $N_{\rm dis}^{(p)}$ is useful because a quadratic $qq$ term has precisely the form of a Keldysh noise kernel.  The underlying randomness here is nevertheless quenched rather than dynamical: $h(\bm x)$ is fixed, and no thermal bath or fluctuation--dissipation relation is being assumed.

The temporal structure of \cref{eq:noisekernel} is especially simple.  If
\begin{equation}
  v_p(\eta)\equiv a^4(\eta)(-H\eta)^p,
\end{equation}
then
\begin{equation}
  N_{\rm dis}^{(p)}(k;\eta,\eta')
  =f(k)v_p(\eta)v_p(\eta').
  \label{eq:rankonetime}
\end{equation}
Hence the disorder kernel is rank one in time: the same random variable $h_{\bm k}$ acts throughout the entire cosmological history.  This is the real-time manifestation of the temporal nonlocality characteristic of quenched disorder~\cite{Aharony:2018mjm}.

The Bunch--Davies initial density matrix supplies the clean statistical fluctuations of the field.  We leave the corresponding initial-state kernel implicit below and denote the resulting clean statistical correlator by $F_0$.  The additional term in \cref{eq:noisekernel} is the only new quadratic kernel generated by the random field.

\subsection{Retarded, advanced and statistical propagators}
\label{subsec:Keldyshpropagators}

To make the physical content of the Keldysh basis explicit, it is convenient to parametrize the two-point functions in terms of the statistical and spectral correlators.
\begin{equation}
  F(k;\eta,\eta')
  \equiv
  \frac12\avg{\vev{
  \{\sigma_{\bm k}(\eta),\sigma_{-\bm k}(\eta')\}
  }_h},
  \label{eq:Fdef}
\end{equation}
\begin{equation}
  \rho(k;\eta,\eta')
  \equiv
  \ii\,\avg{\vev{
  [\sigma_{\bm k}(\eta),\sigma_{-\bm k}(\eta')]
  }_h}.
  \label{eq:rhodef}
\end{equation}
With the convention already used in \cref{eq:GR},
\begin{equation}
  G_R(k;\eta,\eta')
  =\theta(\eta-\eta')\rho(k;\eta,\eta'),
  \qquad
  G_A(k;\eta,\eta')
  =-\theta(\eta'-\eta)\rho(k;\eta,\eta').
  \label{eq:GRGAfromrho}
\end{equation}
The spectral function controls causal propagation and the location of the homogeneous poles, whereas $F$ contains the statistical information about the state and, in the present problem, the realization-to-realization fluctuations induced by the disorder.

There is an immediate way to see which of these objects can change.  For a fixed realization, \cref{eq:Rdef} gives
\begin{equation}
  \sigma_{\bm k}(\eta)
  =\sigma^{(0)}_{\bm k}(\eta)+h_{\bm k}R_{k,p}(\eta).
  \label{eq:fixedhshiftSK}
\end{equation}
The second term is a $c$-number.  Therefore it drops out of the commutator,
\begin{equation}
  [\sigma_{\bm k}(\eta),\sigma_{-\bm k}(\eta')]
  =
  [\sigma^{(0)}_{\bm k}(\eta),\sigma^{(0)}_{-\bm k}(\eta')],
\end{equation}
and hence
\begin{equation}
  \boxed{
  \rho_{\rm dis}=\rho_0,
  \qquad
  G_R^{\rm dis}=G_R^{(0)},
  \qquad
  G_A^{\rm dis}=G_A^{(0)}.}
  \label{eq:unchangedRA}
\end{equation}
This gives a direct real-time explanation of why a linear Gaussian random field does not shift the poles of the massive retarded propagator.

The statistical correlator does change.  Using \cref{eq:fixedhshiftSK} and $\avg{h}=0$,
\begin{equation}
  \boxed{
  F_{\rm dis}(k;\eta,\eta')
  =F_0(k;\eta,\eta')
  +f(k)R_{k,p}(\eta)R_{k,p}(\eta').}
  \label{eq:Fdisdirect}
\end{equation}
The same result follows directly from the Gaussian Keldysh kernel.  A quadratic $qq$ noise kernel generates the standard statistical correction
\begin{equation}
  \Delta F
  =G_R\circ N_{\rm dis}^{(p)}\circ G_A,
  \label{eq:GRNGA}
\end{equation}
where $\circ$ denotes integration over intermediate times.  In the present case,
\begin{align}
  \Delta F(k;\eta,\eta')
  ={}&f(k)
  \int\dd\tau\,G_R(k;\eta,\tau)a^4(\tau)(-H\tau)^p
  \nonumber\\
  &\times
  \int\dd\tau'\,a^4(\tau')(-H\tau')^pG_A(k;\tau',\eta').
  \label{eq:GRNGAexpanded}
\end{align}
Using
\begin{equation}
  R_{k,p}(\eta)
  =\int\dd\tau\,a^4(\tau)(-H\tau)^pG_R(k;\eta,\tau),
\end{equation}
and $G_A(k;\tau,\eta')=G_R(k;\eta',\tau)$ for the real scalar field, one obtains
\begin{equation}
  \boxed{
  \Delta F(k;\eta,\eta')
  =f(k)R_{k,p}(\eta)R_{k,p}(\eta').}
  \label{eq:FdisKeldysh}
\end{equation}
Thus the fixed-realization solution and the Keldysh kernel inversion give exactly the same result,
\begin{equation}
  \boxed{
  F_{\rm dis}
  =F_0+G_R\circ N_{\rm dis}^{(p)}\circ G_A
  =F_0+fR_pR_p.}
  \label{eq:Keldyshmaster}
\end{equation}
The factorization in the last expression is not accidental: it follows from the rank-one temporal structure \cref{eq:rankonetime}.

For later comparison with the contour basis, it is useful to collect the correlators of the $c/q$ fields.  With the normalization in \cref{eq:Keldyshrotation}, the contour identities imply
\begin{equation}
  \boxed{
  \begin{pmatrix}
  \vev{\sigma_c(\eta)\sigma_c(\eta')}
  &
  \vev{\sigma_c(\eta)\sigma_q(\eta')}
  \\
  \vev{\sigma_q(\eta)\sigma_c(\eta')}
  &
  \vev{\sigma_q(\eta)\sigma_q(\eta')}
  \end{pmatrix}
  =
  \begin{pmatrix}
  F(\eta,\eta') & -\ii G_R(\eta,\eta')\\
  -\ii G_A(\eta,\eta') & 0
  \end{pmatrix}.}
  \label{eq:Keldyshmatrix}
\end{equation}
Equation~\eqref{eq:Keldyshmatrix} makes the separation transparent: the random field modifies the $cc$ or statistical component through \cref{eq:FdisKeldysh}, while the causal $cq$ and $qc$ components remain unchanged.

Replicas are not required to perform the basic Gaussian average of a normalized in-in correlator, because $Z_h[J,J]=1$ for each fixed realization.  They are nevertheless useful as an organizational device for quenched observables, especially for separating the disorder-averaged connected correlator from products of realization-dependent one-point functions.

Introduce $n$ copies labelled by $r=1,\ldots,n$ and average all of them over the same $h$.  In the Keldysh basis the influence factor is
\begin{equation}
  \mathcal F_{\rm dis}
  =\exp\left[
  -\frac12\sum_{r,r'=1}^n
  \int_{\bm k}\int\dd\eta\,\dd\eta'\,
  \sigma_{q,r}(\eta,\bm k)
  N^{rr'}_{{\rm dis},p}(k;\eta,\eta')
  \sigma_{q,r'}(\eta',-\bm k)
  \right],
  \label{eq:replicanoiseaction}
\end{equation}
with
\begin{equation}
  \boxed{
  N^{rr'}_{{\rm dis},p}(k;\eta,\eta')
  =f(k)a^4(\eta)a^4(\eta')(-H\eta)^p(-H\eta')^p
  \equiv N_{\rm dis}^{(p)}(k;\eta,\eta').}
  \label{eq:replicanoisekernel}
\end{equation}
Every matrix element in replica space is identical.  Thus
\begin{equation}
  \mathbb N_{{\rm dis},p}
  =N_{\rm dis}^{(p)}\,\mathbb J_n,
  \label{eq:noiseJn}
\end{equation}
where $(\mathbb J_n)_{rr'}=1$.  Since $\mathbb J_n$ has one eigenvalue $n$ and $n-1$ vanishing eigenvalues, only the normalized singlet
\begin{equation}
  \sigma_{q,0}
  =\frac1{\sqrt n}\sum_{r=1}^n\sigma_{q,r}
  \label{eq:replicasingletq}
\end{equation}
feels the disorder.  All replica-transverse directions remain clean.  In other words, the disorder acts only in the sector
\begin{equation}
  \boxed{
  \text{replica singlet}\ \otimes\ \text{Keldysh quantum mode}.}
  \label{eq:rankoneSK}
\end{equation}

The propagators make this statement more concrete.  In the singlet/transverse basis,
\begin{equation}
  F_{00}=F_0+n f(k)R_{k,p}(\eta)R_{k,p}(\eta'),
  \qquad
  F_{\perp}=F_0,
  \label{eq:singletF}
\end{equation}
while the retarded and advanced propagators are the clean ones in every replica channel.  Transforming back to the original replica basis gives
\begin{equation}
  \boxed{
  F^{rr'}(k;\eta,\eta')
  =\delta^{rr'}F_0(k;\eta,\eta')
  +f(k)R_{k,p}(\eta)R_{k,p}(\eta'),}
  \label{eq:replicaF}
\end{equation}
\begin{equation}
  \boxed{
  G_R^{rr'}=\delta^{rr'}G_R^{(0)},
  \qquad
  G_A^{rr'}=\delta^{rr'}G_A^{(0)}.}
  \label{eq:replicaRA}
\end{equation}
The diagonal element $r=r'$ is the disorder-averaged physical two-point function.  For $r\neq r'$, the clean quantum contraction is absent and
\begin{equation}
  F^{r\neq r'}(k;\eta,\eta')
  =f(k)R_{k,p}(\eta)R_{k,p}(\eta')
  =\avg{
  \vev{\sigma_{\bm k}(\eta)}_h
  \vev{\sigma_{-\bm k}(\eta')}_h},
  \label{eq:offdiagonalreplica}
\end{equation}
so the off-diagonal replica correlator isolates the sample-to-sample fluctuations induced by the random field.

\subsection{Back to the $+ / -$ contour basis}
\label{subsec:backtoSK}

The Keldysh basis is optimal for identifying the noise and response structure, but cosmological in-in diagrams are usually written in the original $\pm$ contour basis.  We therefore rotate back before entering the collider calculation.  Define the Wightman correlators
\begin{equation}
  G^>(k;\eta,\eta')
  \equiv
  \avg{\vev{\sigma_{\bm k}(\eta)\sigma_{-\bm k}(\eta')}_h},
  \qquad
  G^<(k;\eta,\eta')
  \equiv
  \avg{\vev{\sigma_{-\bm k}(\eta')\sigma_{\bm k}(\eta)}_h}.
  \label{eq:WightmanDefs}
\end{equation}
In terms of $F$ and $\rho$,
\begin{equation}
  G^>=F-\frac{\ii}{2}\rho,
  \qquad
  G^<=F+\frac{\ii}{2}\rho.
  \label{eq:WightmanFrho}
\end{equation}
The four contour propagators are
\begin{align}
  G_{++}(\eta,\eta')
  &\equiv
  \theta(\eta-\eta')G^>
  +\theta(\eta'-\eta)G^<,
  \\
  G_{--}(\eta,\eta')
  &\equiv
  \theta(\eta'-\eta)G^>
  +\theta(\eta-\eta')G^<,
  \\
  G_{-+}(\eta,\eta')&\equiv G^>,
  \qquad
  G_{+-}(\eta,\eta')\equiv G^<.
  \label{eq:fourSKdefs}
\end{align}
Equivalently,
\begin{equation}
  G_{++}
  =F-\frac{\ii}{2}\operatorname{sgn}(\eta-\eta')\rho,
  \qquad
  G_{--}
  =F+\frac{\ii}{2}\operatorname{sgn}(\eta-\eta')\rho.
  \label{eq:timeorderedFrho}
\end{equation}
Since the random field changes $F$ but leaves $\rho$ unchanged, the same $c$-number term is added to every contour component:
\begin{equation}
  \boxed{
  \Delta G_{++}
  =\Delta G_{+-}
  =\Delta G_{-+}
  =\Delta G_{--}
  =f(k)R_{k,p}(\eta)R_{k,p}(\eta').}
  \label{eq:allSKcorrections}
\end{equation}
More generally, including replica labels,
\begin{equation}
  \boxed{
  G_{ab}^{\,rr'}(k;\eta,\eta')
  =\delta^{rr'}G_{ab}^{(0)}(k;\eta,\eta')
  +f(k)R_{k,p}(\eta)R_{k,p}(\eta'),
  \qquad a,b=\pm1 .}
  \label{eq:fullreplicaSKpropagator}
\end{equation}
This is the contour-basis form of the master result \cref{eq:Keldyshmaster}.  The Keldysh rotation was useful because it made the separation $\Delta F\neq0$ and $\Delta\rho=0$ manifest; the return to the $\pm$ contour basis is useful because the vertices of the cosmological in-in expansion carry explicit contour labels.  Equation~\eqref{eq:allSKcorrections} is therefore the form that will be inserted into the collider diagrams below.

\section{Exact random-field cosmological-collider seed}
\label{sec:collider}

The disorder correction has the same separable form in every contour component,
\begin{equation}
  \Delta F_{\rm quenched}(s;\eta_1,\eta_2)
  =f(s)R_{s,p}(\eta_1)R_{s,p}(\eta_2),
  \label{eq:quenchedstatisticalsector}
\end{equation}
while the spectral function remains unchanged.  The power-law family therefore preserves the exact factorization of the random-field exchange.  From this point onward, for the collider seed and phenomenology, we specialize to the spatial white-noise choice $f(s)=f_0H^3$ introduced in \cref{eq:whiteNoiseNormalization}.

\subsection{Exact factorized seed}
\label{subsec:RFseed}

Define
\begin{align}
  \Delta\widehat{\mathcal I}^{\,\ell_1\ell_2}_{{\rm RF},ab}(u,v,s,p)
  \equiv{}&
  -ab\,\frac{s^{5+\ell_1+\ell_2}}{H^2}
  \int_{-\infty_a}^{0}\dd\eta_1
  \int_{-\infty_b}^{0}\dd\eta_2\,
  (-\eta_1)^{\ell_1}(-\eta_2)^{\ell_2}
  \nonumber\\
  &\times
  e^{\ii a k_{12}\eta_1+\ii b k_{34}\eta_2}
  \Delta G_{ab}(s;\eta_1,\eta_2),
  \nonumber\\[-1mm]
  &\qquad
  u\equiv\frac{s}{k_{12}},\quad
  v\equiv\frac{s}{k_{34}},\quad
  a,b=\pm1.
  \label{eq:RFseeddef}
\end{align}
Using \cref{eq:Rlommel}, introduce
\begin{equation}
  \mathcal R_{a,p}^{\ell}(u)
  \equiv
  \int_0^\infty\dd x\,
  x^{\ell+\frac32}
  e^{-\ii a x/u}
  S_{p-\frac52,\nu}(x).
  \label{eq:Rvertexdef}
\end{equation}
The original time integral is
\begin{equation}
  \int_{-\infty_a}^{0}\dd\eta\,
  (-\eta)^\ell e^{\ii aK\eta}R_{s,p}(\eta)
  =H^{p-2}s^{-\ell-p-1}
  \mathcal R_{a,p}^{\ell}\!\left(\frac{s}{K}\right).
  \label{eq:Rvertexchange}
\end{equation}
Consequently
\begin{equation}
  \Delta\widehat{\mathcal I}^{\,\ell_1\ell_2}_{{\rm RF},ab}(u,v,s,p)
  =-ab\,f_0
  \left(\frac{s}{H}\right)^{3-2p}
  \mathcal R_{a,p}^{\ell_1}(u)
  \mathcal R_{b,p}^{\ell_2}(v).
  \label{eq:RFseedfactorized}
\end{equation}
With this white-noise normalization, unlike the clean de Sitter seed, the disorder contribution therefore depends generically not only on the ratios $u$ and $v$, but also explicitly on the exchanged scale $s$.  All of this additional scale dependence is carried by
\begin{equation}
  f_0\left(\frac{s}{H}\right)^{3-2p}.
  \label{eq:RFexplicitScaleFactor}
\end{equation}
The value $p=3/2$ is consequently distinguished: the explicit dependence on $s/H$ disappears and, up to the overall dimensionless amplitude $f_0$, the seed becomes a function only of the kinematic ratios.  Schematically,
\begin{equation}
  \Delta\widehat{\mathcal I}_{\rm RF}(u,v,s,p)
  \xrightarrow{\,p=3/2\,}
  \Delta\widehat{\mathcal I}_{\rm RF}(u,v).
  \label{eq:RFp32ScaleFree}
\end{equation}
Here and below the right-hand side is understood to retain the overall amplitude $f_0$.  The persistent normalization is recovered at $p=0$.  Before evaluating the one-vertex transform, it is useful to make the contour structure completely explicit.  Equation~\eqref{eq:RFseedfactorized} gives
\begin{align}
  \Delta\widehat{\mathcal I}^{\,\ell_1\ell_2}_{{\rm RF},++}(u,v,s,p)
  ={}&-\mathcal P_p(s)\,
  \mathcal R_{+,p}^{\ell_1}(u)\mathcal R_{+,p}^{\ell_2}(v),
  \\
  \Delta\widehat{\mathcal I}^{\,\ell_1\ell_2}_{{\rm RF},+-}(u,v,s,p)
  ={}&+\mathcal P_p(s)\,
  \mathcal R_{+,p}^{\ell_1}(u)\mathcal R_{-,p}^{\ell_2}(v),
  \\
  \Delta\widehat{\mathcal I}^{\,\ell_1\ell_2}_{{\rm RF},-+}(u,v,s,p)
  ={}&+\mathcal P_p(s)\,
  \mathcal R_{-,p}^{\ell_1}(u)\mathcal R_{+,p}^{\ell_2}(v),
  \\
  \Delta\widehat{\mathcal I}^{\,\ell_1\ell_2}_{{\rm RF},--}(u,v,s,p)
  ={}&-\mathcal P_p(s)\,
  \mathcal R_{-,p}^{\ell_1}(u)\mathcal R_{-,p}^{\ell_2}(v),
  \label{eq:RFfourSKsectors}
\end{align}
where
\begin{equation}
  \mathcal P_p(s)
  \equiv
  f_0\left(\frac{s}{H}\right)^{3-2p}.
  \label{eq:RFprefactorP}
\end{equation}
Thus the four Schwinger--Keldysh sectors only determine how the one-vertex objects are multiplied and combined.  The three hypergeometric functions encountered below arise instead from the three terms in the Lommel decomposition of a \emph{single} one-vertex transform.

\subsubsection{Closed-form evaluation}
\label{subsec:RFclosed}

The physical disorder seed is controlled by the contour difference
\begin{equation}
  \mathcal D_{\ell,p}(u)
  \equiv
  \mathcal R_{+,p}^{\ell}(u)-\mathcal R_{-,p}^{\ell}(u),
  \label{eq:physicalContourDifference}
\end{equation}
which can equivalently be written directly as the sine transform
\begin{equation}
  \mathcal D_{\ell,p}(u)
  =-2\ii\int_0^\infty\dd x\,
  x^{\ell+\frac32}\sin\!\left(\frac{x}{u}\right)
  S_{p-\frac52,\nu}(x).
  \label{eq:physicalContourSine}
\end{equation}
This representation makes the improved future-endpoint behavior of the physical contour combination manifest: the contour subtraction supplies an additional power of $x$ as $x\to0$.  Its folded value is defined by applying the same finite-part prescription to the complete difference,
\begin{equation}
  \mathcal D_{\ell,p}^{\mathrm{fold}}
  =\operatorname{Fin}_{u\to1}
  \left[
  \mathcal R_{+,p}^{\ell}(u)-\mathcal R_{-,p}^{\ell}(u)
  \right].
  \label{eq:physicalFoldedDifference}
\end{equation}
The folded continuation must be applied to the complete retarded combination.  We first display the exact seed and its $v=1$ specialization, and then return to the independent late-time endpoint question.

Summing the four contour assignments in \cref{eq:RFseedfactorized} gives the exact physical random-field seed,
\begin{equation}
  \Delta\widehat{\mathcal I}^{\,\ell_1\ell_2}_{\rm RF}(u,v,s,p)
  =- f_0\left(\frac{s}{H}\right)^{3-2p}
  \left[\mathcal R_{+,p}^{\ell_1}(u)-\mathcal R_{-,p}^{\ell_1}(u)\right]
  \left[\mathcal R_{+,p}^{\ell_2}(v)-\mathcal R_{-,p}^{\ell_2}(v)\right].
\end{equation}
Thus, the complete Schwinger--Keldysh sum is still a product of two one-variable functions.~For the bispectrum, we take the soft limit \(k_4\to0\) of the parent four-point diagram, which sets the second seed variable to its folded value \(v=1\). We then use the formulas derived in appendix \ref{app:hyperinetgrals}.
\begin{equation}
  \Delta\widehat{\mathcal I}^{\,\ell_1\ell_2}_{\rm RF}(u,1,s,p)
  =- f_0\left(\frac{s}{H}\right)^{3-2p}
  \left[\mathcal R_{+,p}^{\ell_1}(u)-\mathcal R_{-,p}^{\ell_1}(u)\right]
  \bigg\{\operatorname{Fin}_{v\to1}\left[\mathcal R_{+,p}^{\ell_2}(v)-\mathcal R_{-,p}^{\ell_2}(v)\right]\bigg\}.
\end{equation}
The one-vertex integrals can be evaluated analytically.  The direct hypergeometric calculation is given in appendix~\ref{app:hyperinetgrals}, while appendix~\ref{app:RFbootstrap} derives the same result independently from the bootstrap equations.\footnote{The individual hypergeometric integrals can also be evaluated directly with symbolic software such as Wolfram Mathematica. We include both analytic derivations in the appendices in order to keep the origin of the three independent blocks explicit.}

Using the closed form of \cref{eq:Rvertexclosed} and performing the contour subtraction in each of its three terms, the complete random-field seed becomes
\begin{equation}
\boxed{
\begin{aligned}
\Delta\widehat{\mathcal I}^{\,\ell_1\ell_2}_{\rm RF}(u,v,s,p)
={}&4 f_0\left(\frac{s}{H}\right)^{3-2p}
\Bigg\{
\frac{\Gamma(\ell_1+p+1)}{\Delta_p}
\sin\!\left[\frac{\pi}{2}(\ell_1+p+1)\right]
 u^{\ell_1+p+1}
\\[-1mm]
&\hspace{5mm}\times
{}_3F_2\!\left(
\begin{matrix}
1,\frac{\ell_1+p+1}{2},\frac{\ell_1+p+2}{2}\\
\frac p2+\frac14-\frac\nu2,\frac p2+\frac14+\frac\nu2
\end{matrix};u^2\right)
\\[1mm]
&\quad+
\mathcal C_{p,\nu}
\left[\sin\theta_{p,\nu}-\cos\theta_{p,\nu}\cot(\pi\nu)\right]
\frac{2^{-\nu}\Gamma\!\left(\ell_1+\frac52+\nu\right)}{\Gamma(1+\nu)}
\\[-1mm]
&\hspace{5mm}\times
\sin\!\left[\frac{\pi}{2}\left(\ell_1+\frac52+\nu\right)\right]
 u^{\ell_1+\frac52+\nu}
{}_2F_1\!\left(
\begin{matrix}
\frac{\ell_1}{2}+\frac54+\frac\nu2,
\frac{\ell_1}{2}+\frac74+\frac\nu2\\
1+\nu
\end{matrix};u^2\right)
\\[1mm]
&\quad+
\mathcal C_{p,\nu}\cos\theta_{p,\nu}\csc(\pi\nu)
\frac{2^{\nu}\Gamma\!\left(\ell_1+\frac52-\nu\right)}{\Gamma(1-\nu)}
\\[-1mm]
&\hspace{5mm}\times
\sin\!\left[\frac{\pi}{2}\left(\ell_1+\frac52-\nu\right)\right]
 u^{\ell_1+\frac52-\nu}
{}_2F_1\!\left(
\begin{matrix}
\frac{\ell_1}{2}+\frac54-\frac\nu2,
\frac{\ell_1}{2}+\frac74-\frac\nu2\\
1-\nu
\end{matrix};u^2\right)
\Bigg\}
\\[2mm]
&\times
\Bigg\{
\frac{\Gamma(\ell_2+p+1)}{\Delta_p}
\sin\!\left[\frac{\pi}{2}(\ell_2+p+1)\right]
 v^{\ell_2+p+1}
\\[-1mm]
&\hspace{5mm}\times
{}_3F_2\!\left(
\begin{matrix}
1,\frac{\ell_2+p+1}{2},\frac{\ell_2+p+2}{2}\\
\frac p2+\frac14-\frac\nu2,\frac p2+\frac14+\frac\nu2
\end{matrix};v^2\right)
\\[1mm]
&\quad+
\mathcal C_{p,\nu}
\left[\sin\theta_{p,\nu}-\cos\theta_{p,\nu}\cot(\pi\nu)\right]
\frac{2^{-\nu}\Gamma\!\left(\ell_2+\frac52+\nu\right)}{\Gamma(1+\nu)}
\\[-1mm]
&\hspace{5mm}\times
\sin\!\left[\frac{\pi}{2}\left(\ell_2+\frac52+\nu\right)\right]
 v^{\ell_2+\frac52+\nu}
{}_2F_1\!\left(
\begin{matrix}
\frac{\ell_2}{2}+\frac54+\frac\nu2,
\frac{\ell_2}{2}+\frac74+\frac\nu2\\
1+\nu
\end{matrix};v^2\right)
\\[1mm]
&\quad+
\mathcal C_{p,\nu}\cos\theta_{p,\nu}\csc(\pi\nu)
\frac{2^{\nu}\Gamma\!\left(\ell_2+\frac52-\nu\right)}{\Gamma(1-\nu)}
\\[-1mm]
&\hspace{5mm}\times
\sin\!\left[\frac{\pi}{2}\left(\ell_2+\frac52-\nu\right)\right]
 v^{\ell_2+\frac52-\nu}
{}_2F_1\!\left(
\begin{matrix}
\frac{\ell_2}{2}+\frac54-\frac\nu2,
\frac{\ell_2}{2}+\frac74-\frac\nu2\\
1-\nu
\end{matrix};v^2\right)
\Bigg\}.
\end{aligned}}
\label{eq:RFfullseed}
\end{equation}
The overall factor $4$ follows from the two contour subtractions, since each one-vertex difference contributes a factor $-2\ii$.  The first term in each bracket is the analytic forced sector, whereas the remaining two terms are the homogeneous collider branches.

For the bispectrum kinematics the second vertex is evaluated at the folded point, $v=1$.  The $v$-dependent bracket in \cref{eq:RFfullseed} must then be replaced by the regular finite part of the \emph{complete} contour difference; one must not set $v=1$ separately in the three hypergeometric blocks because their nonanalytic folded pieces cancel only after the retarded combination has been formed.  It is therefore most transparent to keep the main-text result in the factorized form
\begin{equation}
  \Delta\widehat{\mathcal I}^{\,\ell_1\ell_2}_{\rm RF}(u,1,s,p)
  =-f_0\left(\frac{s}{H}\right)^{3-2p}
  \mathcal D_{\ell_1,p}(u)\,
  \mathcal D^{\rm fold}_{\ell_2,p}.
  \label{eq:RFbispectrumvone}
\end{equation}
The manifestly regular analytic expression for $\mathcal D^{\rm fold}_{\ell,p}$ is derived in appendix~\ref{app:hyperinetgrals}, see in particular \cref{eq:Rvertexfoldedregular}.  In that form the apparent $\Gamma(-\ell-2)$ poles of the separate finite parts have already cancelled before the physical integer values of $\ell$ are taken.

\medskip
\noindent\emph{Late-time endpoint behavior.}
The regularity of the folded continuation does not by itself guarantee convergence at the future endpoint.  The folded limit probes the large-$x$ region of the one-vertex transform, whereas a late-time endpoint divergence comes from $x\to0$ and can therefore be present for every finite value of $u$.  From \cref{eq:smallLommelHyper}, the forced part of the Lommel function behaves as
\begin{equation}
  S_{p-\frac52,\nu}(x)
  =\frac{x^{p-3/2}}{\Delta_p}
  \left[1+\mathcal O(x^2)\right]
  +\text{homogeneous massive terms},
  \qquad x\to0.
  \label{eq:smallXLommelEndpoint}
\end{equation}
Hence the forced contribution to an individual contour integral scales as
\begin{equation}
  \left.\mathcal R_{a,p}^{\ell}(u)\right|_{x\to0,\mathrm{forced}}
  \sim
  \frac{1}{\Delta_p}\int_0\dd x\,x^{\ell+p},
  \label{eq:individualEndpointScaling}
\end{equation}
and is directly convergent only for
\begin{equation}
  \operatorname{Re}(\ell+p)>-1.
  \label{eq:individualEndpointCondition}
\end{equation}
For $\ell=-2$, the separate $+$ and $-$ contour integrals therefore require analytic continuation when $p\leq1$.  The physical Schwinger--Keldysh difference is better behaved because
\begin{equation}
  e^{-\ii x/u}-e^{+\ii x/u}
  =-\frac{2\ii x}{u}+\mathcal O(x^3),
\end{equation}
so that
\begin{equation}
  \left.\mathcal D_{\ell,p}(u)\right|_{x\to0,\mathrm{forced}}
  \sim
  -\frac{2\ii}{u\Delta_p}
  \int_0\dd x\,x^{\ell+p+1}.
  \label{eq:physicalEndpointScaling}
\end{equation}
Thus the physical difference is directly endpoint-convergent provided
\begin{equation}
  \operatorname{Re}(\ell+p)>-2.
  \label{eq:physicalEndpointCondition}
\end{equation}
For the bispectrum weight $\ell=-2$ this gives
\begin{equation}
  \begin{array}{ccl}
  p>1 &:& \mathcal R_{+,p}^{-2}\ \text{and}\ \mathcal R_{-,p}^{-2}\ \text{are separately endpoint-convergent},\\[1mm]
  0<p\leq1 &:& \text{the separate contours require analytic continuation, but }\mathcal D_{-2,p}\text{ is finite},\\[1mm]
  p=0 &:& \mathcal D_{-2,0}\text{ has a local logarithmic endpoint divergence}.
  \end{array}
  \label{eq:ellminus2EndpointRegimes}
\end{equation}
Indeed, at $p=0$,
\begin{equation}
  \left.\mathcal D_{-2,0}(u)\right|_{x\to0,\mathrm{forced}}
  \sim
  -\frac{2\ii}{uM^2}\int_0\frac{\dd x}{x}.
  \label{eq:ellminus2EndpointLogX}
\end{equation}
This divergence exists for every finite $u$ and is therefore not generated by the folded limit $u\to1$.  Conversely, the folded branch cancellation in \cref{eq:RFfoldedcancel} holds for arbitrary $p$, independently of whether the $x\to0$ endpoint requires a prescription. It is useful to isolate the forced contribution to a single physical contour difference.\footnote{For the particular contribution ,
$e^{-\ii\phi}-e^{+\ii\phi}=-2\ii\sin\phi$, giving
$\left[\mathcal R_{+,p}^{\ell}-\mathcal R_{-,p}^{\ell}\right]_{\rm part}
\propto \sin[\pi(\ell+p+1)/2]$ and hence the generalized selection rule
$\ell+p+1\in2\mathbb Z\Rightarrow
[\mathcal R_{+,p}^{\ell}-\mathcal R_{-,p}^{\ell}]_{\rm part}=0$,
which reduces to the persistent odd-$\ell$ rule at $p=0$.
For principal-series fields and real $p$,
$\mathcal R_{-,p}^{\ell}=(\mathcal R_{+,p}^{\ell})^*$.
For $\ell=-2$ and $0<p\leq1$, the contour difference must be formed before separating the individually endpoint-divergent integrals.}

\clearpage

\subsection{Phenomenological structure and soft limits}
\label{subsec:RFpheno}

To connect the exact seed derived above with inflationary observables, it is useful to regard the scalar-exchange four-point function as the parent object from which lower-point correlators can be obtained.  This organization is standard in the cosmological-collider and bootstrap literature~\cite{Arkani-Hamed:2015bza,Arkani-Hamed:2018kmz,Qin:2023closed,Aoki:2023dynamical}.  In particular, the inflationary bispectrum may be obtained by taking one external momentum soft, or equivalently by evaluating one leg of the corresponding de Sitter four-point function on the inflationary background~\cite{Arkani-Hamed:2018kmz}.  In the seed-integral language, the three- and two-point correlators arise as successive folded limits of a four-point seed~\cite{Qin:2023closed}, while the same reduction has been implemented for the exchange of a scalar with a time-dependent bulk mass in Ref.~\cite{Aoki:2023dynamical}.  We will use this hierarchy below, beginning with the four-point function itself.

Throughout this phenomenological discussion, ``exact'' refers to the Gaussian disorder average and the linear response of the free spectator.  Once the spectator is coupled to the inflaton through the mixing and interaction vertices below, we continue to assume that the spectator energy density and disorder-induced fluctuations remain subdominant to the inflationary background and that the interaction couplings are perturbative.  The plots should therefore be read as diagnostics of the momentum-space shape within this regime.  In particular, the soft enhancement that occurs for $p>3/2$ cannot be extrapolated to arbitrarily small momentum ratios at fixed $f_0$ without eventually checking backreaction, the finite duration of the disordered stage, and the validity of the effective description.

Before specializing the correlators, it is useful to clarify the role of the time-profile parameter $p$. In the general case, the spatial scale dependence of the quenched disorder is encoded in $f(k)$, while $p$ controls the deterministic time dependence of the source. For spatial white noise, $f(k)=f_0H^3$, the remaining explicit dependence of the disorder seed on the exchanged scale is isolated in the factor $(s/H)^{3-2p}$ derived in \cref{eq:RFexplicitScaleFactor}.

It is therefore useful to define the dimensionless combination,
\begin{equation}
\epsilon_{\rm dis}(s,p)
\equiv
f_0\left(\frac{s}{H}\right)^{3-2p},
\end{equation}
which measures the explicit scale-dependent weighting of the random-field contribution. A conservative criterion for keeping this explicit prefactor under control is
\begin{equation}
\epsilon_{\rm dis}(s,p)\lesssim 1.
\end{equation}
For $p>3/2$ and $s/H<1$, this implies
\begin{equation}
f_0\lesssim
\left(\frac{s}{H}\right)^{2p-3}.
\end{equation}
This criterion therefore provides a simple estimate of the upper bound on the disorder amplitude, obtained by requiring that the explicit scale-dependent weighting remain under control. In particular, for fixed $f_0$ and $s/H<1$, increasing $p$ sufficiently far above $3/2$ eventually drives the system outside this controlled regime.

The parameter $p$ does not modify the massive collider exponents $3/2\pm\ii\mu$ or the clock frequency $\mu$. Instead, it changes the analytic forced sector, the amplitudes and phases of the homogeneous response, and their relative importance in soft limits. For the $\ell=-2$ block relevant to the bispectrum, the forced contribution scales as $u^{p-1}$ after restoring the external momentum factors, while the nonanalytic clock contribution scales as $u^{1/2\pm\ii\mu}$. The value $p=3/2$ therefore marks the crossover at which the analytic and nonanalytic contributions have the same power-law envelope. At this value, the explicit exchanged-scale dependence also disappears for spatial white noise, yielding a particularly simple analytic soft limit and cleanly isolating the intrinsic modification of the cosmological-collider signal by the disorder.

At the crossover $p=3/2$ the distinction can be stated without committing to a particular vertex weight. The collapsed limit for $\ell_1=\ell_2=-2$ takes the following form,
\begin{equation}
   \lim_{u,v\to 0} \mathcal{\hat{I}}^{-2,-2}_{\rm Total} = \widehat{\mathcal I}_{\rm dS}^{-2,-2}(u,v)\big|_{u,v \to 0} + \Delta\widehat{\mathcal I}_{\mathrm{RF}}^{-2,-2}
\left(u,v\right)\big|_{u,v \to 0}
\end{equation}
where,
\begin{align}
\begin{aligned}
\widehat{\mathcal I}_{\rm dS}^{-2,-2}(u,v)
\big|_{u,v \to 0}
={}&
\frac{\pi\,\operatorname{sech}(\pi\mu)}{2\mu}
(uv)^{\frac12-i\mu}
\left[
u^{2i\mu}\left(\operatorname{csch}(\pi\mu)+i\right)
+
v^{2i\mu}\left(\operatorname{csch}(\pi\mu)-i\right)
\right]
\\[1mm]
&+
\frac{
2^{-1-2i\mu}
\left(1+i\sinh(\pi\mu)\right)
\Gamma(-i\mu)^2
\Gamma\!\left(\frac12+i\mu\right)^2
}{
\pi
}
(uv)^{\frac12+i\mu}
\\[1mm]
&+
\frac{
2^{-1+2i\mu}
\left(1-i\sinh(\pi\mu)\right)
\Gamma(i\mu)^2
\Gamma\!\left(\frac12-i\mu\right)^2
}{
\pi
}
(uv)^{\frac12-i\mu};
\end{aligned}
\label{eq:dS_seed_collapsed_limit}
\end{align}
and
\begin{equation}
\begin{aligned}
\Delta\widehat{\mathcal I}_{\mathrm{RF}}^{-2,-2}
\left(u,v\right)\bigg|_{u,v\to0}
={}&
-\frac{f_0\,2^{-1-2i\mu}}{\mu^2}
\left[1+i\sinh(\pi\mu)\right]
\Gamma(-i\mu)^2
\Gamma\!\left(\frac12+i\mu\right)^2
(uv)^{\frac12+i\mu}
\\[1mm]
&-
\frac{f_0\,2^{-1+2i\mu}}{\mu^2}
\left[1-i\sinh(\pi\mu)\right]
\Gamma(i\mu)^2
\Gamma\!\left(\frac12-i\mu\right)^2
(uv)^{\frac12-i\mu}
\\[2mm]
&+
\frac{2\pi f_0}{\mu^4}\sqrt{uv}
\\[2mm]
&-
\frac{2\pi f_0\sqrt{\pi/2}}
{\mu^3\sinh(\pi\mu)}
\sqrt{uv}
\Bigg[
2^{-i\mu}
\frac{
\Gamma\!\left(\frac12+i\mu\right)
}{
\Gamma(1+i\mu)
}
\sin\!\left(
\frac{\pi}{4}+\frac{i\pi\mu}{2}
\right)
\left(u^{i\mu}+v^{i\mu}\right)
\\
&\hspace{37mm}
+
2^{i\mu}
\frac{
\Gamma\!\left(\frac12-i\mu\right)
}{
\Gamma(1-i\mu)
}
\sin\!\left(
\frac{\pi}{4}-\frac{i\pi\mu}{2}
\right)
\left(u^{-i\mu}+v^{-i\mu}\right)
\Bigg]
\\[2mm]
&+
\frac{\pi^2 f_0}
{2\mu^3\sinh(\pi\mu)}
\sqrt{uv}
\left[
\left(\frac{u}{v}\right)^{i\mu}
+
\left(\frac{v}{u}\right)^{i\mu}
\right].
\end{aligned}
\label{eq:RF_p32_ellminus2_soft}
\end{equation}
Thus the analytic forced response and the two nonanalytic collider branches share the same real power-law envelope; the massive information is carried by the factors $u^{\pm\ii\mu}$.  With the white-noise normalization $f(s)=f_0H^3$, the explicit factor $(s/H)^{3-2p}$ also disappears at $p=3/2$.  This makes the crossover especially useful for isolating interference between the clean exchange and the intrinsic random-field response.  The persistent case $p=0$ remains perfectly well defined for the $\ell_1=\ell_2=0$ four-point seed, even though the corresponding $\ell=-2$ bispectrum block contains the late-time logarithm discussed above.

\subsubsection{Trispectrum}

The four-point exchange correlator is the most direct observable associated with the seed. The trispectrum considered here is generated by the exchange of the massive field through two insertions of the cubic interaction
\begin{equation}
    \mathcal L_{3,\mathrm{int}}=c_3\,\sigma\left(\partial_\eta\varphi\right)^2,
\end{equation}
so that the exchange contribution is proportional to $c_3^2$. For massless external inflaton fluctuations with this derivative scalar-exchange vertex, the external time dependence selects $\ell_1=\ell_2=0$. This is the same seed reduction used in the cosmological-bootstrap treatment of scalar exchange and in the explicit inflationary four-point construction of Refs.~\cite{Qin:2023closed,Aoki:2023dynamical}. In the notation used throughout this work,
\begin{equation}
  s=|\bm k_1+\bm k_2|,
  \qquad
  u=\frac{s}{k_{12}},
  \qquad
  v=\frac{s}{k_{34}},
\end{equation}
and one exchange channel is
\begin{equation}
  \left\langle
  \delta\phi_{\bm k_1}
  \delta\phi_{\bm k_2}
  \delta\phi_{\bm k_3}
  \delta\phi_{\bm k_4}
  \right\rangle_{s}^{\prime}
  =
  \frac{c_3^2 H^6}{4k_1k_2k_3k_4\,s^5}
  \sum_{a,b=\pm1}
  \widehat{\mathcal I}^{\,00}_{ab}(u,v).
  \label{eq:trispectrumSeed00}
\end{equation}
The full four-point function is obtained by adding the crossed permutations. Equation~\eqref{eq:trispectrumSeed00} is the direct analogue, in our notation, of the standard reconstruction of the inflationary four-point function from the $\ell_1=\ell_2=0$ seed~\cite{Aoki:2023dynamical}.

For comparison with the usual dimensionless primordial trispectrum, we use the standard shape convention~\cite{Chen:2018trispectrum,Regan:2010trispectrum}
\begin{equation}
  \left\langle
  \zeta_{\bm k_1}\zeta_{\bm k_2}\zeta_{\bm k_3}\zeta_{\bm k_4}
  \right\rangle_c^{\prime}
  =(2\pi)^6\mathcal P_\zeta^3
  \frac{(k_1+k_2+k_3+k_4)^3}
  {(k_1k_2k_3k_4)^3}
  \mathcal T(\bm k_1,\bm k_2,\bm k_3,\bm k_4).
  \label{eq:trispectrum_shape_definition}
\end{equation}
Using $\zeta=\delta\phi/(\sqrt{2\epsilon}M_{\rm Pl})$ and $\mathcal P_\zeta=H^2/(8\pi^2\epsilon M_{\rm Pl}^2)$, the contribution of one exchange channel is therefore
\begin{equation}
  \mathcal T_s
  =
  \frac{\epsilon}{2}
  (c_3 M_{\rm Pl})^2
  \frac{(k_1k_2k_3k_4)^2}
  {(k_1+k_2+k_3+k_4)^3s^5}
  \sum_{a,b=\pm1}
  \widehat{\mathcal I}^{\,00}_{ab}(u,v),
  \label{eq:trispectrum_shape_seed}
\end{equation}
with the complete trispectrum obtained by adding the five crossed permutations. In the disordered theory, the seed entering this expression is the sum of the clean and random-field pieces,
\begin{equation}
  \widehat{\mathcal I}_{\rm Total}^{00}(u,v;s,p)
  =
  \widehat{\mathcal I}_{\rm dS}^{00}(u,v)
  +
  \Delta\widehat{\mathcal I}_{\rm RF}^{00}(u,v;s,p).
  \label{eq:trispectrumTotalSeed}
\end{equation}
The persistent case $p=0$ is finite for $\ell_1=\ell_2=0$, so the trispectrum can be evaluated directly from the exact two-variable seed without the late-time endpoint prescription required by the $\ell=-2$ block.

\begin{figure}[ht]
    \centering
    \includegraphics[width=0.7\linewidth]{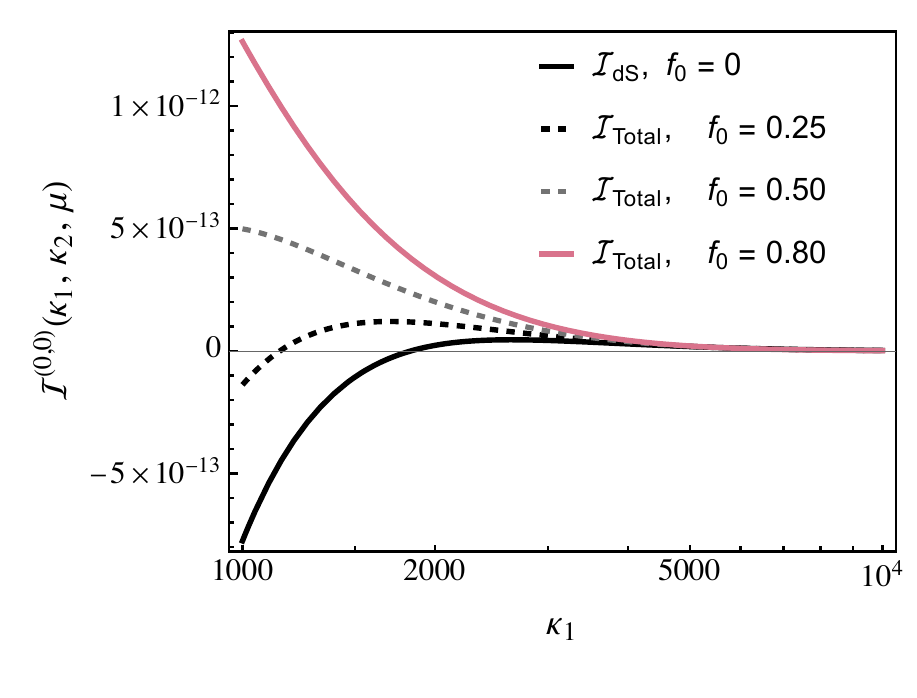}
    \caption{Collapsed trispectrum seed as a function of $\kappa_1=u^{-1}$, for $p=3/2$, $\mu=1.5$, and fixed $\kappa_2=v=0.02$. The black solid curve shows the clean de Sitter result, while the remaining curves include the disorder contribution for increasing values of $f_0$. As the disorder strength grows, the original oscillatory collider pattern is progressively suppressed by destructive interference. For $p=3/2$, the coefficient of the $(uv)^{5/2\pm i\mu}$ collider branches is proportional to $1-\pi f_0/\mu^2$ and therefore vanishes at $f_0=\mu^2/\pi\simeq0.716$ for $\mu=1.5$. The $f_0=0.80$ curve lies beyond the cancellation point, where this branch has changed sign.}
    \label{trispecdisord1}
\end{figure}

We first isolate the effect of the disorder strength at the scale-invariant crossover $p=3/2$. For the one-dimensional collapsed cuts below we use $\kappa_1\equiv u^{-1}$ and $\kappa_2\equiv v$, so that increasing $\kappa_1$ probes progressively smaller $u$. At $p=3/2$ the explicit factor $(s/H)^{3-2p}$ is unity, and the white-noise result depends on the disorder amplitude only through the dimensionless parameter $f_0$. In the collapsed expansion, the part of the random-field seed with exactly the same two-branch structure $(uv)^{5/2\pm i \mu}$ as the original de Sitter collider pair is
\begin{equation}
  \left.\Delta\widehat{\mathcal I}_{\rm RF}^{00}\right|_{(uv)^{5/2\pm\ii\mu}}
  =-\frac{\pi f_0}{\mu^2}
  \left.\widehat{\mathcal I}_{\rm dS}^{00}\right|_{(uv)^{5/2\pm\ii\mu}}.
  \label{eq:trispectrumRFsamebranch}
\end{equation}
Adding the clean contribution therefore gives the particularly simple reorganization
\begin{equation}
  \left.
  \widehat{\mathcal I}_{\rm Total}^{00}
  \right|_{(uv)^{5/2\pm\ii\mu}}
  =
  \left(1-\frac{\pi f_0}{\mu^2}\right)
  \left.
  \widehat{\mathcal I}_{\rm dS}^{00}
  \right|_{(uv)^{5/2\pm\ii\mu}}.
  \label{eq:trispectrumDestructiveInterference}
\end{equation}
Thus the disorder contribution interferes destructively with the original collapsed collider branch. The cancellation occurs at
\begin{equation}
  f_{0,\rm crit}=\frac{\mu^2}{\pi}.
  \label{eq:trispectrumCriticalDisorder}
\end{equation}
For the benchmark $\mu=1.5$ used in \cref{trispecdisord1}, $f_{0,\rm crit}\simeq0.716$. The figure therefore follows the progressive suppression of the original oscillatory pattern as $f_0$ approaches this value and then crosses to the opposite side of the cancellation. This should not be interpreted as the removal of all oscillatory information: the factorized random-field contribution also contains mixed analytic--collider terms with $u^{\pm\ii\mu}$ and $v^{\pm\ii\mu}$. Rather, the disorder reorganizes the harmonic content of the collapsed signal while leaving the heavy-field frequency $\mu$ unchanged.

A second effect becomes visible when the temporal weight $p$ is varied at fixed disorder strength. For white noise, the exact random-field seed carries the explicit factor derived in \cref{eq:RFexplicitScaleFactor},
\begin{equation}
  \Delta\widehat{\mathcal I}_{\rm RF}^{00}(u,v;s,p)
  \propto
  f_0\left(\frac{s}{H}\right)^{3-2p}.
  \label{eq:trispectrumPEnhancement}
\end{equation}
Consequently, for $s/H<1$ the explicit scale weighting enhances the disorder contribution when $p>3/2$. This effect is absent exactly at the crossover, where the exponent vanishes. 

\begin{figure}[ht]
    \centering
    \includegraphics[width=0.7\linewidth]{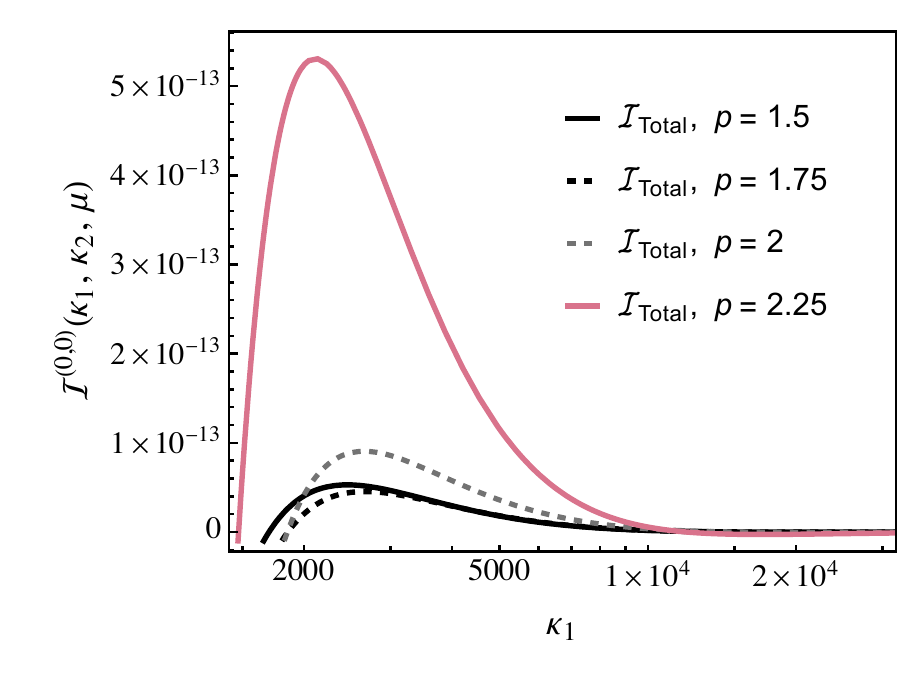}
    \caption{Collapsed trispectrum seed for fixed $\mu=1.5$, $f_0=0.05$, and $s/H=0.1$, shown for increasing values of the temporal-weight parameter $p$. The disorder contribution is strongly enhanced as $p$ increases above the crossover value $p=3/2$. An essential part of this growth originates from the explicit scale dependence of the random-field seed, $\Delta\widehat{\mathcal I}_{\mathrm{RF}}\propto f_0(s/H)^{3-2p}$, which becomes increasingly large for $s/H<1$ and $p>3/2$. The enhancement therefore reflects both the $p$ dependence of the response functions and the explicit scale weighting induced by the time-dependent disorder profile.}
    \label{trispecdisord2}
\end{figure}

For the benchmark $s/H=0.1$ used in \cref{trispecdisord2}, the explicit scale-dependent prefactor becomes
\[
\left(\frac{s}{H}\right)^{3-2p}
=10^{\,2p-3}.
\]
It is therefore equal to $1$ at $p=3/2$, and increases to approximately $3.2$, $10$, and $31.6$ for $p=7/4$, $2$, and $9/4$, respectively. Thus, even before accounting for the intrinsic $p$ dependence of the response functions, the normalization of the random-field contribution grows rapidly for $p>3/2$ when $s/H<1$. This behavior has a simple interpretation. Although increasing $p$ makes the source $(-H\eta)^p h(\mathbf{x})$ decay more rapidly toward the future boundary, it also gives greater weight to earlier times, for which $-H\eta>1$. Modes with $s/H<1$ therefore experience a stronger forcing during the epoch relevant for their excitation as $p$ is increased. The full enhancement is not determined by this prefactor alone, since the response functions $\mathcal R_{a,p}^{0}$ also depend nontrivially on $p$ through the forced and homogeneous components. The growth seen in \cref{trispecdisord2} should consequently be understood as the combined effect of this explicit scale weighting and the intrinsic $p$ dependence of the retarded response.

The two effects displayed above are complementary. At fixed $p=3/2$, changing $f_0$ primarily reorganizes the relative amplitudes of the clean and disorder-induced collider structures and can produce destructive interference. At fixed $f_0$, changing $p$ modifies both the intrinsic retarded response and, away from $p=3/2$, the overall scale dependence of the random-field contribution. These observations motivate the bispectrum analysis below, where the $\ell=-2$ block provides a sharper probe of the intrinsic dependence on the temporal profile.  The persistent case $p=0$ is already contained in the exact formulas above; its qualitatively new feature in the lower-point limit is instead the local $\ell=-2$ endpoint logarithm discussed in \cref{eq:ellminus2EndpointRegimes}.

\subsubsection{Bispectrum}

The bispectrum is generated by combining a linear mixing between the inflaton fluctuation and the massive spectator with the cubic interaction involving two inflaton fluctuations,
\begin{equation}
\delta\mathcal L_{2}
\supset
c_2 a^{3} \partial_{\eta}\varphi \sigma,
\qquad
\delta\mathcal L_{3}
\supset
c_3 a^{2}
\left(\partial_{\eta}\varphi\right)^{2}\sigma,
\label{eq:bispectrum_interactions}
\end{equation}
where $c_2$ and $c_3$ are time-independent couplings. As in the trispectrum discussion, we suppress the overall product of interaction couplings and numerical vertex-normalization factors in what follows, since they only determine the overall amplitude of the final shape.

These interactions generate the tree-level exchange contribution
\begin{align}
\left\langle
\varphi_{\bm k_1}
\varphi_{\bm k_2}
\varphi_{\bm k_3}
\right\rangle_{\sigma}^{\prime}
={}&
-\frac{c_2 c_3 H}{4k_1k_2k^4_3}
\sum_{a,b=\pm1}
\widehat{\mathcal I}_{ab}(u,1)
+2\,{\rm perm.},
\label{eq:bispectrum_exchange_integral}
\end{align}
It is useful to separate the three exchange channels as
\begin{equation}
\left\langle
\varphi_{\bm k_1}
\varphi_{\bm k_2}
\varphi_{\bm k_3}
\right\rangle_{\sigma}^{\prime} =
\mathcal C_{12|3}
+
\mathcal C_{23|1}
+
\mathcal C_{31|2}.
\label{eq:bispectrum_three_channels}
\end{equation}
In the squeezed configuration $k_3\ll k_1\simeq k_2$, the channel $\mathcal C_{12|3}$ is the collider channel. The remaining two channels carry hard internal momenta and contribute to the analytic squeezed-limit background. The soft channel is directly related to the exchange seed introduced above. For the interactions in \eqref{eq:bispectrum_interactions}, the time dependence of the two vertices selects $\ell_1=0,\ell_2=-2$.

To connect with the primordial bispectrum, we use $\zeta=\delta\phi/(\sqrt{2\epsilon}M_{\rm Pl})$, $\mathcal P_\zeta=H^2/(8\pi^2\epsilon M_{\rm Pl}^2)$ and define the dimensionless shape function by
\begin{equation}
\left\langle
\zeta_{\bm k_1}
\zeta_{\bm k_2}
\zeta_{\bm k_3}
\right\rangle^{\prime} \equiv (2\pi)^4
\frac{\mathcal P_{\zeta}^{2}}
{(k_1k_2k_3)^2}
S(k_1,k_2,k_3).
\label{eq:bispectrum_shape_definition}
\end{equation}
Consequently, the momentum dependence of the soft exchange channel is entirely determined by the single-folded seed,
\begin{equation}
S_{12|3}(k_1,k_2,k_3)
\propto -
\frac{k_1k_2}{k_3^2}
\sum_{a,b=\pm1}
\widehat{\mathcal I}^{0,-2}_{ab}
\left(
\frac{k_3}{k_1+k_2},1
\right).
\label{eq:bispectrum_shape_seed}
\end{equation}
In the disordered theory, the only replacement required in the soft channel is
\begin{equation}
\widehat{\mathcal I}_{\rm Total}^{0,-2}(u,1;s,p)
=
\widehat{\mathcal I}_{\rm dS}^{0,-2}(u,1)
+
\Delta\widehat{\mathcal I}_{\rm RF}^{0,-2}(u,1;s,p).
\label{eq:bispectrum_total_seed}
\end{equation}
Thus, just as the trispectrum is reconstructed from the $\ell_1=\ell_2=0$ two-variable seed, the squeezed collider contribution to the bispectrum is completely encoded in the single-folded $\ell_1=0$, $\ell_2=-2$ seed.

The complete bispectrum shape is obtained by summing the three exchange channels,
\begin{equation}
\begin{aligned}
S_{\rm Total}(k_1,k_2,k_3)
=
-\mathcal N_B
\Bigg[
&
\frac{k_1k_2}{k_3^2}\,
\widehat{\mathcal I}_{\rm Total}^{\,0,-2}
\left(
\frac{k_3}{k_1+k_2},1;k_3,p
\right)
\\
&+
\frac{k_2k_3}{k_1^2}\,
\widehat{\mathcal I}_{\rm Total}^{\,0,-2}
\left(
\frac{k_1}{k_2+k_3},1;k_1,p
\right)
\\
&+
\frac{k_3k_1}{k_2^2}\,
\widehat{\mathcal I}_{\rm Total}^{\,0,-2}
\left(
\frac{k_2}{k_3+k_1},1;k_2,p
\right)
\Bigg],
\end{aligned}
\label{eq:full_bispectrum_shape_disorder}
\end{equation}
where $\mathcal N_B$ is a momentum-independent normalization factor fixed by the background and the interaction couplings. The exchanged momentum entering the disorder seed is different in the three channels. This distinction is relevant away from $p=3/2$, since
\begin{equation}
\Delta\widehat{\mathcal I}_{\rm RF}
\propto
f_0
\left(\frac{s}{H}\right)^{3-2p}.
\label{eq:bispectrum_disorder_scaling_s}
\end{equation}
For the one-dimensional family of configurations used below, we set
\begin{equation}
k_1=k_2\equiv k,
\qquad
x\equiv\frac{k_3}{k},
\qquad
0<x\leq 1.
\end{equation}
The three exchange channels then reduce to two independent contributions,
\begin{equation}
u_{12|3}=\frac{x}{2},
\qquad
u_{23|1}=u_{31|2}=\frac{1}{1+x},
\end{equation}
and the complete shape becomes
\begin{equation}
\begin{aligned}
S_{\rm Total}(x)
=
-\mathcal N_B
\Bigg\{
&
\frac{1}{x^2}
\left[
\widehat{\mathcal I}_{\rm dS}^{\,0,-2}
\left(
\frac{x}{2},1
\right)
+
\Delta\widehat{\mathcal I}_{\rm RF}^{\,0,-2}
\left(
\frac{x}{2},1;xk,p
\right)
\right]
\\
&+
2x
\left[
\widehat{\mathcal I}_{\rm dS}^{\,0,-2}
\left(
\frac{1}{1+x},1
\right)
+
\Delta\widehat{\mathcal I}_{\rm RF}^{\,0,-2}
\left(
\frac{1}{1+x},1;k,p
\right)
\right]
\Bigg\}.
\end{aligned}
\label{eq:full_bispectrum_shape_disorder_x}
\end{equation}
The two characteristic kinematic regimes are then contained in a single variable,
\begin{equation}
x\rightarrow0
\quad\text{(squeezed)},
\qquad
x=1
\quad\text{(equilateral)}.
\end{equation}
It is also convenient to introduce $\kappa\equiv x^{-1}=k/k_3$ when focusing on the squeezed regime. In terms of $\kappa$, the equilateral point is at $\kappa=1$, while the deep squeezed limit corresponds to $\kappa\gg1$.

\begin{figure}[ht]
\centering
\begin{minipage}[t]{0.485\linewidth}
    \centering
    \includegraphics[width=\linewidth]{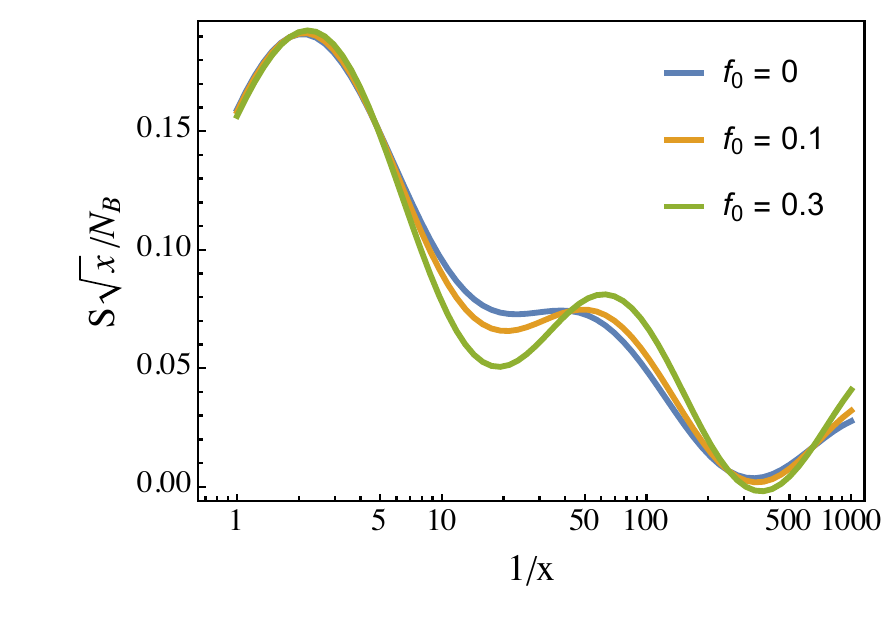}
    \par\smallskip
    \textbf{(a)} $\sqrt{x}\,S(1/x)$.
\end{minipage}
\hfill
\begin{minipage}[t]{0.485\linewidth}
    \centering
    \includegraphics[width=\linewidth]{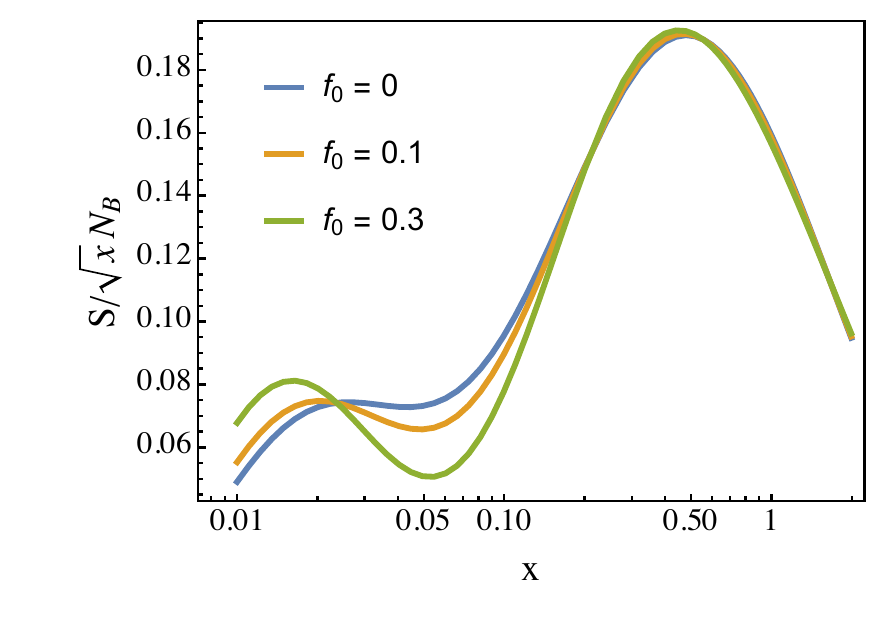}
    \par\smallskip
    \textbf{(b)} $x^{-1/2}S(x)$.
\end{minipage}
\caption{Dependence of the bispectrum on the disorder amplitude for fixed $\mu=2$, $p=3/2$, and $(\ell_1,\ell_2)=(0,-2)$. \textbf{(a)} Rescaled squeezed-limit representation as a function of $\kappa\equiv k/k_3$, for several values of $f_0$. The factor $\sqrt{\kappa}$ removes the standard massive-field envelope and makes the oscillatory collider structure easier to compare. \textbf{(b)} Full one-dimensional shape as a function of $x\equiv k_3/k$, spanning the squeezed and equilateral configurations. Since $p=3/2$ removes the explicit factor $(s/H)^{3-2p}$, the differences between the curves isolate the intrinsic disorder response and its interference with the clean exchange.}
\label{fig:bispectrum_f0}
\end{figure}

The role of the disorder amplitude is particularly transparent at $p=3/2$. At this value the explicit scale factor in \eqref{eq:bispectrum_disorder_scaling_s} becomes unity, so varying $f_0$ does not introduce an additional power of the exchanged momentum. The changes in the shape therefore directly probe the interference between the clean de Sitter exchange and the random-field response. This is illustrated in \cref{fig:bispectrum_f0}: the full shape is modified most visibly in the squeezed and intermediate configurations, whereas the broad equilateral region is comparatively stable. In the $\kappa$ representation the same effect appears as a deformation of the oscillatory collider pattern rather than as a uniform rescaling of the signal.

The dependence on the temporal profile follows directly from the soft expansion of the exact one-vertex response derived above. For $\ell_1=0$, its forced contribution starts as $u^{p+1}$, whereas the two homogeneous massive branches scale as $u^{5/2\pm i\mu}$. In the soft channel $u=x/2$ and $s=xk$. Combining these powers with the explicit random-field factor $(s/H)^{3-2p}$ and the overall $x^{-2}$ factor in \cref{eq:full_bispectrum_shape_disorder}, the forced contribution scales as $x^{2-p}$, while the massive pair scales as $x^{7/2-2p\pm i\mu}$. The latter therefore produces an oscillatory contribution with envelope $x^{7/2-2p}$ and logarithmic frequency $\mu$.

The relative scaling of the massive and forced sectors is thus $x^{3/2-p}$. The value $p=3/2$ separates the two behaviors: for $p<3/2$ the forced contribution becomes relatively more important in the squeezed limit, whereas for $p>3/2$ the nonanalytic massive contribution becomes increasingly prominent. Importantly, changing $p$ affects the envelope, amplitude, and phase of the oscillation, but not its frequency, which remains fixed by the heavy-field mass.

\begin{figure}[ht]
\centering
\begin{minipage}[t]{0.485\linewidth}
    \centering
    \includegraphics[width=\linewidth]{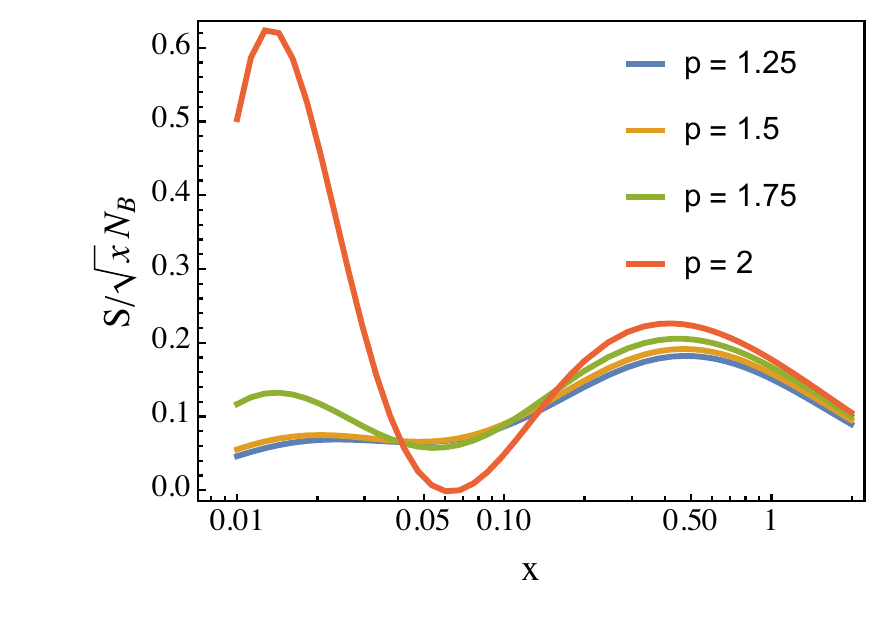}
    \par\smallskip
    \textbf{(a)} Full shape.
\end{minipage}
\hfill
\begin{minipage}[t]{0.485\linewidth}
    \centering
    \includegraphics[width=\linewidth]{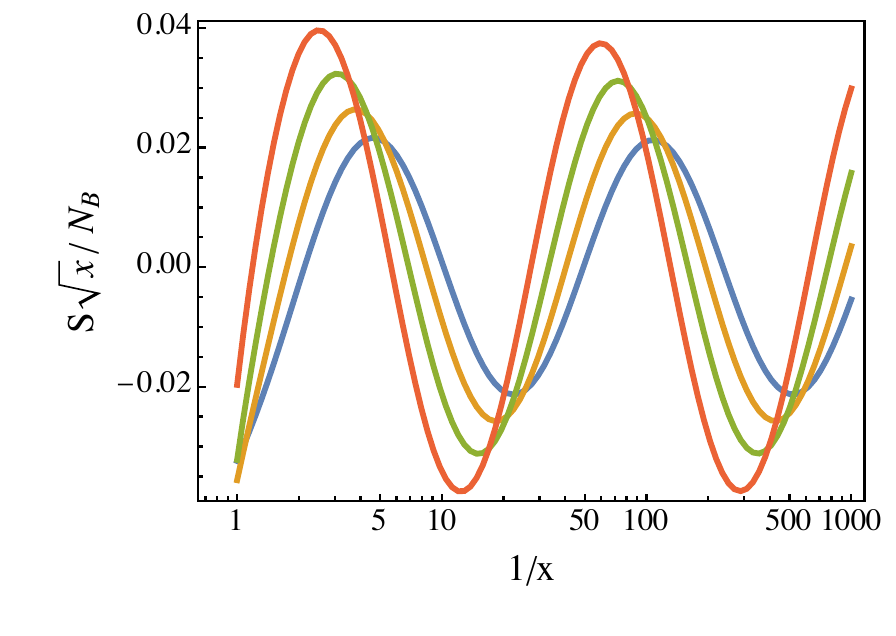}
    \par\smallskip
    \textbf{(b)} Rescaled soft clock.
\end{minipage}
\caption{Dependence of the bispectrum on the temporal-weight exponent $p$, for fixed $\mu=2$, $k/H=1$, $(\ell_1,\ell_2)=(0,-2)$, and $f_0=0.1$. \textbf{(a)} Full shape $x^{-1/2}S(x)$ for several values of $p$. Increasing $p$ preferentially enhances the disorder contribution in the squeezed region, consistently with the explicit factor $x^{3-2p}$ in the soft channel. \textbf{(b)} Isolated soft homogeneous contribution after removing its predicted power-law envelope, $\kappa^{-(2p-3)}\sqrt{\kappa}\,\Delta S_{\rm clock}(1/\kappa)$. The curves display the same logarithmic oscillation frequency, set by $\mu$, while their amplitudes and phases vary with $p$.}
\label{fig:bispectrum_p}
\end{figure}

The numerical behavior is shown in \cref{fig:bispectrum_p}. Panel~$(a)$ illustrates how increasing $p$ preferentially enhances the disorder contribution toward the squeezed region, while leaving the overall shape comparatively less affected away from this limit. The effect is therefore considerably more pronounced in the nonanalytic clock sector than in the complete bispectrum. This becomes explicit in panel~$(b$, where the predicted $p$-dependent power-law envelope has been removed from the isolated homogeneous contribution. The resulting curves oscillate with the same logarithmic frequency fixed by $\mu $, while their relative amplitudes and phases vary with the temporal profile. Therefore, \cref{fig:bispectrum_p} provides a direct numerical check that the random field modifies how the massive collider modes are excited without shifting the heavy-field frequency.

For the largest values of $p$, the curves are deliberately extended into the deep squeezed region beyond the conservative regime $\epsilon_{\rm dis}\lesssim1$. We retain this part of the plot to display the asymptotic tendency induced by the time-dependent disorder and to make the enhancement of the massive clock manifest. The corresponding deep-squeezed portion should therefore be interpreted as an extrapolation of the exact result rather than as a parametrically controlled benchmark.

\section{Conclusion}

We have shown that a spatially quenched random environment can modify cosmological-collider signals without modifying the underlying massive spectrum. In the linear Gaussian problem considered here, the disorder leaves the heavy-field poles and their characteristic de Sitter exponents unchanged, while modifying the statistical sector of the propagator. As a result, the massive modes are excited with different amplitudes and phases and are imprinted differently in inflationary correlators, even though their spectral information is preserved.

The temporal profile of the random source controls the relative importance of the forced and massive sectors. For a persistent source, \(p=0\), the static late-time response is finite in the trispectrum seed but produces a local endpoint logarithm in the \(\ell=-2\) block relevant to the bispectrum. Once the source decays, this local obstruction disappears while the homogeneous massive response survives. The collider oscillation can therefore remain as a memory of a disordered stage after the forcing itself has become negligible.

For spatial white noise, the value \(p=3/2\) is particularly simple because the explicit exchanged-scale dependence of the disorder seed disappears. In the collapsed trispectrum, the coefficient of the original de Sitter collider pair becomes
\[
1-\frac{\pi f_0}{\mu^2},
\]
so that this branch is progressively suppressed by destructive interference, vanishes at
\begin{equation}
f_0=\frac{\mu^2}{\pi},
\end{equation}
and changes sign for larger disorder strength. The cancellation does not remove the complete oscillatory signal, since the factorized disorder contribution also generates mixed analytic--collider structures. The random field therefore reorganizes the relative harmonic content of the collider signal rather than simply damping it.

The bispectrum provides a complementary manifestation of the same effect. In the squeezed limit, the disorder-induced massive and forced contributions behave as
\[
S_{\rm clock}(x)
\sim
x^{\frac72-2p}
\cos\!\left(\mu\log x+\phi_p\right),
\qquad
S_{\rm forced}(x)\sim x^{2-p}.
\]
The value \(p=3/2\) separates the two soft behaviors: for \(p<3/2\) the analytic forced response is relatively more important, whereas for \(p>3/2\) the nonanalytic massive contribution becomes parametrically dominant toward the squeezed limit. This enhancement is concentrated in the oscillatory sector and does not imply an equally large deformation of the complete bispectrum shape. Most importantly, varying \(p\) changes the power-law envelope, amplitude, and phase of the oscillation while leaving its logarithmic frequency \(\mu\) unchanged. The mass therefore continues to determine the clock frequency, while the temporal history of the disorder determines how that clock is excited.

This separation also gives a simple physical interpretation of the late-time signal. The local forced contribution can disappear as the source decays, while the homogeneous massive modes excited during the disordered stage remain. Cosmological-collider correlators can therefore retain information about an earlier environment even when that environment is no longer active. In the present model, this information is carried by the relative weighting and phase of the massive branches rather than by a displacement of the heavy-field spectrum itself.

Several extensions follow naturally from this construction. A particularly interesting possibility is multiplicative disorder, such as a \textit{random mass}. In that case the randomness enters directly in the quadratic operator governing propagation, so the disorder-averaged retarded propagator and spectral function may themselves be modified. Such a deformation could shift or broaden the heavy-field poles and thereby affect the principal-series parameter \(\mu\), leading to a qualitatively different situation in which the environment changes not only how the cosmological clock is excited, but potentially also its spectrum.

A complementary direction is to formulate the problem in Euclidean signature. Quenched disorder admits a natural replica description in AdS~\cite{Fujita:2008zv}, while the relation between late-time Schwinger--Keldysh correlators in de Sitter and boundary correlators in EAdS provides a possible route toward such a construction~\cite{Abhishek:2025oki}. More realistic disorder profiles, including finite spatial correlation lengths and smoothly localized time dependence, would also allow one to study finite-duration disordered epochs and the efficiency with which their history is retained in the late-time massive signal. The power-law family developed here provides an analytic starting point for these generalizations through Mellin superposition.

The additive random field considered in this work thus provides a controlled example in which an inflationary environment changes the observable realization of a cosmological-collider signal while leaving its characteristic mass-dependent frequency intact.

\acknowledgments

The author acknowledges financial support from CNPq. I am especially grateful to my friend Felipe Sobrero, whose suggestion to introduce disorder into this problem sparked the present work; to Nami F. Svaiter for many stimulating and extensive discussions on disorder in quantum field theory; and to Felipe T. Falciano for valuable discussions and comments throughout the development of this work.

\appendix

\section{de Sitter exchange seed}
\label{app:cleanDS}

For the phenomenological comparison with the random-field contribution, we need the complete clean scalar-exchange seed, including the ordered equal-sign Schwinger--Keldysh sectors.  We therefore collect here the global de Sitter result in the same momentum variables and normalization used in the main text.  The direct Mellin--Barnes and bootstrap constructions of these seeds are standard; see Refs.~\cite{Arkani-Hamed:2015bza,Qin:2022pmb,Qin:2023closed,Stefanyszyn:2023factorisation}.  The formulas below are written so that they can be used directly for the trispectrum and single-folded bispectrum comparisons discussed in the main text.

For a principal-series scalar, $\nu=\ii\mu$, define
\begin{equation}
  u=\frac{s}{k_{12}},
  \qquad
  v=\frac{s}{k_{34}},
  \qquad
  0<u,v\leq1,
\end{equation}
and the clean dimensionless seed
\begin{align}
  \widehat{\mathcal I}_{ab}^{\ell_1\ell_2}(u,v)
  \equiv{}&-ab\,\frac{s^{5+\ell_1+\ell_2}}{H^2}
  \int_{-\infty_a}^{0}\dd\eta_1
  \int_{-\infty_b}^{0}\dd\eta_2\,
  (-\eta_1)^{\ell_1}(-\eta_2)^{\ell_2}
  \nonumber\\
  &\times
  \ee^{\ii a k_{12}\eta_1+\ii b k_{34}\eta_2}
  G^{(0)}_{ab}(s;\eta_1,\eta_2),
  \qquad a,b=\pm1.
  \label{eq:cleanseeddef}
\end{align}
The physical clean seed is the sum over the four contour assignments,
\begin{equation}
  \widehat{\mathcal I}_{\rm dS}^{\ell_1\ell_2}(u,v)
  =\sum_{a,b=\pm1}
  \widehat{\mathcal I}_{ab}^{\ell_1\ell_2}(u,v).
  \label{eq:cleanfullSKsum}
\end{equation}
For compactness in the ordered formulas we use
\begin{equation}
  \lambda_i\equiv\ell_i+\frac52,
  \qquad
  P\equiv5+\ell_1+\ell_2=\lambda_1+\lambda_2.
  \label{eq:cleanlambdaP}
\end{equation}

The mixed Wightman sectors factorize into one-vertex transforms.  A convenient homogeneous basis is
\begin{align}
  \Phi_\sigma^{\ell}(u)
  \equiv{}&
  2^{-\sigma\ii\mu}
  u^{\ell+\frac52+\sigma\ii\mu}
  \Gamma\!\left(\ell+\frac52+\sigma\ii\mu\right)
  \Gamma(-\sigma\ii\mu)
  \nonumber\\
  &\times{}_2F_1\!\left(
  \begin{matrix}
  \frac{\ell}{2}+\frac54+\frac{\sigma\ii\mu}{2},
  \frac{\ell}{2}+\frac74+\frac{\sigma\ii\mu}{2}\\
  1+\sigma\ii\mu
  \end{matrix};u^2\right),
  \qquad \sigma=\pm1.
  \label{eq:cleanPhidef}
\end{align}
For real $u$ and $\mu$, $[\Phi_+^{\ell}(u)]^*=\Phi_-^{\ell}(u)$.  With the conventions of \cref{eq:cleanseeddef}, the two mixed-sign seeds are
\begin{align}
  \widehat{\mathcal I}_{-+}^{\ell_1\ell_2}(u,v)
  ={}&\frac{\ee^{+\ii\pi(\ell_1-\ell_2)/2}}{4\pi}
  \left[\Phi_+^{\ell_1}(u)+\Phi_-^{\ell_1}(u)\right]
  \left[\Phi_+^{\ell_2}(v)+\Phi_-^{\ell_2}(v)\right],
  \\
  \widehat{\mathcal I}_{+-}^{\ell_1\ell_2}(u,v)
  ={}&\frac{\ee^{-\ii\pi(\ell_1-\ell_2)/2}}{4\pi}
  \left[\Phi_+^{\ell_1}(u)+\Phi_-^{\ell_1}(u)\right]
  \left[\Phi_+^{\ell_2}(v)+\Phi_-^{\ell_2}(v)\right].
  \label{eq:cleanmixedseeds}
\end{align}

The equal-sign sectors contain the causal ordered contribution.  In the momentum ordering
\begin{equation}
  0<u<v<1,
  \label{eq:cleanordering}
\end{equation}
the exact result separates as
\begin{equation}
  \widehat{\mathcal I}_{aa}^{\ell_1\ell_2}(u,v)
  =\widehat{\mathcal I}_{aa,{\rm F}}^{\ell_1\ell_2}(u,v)
  +\widehat{\mathcal I}_{aa,{\rm TO}}^{\ell_1\ell_2}(u,v),
  \qquad a=\pm1.
  \label{eq:cleanequalsignsplit}
\end{equation}
The first term is the factorized Wightman completion fixed by the Schwinger--Keldysh contour,
\begin{equation}
\begin{aligned}
\widehat{\mathcal I}_{aa,{\rm F}}^{\ell_1\ell_2}(u,v)
={}&
\frac{a\ii\,\ee^{-a\ii\pi(\ell_1+\ell_2)/2}}{4\pi}
\left[
\ee^{\pi\mu}\Phi_a^{\ell_1}(u)
+\ee^{-\pi\mu}\Phi_{-a}^{\ell_1}(u)
\right]
\\
&\times
\left[
\Phi_+^{\ell_2}(v)+\Phi_-^{\ell_2}(v)
\right].
\end{aligned}
\label{eq:cleanfactorizedequalsign}
\end{equation}
The genuinely time-ordered part is most conveniently represented by its exact convergent residue series,
{\small
\begin{equation}
\begin{aligned}
\widehat{\mathcal I}_{aa,{\rm TO}}^{\ell_1\ell_2}(u,v)
={}&-
\frac{\ii\,\ee^{-a\ii\pi(\ell_1+\ell_2)/2}\sinh(\pi\mu)}{2\pi}
\,u^P
\sum_{\sigma=\pm1}\sigma
\sum_{n_1,n_2=0}^{\infty}
\frac{(-1)^{n_1+n_2}}{n_1!\,n_2!}
\left(\frac{u}{2}\right)^{2(n_1+n_2)}
\\
&\times
\Gamma(-n_1-\sigma\ii\mu)
\Gamma(-n_2+\sigma\ii\mu)
\Gamma(\lambda_2+2n_2-\sigma\ii\mu)
\Gamma(P+2n_1+2n_2)
\\
&\times
{}_2\widetilde F_1\!\left(
\begin{matrix}
\lambda_2+2n_2-\sigma\ii\mu,
P+2n_1+2n_2\\
\lambda_2+1+2n_2-\sigma\ii\mu
\end{matrix};-\frac{u}{v}\right),
\end{aligned}
\label{eq:cleanorderedresidues}
\end{equation}
}
where
\begin{equation}
  {}_2\widetilde F_1(A,B;C;z)
  \equiv\frac{{}_2F_1(A,B;C;z)}{\Gamma(C)}.
  \label{eq:cleanregularized2F1}
\end{equation}
For $u>v$, the corresponding representation is obtained by exchanging
$(u,\ell_1)\leftrightarrow(v,\ell_2)$.  Hermiticity of the original in--in integrals gives
\begin{equation}
  \widehat{\mathcal I}_{--}^{\ell_1\ell_2}(u,v)
  =\left[
  \widehat{\mathcal I}_{++}^{\ell_1\ell_2}(u,v)
  \right]^*.
  \label{eq:cleanhermiticity}
\end{equation}
Equations~\eqref{eq:cleanmixedseeds}, \eqref{eq:cleanfactorizedequalsign}, and \eqref{eq:cleanorderedresidues}, together with the exchange rule across $u=v$, give the complete global clean de Sitter seed.

The analytic distinction between the two equal-sign pieces is useful in the collapsed limit.  The factorized contribution contains the collider branches
\begin{equation}
  \Phi_\sigma^{\ell}(u)
  =2^{-\sigma\ii\mu}
  \Gamma\!\left(\ell+\frac52+\sigma\ii\mu\right)
  \Gamma(-\sigma\ii\mu)
  u^{\ell+\frac52+\sigma\ii\mu}
  \left[1+\mathcal O(u^2)\right],
  \label{eq:cleanPhisoft}
\end{equation}
whereas the ordered term contains only integer powers after the overall factor $u^P$ is extracted.  Thus the nonanalytic cosmological-collider oscillations arise from the homogeneous Wightman completion, while the ordered contribution supplies the analytic causal completion required by the Bunch--Davies contour.

For the single-folded kinematics needed for the bispectrum, $v=1$, it is preferable not to evaluate the two pieces of \cref{eq:cleanequalsignsplit} separately from their generic $v<1$ forms.  The folded continuation is most stable when performed analytically first.  The finite value of the second homogeneous factor is
\begin{equation}
\lim_{v\to1}
\left[
\Phi_+^{\ell_2}(v)+\Phi_-^{\ell_2}(v)
\right]
=
2^{-\frac32-\ell_2}\sqrt{\pi}\,
\frac{
\Gamma\!\left(\ell_2+\frac52-\ii\mu\right)
\Gamma\!\left(\ell_2+\frac52+\ii\mu\right)
}{\Gamma(\ell_2+3)}.
\label{eq:cleanPhifoldedvalue}
\end{equation}
Consequently, the folded factorized equal-sign contribution is
\begin{equation}
\begin{aligned}
\widehat{\mathcal I}_{aa,{\rm hom}}^{\ell_1\ell_2}(u,1)
={}&
\frac{a\ii\,\ee^{-a\ii\pi(\ell_1+\ell_2)/2}
2^{-\frac72-\ell_2}}{\sqrt{\pi}}
\frac{
\Gamma(\lambda_2-\ii\mu)
\Gamma(\lambda_2+\ii\mu)
}{\Gamma(\ell_2+3)}
\\
&\times
\left[
\ee^{\pi\mu}\Phi_a^{\ell_1}(u)
+\ee^{-\pi\mu}\Phi_{-a}^{\ell_1}(u)
\right].
\end{aligned}
\label{eq:cleanfoldedhom}
\end{equation}
The ordered contribution on the same edge resums to a single generalized hypergeometric function,
\begin{equation}
\begin{aligned}
\widehat{\mathcal I}_{aa,{\rm part}}^{\ell_1\ell_2}(u,1)
={}&
\ee^{-a\ii\pi(\ell_1+\ell_2)/2}
\frac{\Gamma(P)}{\lambda_2^2+\mu^2}
\left(\frac{u}{1+u}\right)^P
\\
&\times
{}_3F_2\!\left(
\begin{matrix}
1,\,3+\ell_2,\,P\\
\lambda_2+1-\ii\mu,\,
\lambda_2+1+\ii\mu
\end{matrix};\frac{2u}{1+u}\right).
\end{aligned}
\label{eq:cleanfoldedpart}
\end{equation}
The complete equal-sign seed on the single-folded edge is therefore
\begin{equation}
  \widehat{\mathcal I}_{aa}^{\ell_1\ell_2}(u,1)
  =
  \widehat{\mathcal I}_{aa,{\rm hom}}^{\ell_1\ell_2}(u,1)
  +
  \widehat{\mathcal I}_{aa,{\rm part}}^{\ell_1\ell_2}(u,1).
  \label{eq:cleanfoldedequalsign}
\end{equation}
This representation is particularly useful numerically because the first folded continuation has already been carried out analytically.

For the trispectrum channel used in the phenomenological discussion, $\ell_1=\ell_2=0$, one simply takes
\begin{equation}
  \lambda_1=\lambda_2=\frac52,
  \qquad P=5,
\end{equation}
so that
\begin{equation}
  \widehat{\mathcal I}_{\rm dS}^{00}(u,v)
  =
  \widehat{\mathcal I}_{++}^{00}(u,v)
  +\widehat{\mathcal I}_{--}^{00}(u,v)
  +2\widehat{\mathcal I}_{+-}^{00}(u,v).
  \label{eq:cleantrispectrumseed}
\end{equation}
Here the equality of the two mixed-sign terms follows from \cref{eq:cleanmixedseeds}.  This is the clean seed to be compared directly with the disorder correction $\Delta\widehat{\mathcal I}_{\rm RF}^{00}(u,v,s,p)$.

For the single-folded bispectrum channel, $\ell_1=0$, $\ell_2=-2$, and $v=1$, \cref{eq:cleanfoldedpart} simplifies to
\begin{equation}
\begin{aligned}
\widehat{\mathcal I}_{aa,{\rm part}}^{0,-2}(u,1)
={}&-
\frac{2}{\mu^2+\frac14}
\left(\frac{u}{1+u}\right)^3
{}_3F_2\!\left(
\begin{matrix}
1,1,3\\
\frac32-\ii\mu,\frac32+\ii\mu
\end{matrix};\frac{2u}{1+u}\right).
\end{aligned}
\label{eq:cleanbispectrumfoldedpart}
\end{equation}
Together with \cref{eq:cleanfoldedhom,eq:cleanmixedseeds,eq:cleanPhifoldedvalue}, this gives a stable analytic representation of the complete clean de Sitter seed at the kinematics required for the bispectrum.  Upon the quadratic change of variables $u_{\rm Aoki}=2u/(1+u)$, the single-folded expression agrees with the constant-mass limit of the inflationary correlator derived in Ref.~\cite{Aoki:2023dynamical}; this provides an additional analytic check of the normalization and folded continuation used here.

\section{Hypergeometric Integrals}
\label{app:hyperinetgrals}

This appendix collects the hypergeometric manipulations entering the exact one-vertex transform and its folded continuation.  We first derive the closed form of \eqref{eq:Rvertexdef} directly from the Lommel decomposition, keeping the three contributions separate so that their analytic origin remains explicit.  We then continue the resulting ${}_3F_2$ and ${}_2F_1$ blocks to the folded point, show the cancellation of their common nonanalytic branch, and finally rewrite the finite remainder in a form in which the apparent poles at integer $\ell$ are removed algebraically.  No additional boundary condition is imposed in this appendix; all phases and homogeneous coefficients are those already fixed by the retarded prescription in the main text.

We now evaluate \eqref{eq:Rvertexdef} without skipping the intermediate steps.  First use the Bessel connection formula
\begin{equation}
  Y_\nu(x)
  =\cot(\pi\nu)J_\nu(x)
  -\csc(\pi\nu)J_{-\nu}(x),
  \label{eq:YtoJpmRF}
\end{equation}
so that \eqref{eq:SminusS} becomes
\begin{align}
  S_{p-\frac52,\nu}(x)
  ={}&s_{p-\frac52,\nu}(x)
  \nonumber\\
  &+\mathcal C_{p,\nu}
  \left[\sin\theta_{p,\nu}-\cos\theta_{p,\nu}\cot(\pi\nu)\right]J_\nu(x)
  \nonumber\\
  &+\mathcal C_{p,\nu}\cos\theta_{p,\nu}\csc(\pi\nu)J_{-\nu}(x).
  \label{eq:SthreeblocksRF}
\end{align}
Hence the single contour transform separates into three pieces,
\begin{equation}
  \mathcal R_{a,p}^{\ell}(u)
  =\mathcal R_{a,p;\mathrm{part}}^{\ell}(u)
  +\mathcal R_{a,p;+}^{\ell}(u)
  +\mathcal R_{a,p;-}^{\ell}(u).
  \label{eq:RthreeblocksRF}
\end{equation}
The only integral identity required in all three cases is the oscillatory Mellin transform
\begin{equation}
  \int_0^\infty\dd x\,x^{q-1}e^{-\ii a x/u}
  =e^{-\ii a\pi q/2}\Gamma(q)u^q,
  \qquad a=\pm1,
  \label{eq:oscillatorytransformRF}
\end{equation}
initially in its convergence domain and elsewhere by analytic continuation with the inherited Bunch--Davies $\ii\epsilon$ prescription.

\paragraph{Particular Lommel block: ${}_1F_2\rightarrow{}_3F_2$.}
Using \eqref{eq:smallLommelHyper}, the particular contribution is
\begin{align}
  \mathcal R_{a,p;\mathrm{part}}^{\ell}(u)
  ={}&\frac{1}{\Delta_p}
  \int_0^\infty\dd x\,
  x^{\ell+p}e^{-\ii a x/u}
  \nonumber\\[-1mm]
  &\times{}_1F_2\!\left(
  1;
  \frac p2+\frac14-\frac\nu2,
  \frac p2+\frac14+\frac\nu2;
  -\frac{x^2}{4}
  \right).
  \label{eq:RpartIntegralRF}
\end{align}
Since $(1)_n/n!=1$, the hypergeometric series reads
\begin{align}
  {}_1F_2\!\left(
  1;
  \frac p2+\frac14-\frac\nu2,
  \frac p2+\frac14+\frac\nu2;
  -\frac{x^2}{4}
  \right)
  ={}&\sum_{n=0}^{\infty}
  \frac{(-1)^n x^{2n}}{4^n}
  \nonumber\\[-1mm]
  &\times
  \frac{1}{
  \left(\frac p2+\frac14-\frac\nu2\right)_n
  \left(\frac p2+\frac14+\frac\nu2\right)_n}.
  \label{eq:1F2seriesRF}
\end{align}
Term-by-term use of \eqref{eq:oscillatorytransformRF}, with
$q=\ell+p+2n+1$, gives
\begin{align}
  \mathcal R_{a,p;\mathrm{part}}^{\ell}(u)
  ={}&\frac{1}{\Delta_p}
  \sum_{n=0}^{\infty}
  \frac{(-1)^n4^{-n}}{
  \left(\frac p2+\frac14-\frac\nu2\right)_n
  \left(\frac p2+\frac14+\frac\nu2\right)_n}
  \nonumber\\
  &\times
  e^{-\frac{\ii a\pi}{2}(\ell+p+2n+1)}
  \Gamma(\ell+p+2n+1)
  u^{\ell+p+2n+1}.
  \label{eq:RpartAfterIntegralRF}
\end{align}
For $a=\pm1$ and integer $n$,
\begin{equation}
  e^{-\ii a\pi n}=(-1)^n,
  \label{eq:phaseCancellationRF}
\end{equation}
so the two factors $(-1)^n$ cancel.  We then use
\begin{align}
  \Gamma(\ell+p+2n+1)
  ={}&\Gamma(\ell+p+1)(\ell+p+1)_{2n}
  \nonumber\\
  ={}&\Gamma(\ell+p+1)4^n
  \left(\frac{\ell+p+1}{2}\right)_n
  \left(\frac{\ell+p+2}{2}\right)_n,
  \label{eq:duplicationPochhammerRF}
\end{align}
where the second line is the duplication formula for the Pochhammer symbol.  The factor $4^n$ cancels $4^{-n}$ in \eqref{eq:RpartAfterIntegralRF}.  Restoring $(1)_n/n!=1$, the remaining series is precisely a ${}_3F_2$,
\begin{equation}
  \begin{aligned}
  \mathcal R_{a,p;\mathrm{part}}^{\ell}(u)
  ={}&
  \frac{\Gamma(\ell+p+1)}{\Delta_p}
  e^{-\frac{\ii a\pi}{2}(\ell+p+1)}
  u^{\ell+p+1}
  \\[-1mm]
  &\times{}_3F_2\!\left(
  \begin{matrix}
  1,\frac{\ell+p+1}{2},\frac{\ell+p+2}{2}\\
  \frac p2+\frac14-\frac\nu2,\frac p2+\frac14+\frac\nu2
  \end{matrix};u^2\right).
  \end{aligned}
  \label{eq:Rpart3F2RF}
\end{equation}

\paragraph{Homogeneous Bessel blocks: $J_{\pm\nu}\rightarrow{}_2F_1$.}
For the $J_\nu$ branch, expand
\begin{equation}
  J_\nu(x)
  =\sum_{n=0}^{\infty}
  \frac{(-1)^n}{n!\,\Gamma(n+1+\nu)}
  \left(\frac{x}{2}\right)^{2n+\nu}.
  \label{eq:JplusSeriesRF}
\end{equation}
Substitution into the one-vertex integral gives
\begin{align}
  &\int_0^\infty\dd x\,
  x^{\ell+\frac32}e^{-\ii a x/u}J_\nu(x)
  \nonumber\\
  &\quad=
  \sum_{n=0}^{\infty}
  \frac{(-1)^n2^{-2n-\nu}}{n!\,\Gamma(n+1+\nu)}
  e^{-\frac{\ii a\pi}{2}(\ell+\frac52+\nu+2n)}
  \nonumber\\[-1mm]
  &\hspace{38mm}\times
  \Gamma\!\left(\ell+\frac52+\nu+2n\right)
  u^{\ell+\frac52+\nu+2n}.
  \label{eq:JplusAfterIntegralRF}
\end{align}
Again $e^{-\ii a\pi n}=(-1)^n$ cancels the sign of the Bessel series.  Using
\begin{align}
  \Gamma(n+1+\nu)
  ={}&\Gamma(1+\nu)(1+\nu)_n,
  \\
  \Gamma\!\left(\ell+\frac52+\nu+2n\right)
  ={}&\Gamma\!\left(\ell+\frac52+\nu\right)4^n
  \left(\frac\ell2+\frac54+\frac\nu2\right)_n
  \left(\frac\ell2+\frac74+\frac\nu2\right)_n,
  \label{eq:JplusGammaRF}
\end{align}
the factor $4^n$ cancels $2^{-2n}$ and the remaining series resums to
\begin{equation}
  \begin{aligned}
  &\int_0^\infty\dd x\,
  x^{\ell+\frac32}e^{-\ii a x/u}J_\nu(x)
  \\
  &\quad=
  \frac{2^{-\nu}\Gamma\!\left(\ell+\frac52+\nu\right)}{\Gamma(1+\nu)}
  e^{-\frac{\ii a\pi}{2}(\ell+\frac52+\nu)}
  u^{\ell+\frac52+\nu}
  \\[-1mm]
  &\hspace{26mm}\times
  {}_2F_1\!\left(
  \frac\ell2+\frac54+\frac\nu2,
  \frac\ell2+\frac74+\frac\nu2;
  1+\nu;u^2\right).
  \end{aligned}
  \label{eq:Jplus2F1RF}
\end{equation}
The $J_{-\nu}$ transform follows by $\nu\rightarrow-\nu$,
\begin{equation}
  \begin{aligned}
  &\int_0^\infty\dd x\,
  x^{\ell+\frac32}e^{-\ii a x/u}J_{-\nu}(x)
  \\
  &\quad=
  \frac{2^{\nu}\Gamma\!\left(\ell+\frac52-\nu\right)}{\Gamma(1-\nu)}
  e^{-\frac{\ii a\pi}{2}(\ell+\frac52-\nu)}
  u^{\ell+\frac52-\nu}
  \\[-1mm]
  &\hspace{26mm}\times
  {}_2F_1\!\left(
  \frac\ell2+\frac54-\frac\nu2,
  \frac\ell2+\frac74-\frac\nu2;
  1-\nu;u^2\right).
  \end{aligned}
  \label{eq:Jminus2F1RF}
\end{equation}
Finally, multiplying \cref{eq:Jplus2F1RF,eq:Jminus2F1RF} by the coefficients in \eqref{eq:SthreeblocksRF} and adding the particular contribution \eqref{eq:Rpart3F2RF}, one obtains
\begin{align}
  \mathcal R_{a,p}^{\ell}(u)
  ={}&
  \frac{\Gamma(\ell+p+1)}{\Delta_p}
  e^{-\frac{\ii a\pi}{2}(\ell+p+1)}
  u^{\ell+p+1}
  {}_3F_2\!\left(
  \begin{matrix}
  1,\frac{\ell+p+1}{2},\frac{\ell+p+2}{2}\\
  \frac p2+\frac14-\frac\nu2,\frac p2+\frac14+\frac\nu2
  \end{matrix};u^2\right)
  \nonumber\\[1mm]
  &+\mathcal C_{p,\nu}
  \left[\sin\theta_{p,\nu}-\cos\theta_{p,\nu}\cot(\pi\nu)\right]
  \frac{2^{-\nu}\Gamma\!\left(\ell+\frac52+\nu\right)}{\Gamma(1+\nu)}
  e^{-\frac{\ii a\pi}{2}(\ell+\frac52+\nu)}
  u^{\ell+\frac52+\nu}
  \nonumber\\[-1mm]
  &\hspace{12mm}\times
  {}_2F_1\!\left(
  \frac\ell2+\frac54+\frac\nu2,
  \frac\ell2+\frac74+\frac\nu2;
  1+\nu;u^2\right)
  \nonumber\\[1mm]
  &+\mathcal C_{p,\nu}\cos\theta_{p,\nu}\csc(\pi\nu)
  \frac{2^{\nu}\Gamma\!\left(\ell+\frac52-\nu\right)}{\Gamma(1-\nu)}
  e^{-\frac{\ii a\pi}{2}(\ell+\frac52-\nu)}
  u^{\ell+\frac52-\nu}
  \nonumber\\[-1mm]
  &\hspace{12mm}\times
  {}_2F_1\!\left(
  \frac\ell2+\frac54-\frac\nu2,
  \frac\ell2+\frac74-\frac\nu2;
  1-\nu;u^2\right).
  \label{eq:Rvertexclosed}
\end{align}
The effect of $p$ is highly structured: it shifts the analytic forced block from $u^{\ell+1}$ to $u^{\ell+p+1}$ and changes its hypergeometric parameters, while the nonanalytic collider exponents remain $u^{\ell+5/2\pm\nu}$.  The amplitudes of the latter nevertheless depend on $p$ through $\mathcal C_{p,\nu}$ and $\theta_{p,\nu}$.  An independent derivation from the de Sitter bootstrap equations is given in appendix~\ref{app:RFbootstrap}, where the ordinary de Sitter contact source and the additional random-field source are treated simultaneously.

For $p\leq1$ and $\ell=-2$, the individual contour integrals in \cref{eq:Rvertexdef} need analytic continuation at the future endpoint.  The physical contour difference below is better behaved: for every $p>0$ it can be defined directly before separating the two branches.

\subsection{Folded limit}
We now take the folded limit $u\to1$ directly in the three hypergeometric functions appearing in \cref{eq:Rvertexclosed}.  The continuation formulas used below are standard: for ${}_2F_1$ we use the connection formula at unit argument, while for ${}_3F_2$ we use the finite-part formula employed in Ref.~\cite{Qin:2023closed}, which follows from the continuation analysis of B\"uhring~\cite{Buhring:1987unit}; see also Ref.~\cite{NIST:DLMF}.  We keep $\ell$ generic throughout the continuation and only afterwards take the integer values relevant for the physical vertices.

\medskip
\noindent\textbf{1. Common folded balance.}
The ${}_3F_2$ block in \cref{eq:Rvertexclosed} has parametric excess
\begin{align}
&\left(\frac p2+\frac14-\frac\nu2\right)
+\left(\frac p2+\frac14+\frac\nu2\right)
-1-\frac{\ell+p+1}{2}-\frac{\ell+p+2}{2}
=-\ell-2.
\label{eq:foldedBalanceP}
\end{align}
The same quantity is obtained for each Gauss function,
\begin{align}
&(1\pm\nu)
-\left(\frac\ell2+\frac54\pm\frac\nu2\right)
-\left(\frac\ell2+\frac74\pm\frac\nu2\right)
=-\ell-2.
\end{align}
Hence all three blocks contain the same possible nonanalytic folded sector,
\begin{equation}
(1-u^2)^{-\ell-2}.
\label{eq:foldedBranchGeneric}
\end{equation}
The temporal weight $p$ changes its coefficient but not its exponent.

\medskip
\noindent\textbf{2. Folded continuation of the $+\nu$ Gauss block.}
For noninteger balance, the finite part of the Gauss function at unit argument is~\cite{Qin:2023closed,NIST:DLMF}
\begin{align}
&\operatorname{Fin}_{u\to1}
{}_2F_1\!\left(
\frac\ell2+\frac54+\frac\nu2,
\frac\ell2+\frac74+\frac\nu2;
1+\nu;u^2\right)
\nonumber\\
&\hspace{12mm}=
\frac{\Gamma(1+\nu)\Gamma(-\ell-2)}
{\Gamma\!\left(-\frac\ell2-\frac34+\frac\nu2\right)
 \Gamma\!\left(-\frac\ell2-\frac14+\frac\nu2\right)}.
\label{eq:foldedFinitePlus}
\end{align}
The accompanying nonanalytic branch is
\begin{align}
&\frac{\Gamma(1+\nu)\Gamma(\ell+2)}
{\Gamma\!\left(\frac\ell2+\frac54+\frac\nu2\right)
 \Gamma\!\left(\frac\ell2+\frac74+\frac\nu2\right)}
(1-u^2)^{-\ell-2}.
\label{eq:foldedNonanalyticPlus}
\end{align}

\medskip
\noindent\textbf{3. Folded continuation of the $-\nu$ Gauss block.}
Similarly,
\begin{align}
&\operatorname{Fin}_{u\to1}
{}_2F_1\!\left(
\frac\ell2+\frac54-\frac\nu2,
\frac\ell2+\frac74-\frac\nu2;
1-\nu;u^2\right)
\nonumber\\
&\hspace{12mm}=
\frac{\Gamma(1-\nu)\Gamma(-\ell-2)}
{\Gamma\!\left(-\frac\ell2-\frac34-\frac\nu2\right)
 \Gamma\!\left(-\frac\ell2-\frac14-\frac\nu2\right)},
\label{eq:foldedFiniteMinus}
\end{align}
while its nonanalytic branch is
\begin{align}
&\frac{\Gamma(1-\nu)\Gamma(\ell+2)}
{\Gamma\!\left(\frac\ell2+\frac54-\frac\nu2\right)
 \Gamma\!\left(\frac\ell2+\frac74-\frac\nu2\right)}
(1-u^2)^{-\ell-2}.
\label{eq:foldedNonanalyticMinus}
\end{align}

\medskip
\noindent\textbf{4. Folded continuation of the ${}_3F_2$ block.}
For a generic ${}_3F_2$ with noninteger balance, Ref.~\cite{Qin:2023closed} uses the finite-part formula derived from B\"uhring's continuation at unit argument~\cite{Buhring:1987unit}.  Applied directly to the parameters of \cref{eq:Rvertexclosed}, it gives
\begin{align}
&\operatorname{Fin}_{u\to1}
{}_3F_2\!\left(
\begin{matrix}
1,\frac{\ell+p+1}{2},\frac{\ell+p+2}{2}\\
\frac p2+\frac14-\frac\nu2,\frac p2+\frac14+\frac\nu2
\end{matrix};u^2\right)
\nonumber\\[1mm]
&=
\frac{
\Gamma\!\left(\frac p2+\frac14-\frac\nu2\right)
\Gamma\!\left(\frac p2+\frac14+\frac\nu2\right)
\Gamma(-\ell-2)}
{\Gamma\!\left(\frac{p-\ell-3}{2}\right)
 \Gamma\!\left(\frac{p-\ell-2}{2}\right)}
\nonumber\\[1mm]
&\quad\times
{}_3F_2\!\left(
\begin{matrix}
\frac p2-\frac34-\frac\nu2,
\frac p2-\frac34+\frac\nu2,
-\ell-2\\
\frac{p-\ell-3}{2},
\frac{p-\ell-2}{2}
\end{matrix};1\right).
\label{eq:foldedFinite3F2}
\end{align}
The transformed ${}_3F_2(1)$ is convergent because its parametric excess is
\begin{align}
&\frac{p-\ell-3}{2}+\frac{p-\ell-2}{2}
-\left(\frac p2-\frac34-\frac\nu2\right)
-\left(\frac p2-\frac34+\frac\nu2\right)
-(-\ell-2)
=1.
\label{eq:foldedTransformedBalance}
\end{align}
The nonanalytic branch of the original ${}_3F_2$ is~\cite{Buhring:1987unit}
\begin{align}
&\frac{
\Gamma\!\left(\frac p2+\frac14-\frac\nu2\right)
\Gamma\!\left(\frac p2+\frac14+\frac\nu2\right)
\Gamma(\ell+2)}
{\Gamma\!\left(\frac{\ell+p+1}{2}\right)
 \Gamma\!\left(\frac{\ell+p+2}{2}\right)}
(1-u^2)^{-\ell-2}.
\label{eq:foldedNonanalytic3F2}
\end{align}

\medskip
\noindent\textbf{5. Restoring the prefactors of \cref{eq:Rvertexclosed}.}
Using the duplication formula for the Gamma functions, the three nonanalytic contributions to the one-vertex response are
\begin{align}
\left.\mathcal R_{a,p,\mathrm{part}}^\ell\right|_{\mathrm{folded\,nonan}}
={}&
\frac{2^{\ell+3/2}\mathcal C_{p,\nu}}{\sqrt\pi}
\Gamma(\ell+2)
 e^{-\frac{\ii a\pi}{2}(\ell+p+1)}
(1-u^2)^{-\ell-2},
\\
\left.\mathcal R_{a,p,+}^\ell\right|_{\mathrm{folded\,nonan}}
={}&
\frac{2^{\ell+3/2}\mathcal C_{p,\nu}}{\sqrt\pi}
\Gamma(\ell+2)
\left[\sin\theta_{p,\nu}-\cos\theta_{p,\nu}\cot(\pi\nu)\right]
\nonumber\\
&\times e^{-\frac{\ii a\pi}{2}(\ell+\frac52+\nu)}
(1-u^2)^{-\ell-2},
\\
\left.\mathcal R_{a,p,-}^\ell\right|_{\mathrm{folded\,nonan}}
={}&
\frac{2^{\ell+3/2}\mathcal C_{p,\nu}}{\sqrt\pi}
\Gamma(\ell+2)
\cos\theta_{p,\nu}\csc(\pi\nu)
\nonumber\\
&\times e^{-\frac{\ii a\pi}{2}(\ell+\frac52-\nu)}
(1-u^2)^{-\ell-2}.
\label{eq:foldedThreeNonanalytic}
\end{align}

\medskip
\noindent\textbf{6. Cancellation of the nonanalytic folded branch.}
Factoring out the common prefactor of the three terms above, the coefficient multiplying $(1-u^2)^{-\ell-2}$ is
\begin{align}
1
&+\left[\sin\theta_{p,\nu}-\cos\theta_{p,\nu}\cot(\pi\nu)\right]
 e^{-\ii a\pi(\frac34-\frac p2+\frac\nu2)}
\nonumber\\
&+\cos\theta_{p,\nu}\csc(\pi\nu)
 e^{-\ii a\pi(\frac34-\frac p2-\frac\nu2)}.
\end{align}
Using
\begin{equation}
\theta_{p,\nu}=\frac\pi2\left(p-\frac52-\nu\right),
\end{equation}
this coefficient vanishes identically for both $a=+1$ and $a=-1$.  Therefore
\begin{equation}
\left.\mathcal R_{a,p}^{\ell}(u)\right|_{\mathrm{folded\,nonan}}=0,
\qquad a=\pm1,
\label{eq:RFfoldedcancel}
\end{equation}
for arbitrary $p$.  This is the same folded regularity mechanism emphasized in Ref.~\cite{Qin:2023closed}: the singular pieces of the separate hypergeometric blocks cancel in the Bunch--Davies combination.  In the present problem the same statement follows directly from the retarded large-$x$ behavior $S_{p-5/2,\nu}(x)\sim x^{p-7/2}$, which contains no residual $e^{\pm\ii x}$ wave.

\medskip
\noindent\textbf{7. Finite folded.}
After the nonanalytic branch has cancelled, the folded value is obtained by adding the finite parts in \cref{eq:foldedFinitePlus,eq:foldedFiniteMinus,eq:foldedFinite3F2} with the prefactors of \cref{eq:Rvertexclosed}.  For generic noninteger $\ell$ this gives
\begin{equation}
\begin{aligned}
\operatorname{Fin}_{u\to1}\mathcal R_{a,p}^{\ell}(u)
={}&
\frac{\Gamma(\ell+p+1)}{\Delta_p}
 e^{-\frac{\ii a\pi}{2}(\ell+p+1)}
\frac{
\Gamma\!\left(\frac p2+\frac14-\frac\nu2\right)
\Gamma\!\left(\frac p2+\frac14+\frac\nu2\right)
\Gamma(-\ell-2)}
{\Gamma\!\left(\frac{p-\ell-3}{2}\right)
 \Gamma\!\left(\frac{p-\ell-2}{2}\right)}
\\
&\times
{}_3F_2\!\left(
\begin{matrix}
\frac p2-\frac34-\frac\nu2,
\frac p2-\frac34+\frac\nu2,
-\ell-2\\
\frac{p-\ell-3}{2},
\frac{p-\ell-2}{2}
\end{matrix};1\right)
\\[1mm]
&+\mathcal C_{p,\nu}
\left[\sin\theta_{p,\nu}-\cos\theta_{p,\nu}\cot(\pi\nu)\right]
2^{-\nu}
\Gamma\!\left(\ell+\frac52+\nu\right)
 e^{-\frac{\ii a\pi}{2}(\ell+\frac52+\nu)}
\\
&\times
\frac{\Gamma(-\ell-2)}
{\Gamma\!\left(-\frac\ell2-\frac34+\frac\nu2\right)
 \Gamma\!\left(-\frac\ell2-\frac14+\frac\nu2\right)}
\\[1mm]
&+\mathcal C_{p,\nu}\cos\theta_{p,\nu}\csc(\pi\nu)
2^{\nu}
\Gamma\!\left(\ell+\frac52-\nu\right)
 e^{-\frac{\ii a\pi}{2}(\ell+\frac52-\nu)}
\\
&\times
\frac{\Gamma(-\ell-2)}
{\Gamma\!\left(-\frac\ell2-\frac34-\frac\nu2\right)
 \Gamma\!\left(-\frac\ell2-\frac14-\frac\nu2\right)}.
\end{aligned}
\label{eq:Rvertexfoldedgeneric}
\end{equation}
The crucial difference from simply substituting $u=1$ in \cref{eq:Rvertexclosed} is that the ${}_3F_2$ appearing here is the convergent finite-part continuation of Ref.~\cite{Qin:2023closed,Buhring:1987unit}, not the original ${}_3F_2$ evaluated naively at unit argument.

The expression in \cref{eq:Rvertexfoldedgeneric} is correct for generic noninteger $\ell$, but its regularity at the integer values relevant for the physical vertices is not manifest because all three terms contain $\Gamma(-\ell-2)$.  The apparent poles can be removed before any integer limit is taken.  The necessary step is to apply, directly to the ${}_3F_2(1)$ in the first line of \cref{eq:Rvertexfoldedgeneric}, the unit-argument transformation used in Ref.~\cite{Qin:2023closed}.  For the parameters appearing here, that transformation reads
\begin{align}
&{}_3F_2\!\left(
\begin{matrix}
\frac p2-\frac34-\frac\nu2,
\frac p2-\frac34+\frac\nu2,
-\ell-2\\
\frac{p-\ell-3}{2},
\frac{p-\ell-2}{2}
\end{matrix};1\right)
\nonumber\\
={}&
\frac{
\Gamma\!\left(\frac{p-\ell-3}{2}\right)
\Gamma\!\left(\frac{p-\ell-2}{2}\right)
\Gamma\!\left(\frac\ell2+\frac54+\frac\nu2\right)
\Gamma\!\left(\frac\ell2+\frac74+\frac\nu2\right)
}{
\Gamma\!\left(-\frac\ell2-\frac34+\frac\nu2\right)
\Gamma\!\left(\frac{\ell+p+1}{2}\right)
\Gamma\!\left(-\frac\ell2-\frac14+\frac\nu2\right)
\Gamma\!\left(\frac{\ell+p+2}{2}\right)
}
\nonumber\\
&-
\frac{2}{\ell+\frac52+\nu}
\frac{
\Gamma\!\left(\frac{p-\ell-3}{2}\right)
\Gamma\!\left(\frac{p-\ell-2}{2}\right)
}{
\Gamma\!\left(\frac p2-\frac34-\frac\nu2\right)
\Gamma(-\ell-2)
\Gamma\!\left(\frac p2+\frac14+\frac\nu2\right)
}
\nonumber\\
&\qquad\times
{}_3F_2\!\left(
\begin{matrix}
1,
\frac{\ell+p+1}{2},
-\frac\ell2-\frac34+\frac\nu2\\
\frac p2+\frac14+\frac\nu2,
\frac\ell2+\frac94+\frac\nu2
\end{matrix};1\right).
\label{eq:folded3F2secondtransformation}
\end{align}
The equality follows because the third upper parameter required by the transformation is precisely
\begin{equation}
\frac{p-\ell-3}{2}
+\frac{p-\ell-2}{2}
-\left(\frac p2-\frac34-\frac\nu2\right)
-(-\ell-2)-1
=
\frac p2-\frac34+\frac\nu2,
\end{equation}
while the remaining combinations reduce to
\begin{align}
\frac{p-\ell-3}{2}
-\left(\frac p2-\frac34-\frac\nu2\right)
&=-\frac\ell2-\frac34+\frac\nu2,
\\
\frac{p-\ell-3}{2}-(-\ell-2)
&=\frac{\ell+p+1}{2},
\\
\frac{p-\ell-2}{2}
-\left(\frac p2-\frac34-\frac\nu2\right)
&=-\frac\ell2-\frac14+\frac\nu2,
\\
\frac{p-\ell-2}{2}-(-\ell-2)
&=\frac{\ell+p+2}{2},
\\
\frac{p-\ell-3}{2}
-\left(\frac p2-\frac34-\frac\nu2\right)
-(-\ell-2)
&=\frac\ell2+\frac54+\frac\nu2,
\\
\frac{p-\ell-2}{2}
-\left(\frac p2-\frac34-\frac\nu2\right)
-(-\ell-2)
&=\frac\ell2+\frac74+\frac\nu2.
\end{align}
Thus no auxiliary notation is needed: every factor in \cref{eq:folded3F2secondtransformation} is already one of the parameters appearing in \cref{eq:Rvertexfoldedgeneric}.

Consider first the second term of \cref{eq:folded3F2secondtransformation}.  Multiplying it by the prefactor of the ${}_3F_2$ contribution in \cref{eq:Rvertexfoldedgeneric} cancels
\begin{equation}
\Gamma(-\ell-2)\,\frac{1}{\Gamma(-\ell-2)}=1,
\end{equation}
as well as the factors $\Gamma((p-\ell-3)/2)$ and $\Gamma((p-\ell-2)/2)$.  The Gamma-function recurrence relation gives directly
\begin{equation}
\frac{
\Gamma\!\left(\frac p2+\frac14-\frac\nu2\right)
}{
\Gamma\!\left(\frac p2-\frac34-\frac\nu2\right)
}
=
\frac12\left(p-\frac32-\nu\right).
\end{equation}
Together with
\begin{equation}
\Delta_p
=
\left(p-\frac32-\nu\right)
\left(p-\frac32+\nu\right),
\end{equation}
this contribution becomes
\begin{align}
&-
\frac{
\Gamma(\ell+p+1)
e^{-\frac{\ii a\pi}{2}(\ell+p+1)}
}{
\left(p-\frac32+\nu\right)
\left(\ell+\frac52+\nu\right)
}
\nonumber\\
&\qquad\times
{}_3F_2\!\left(
\begin{matrix}
1,
\frac{\ell+p+1}{2},
-\frac\ell2-\frac34+\frac\nu2\\
\frac p2+\frac14+\frac\nu2,
\frac\ell2+\frac94+\frac\nu2
\end{matrix};1\right).
\label{eq:foldedRegular3F2piece}
\end{align}
This part is therefore already free of the apparent factor $\Gamma(-\ell-2)$.

The first term of \cref{eq:folded3F2secondtransformation} must instead be combined with the two Gauss finite parts in \cref{eq:Rvertexfoldedgeneric}.  The duplication formula, applied directly to the Gamma functions appearing here, gives
\begin{equation}
\Gamma\!\left(\frac{\ell+p+1}{2}\right)
\Gamma\!\left(\frac{\ell+p+2}{2}\right)
=
2^{-\ell-p}\sqrt{\pi}\,\Gamma(\ell+p+1),
\end{equation}
and therefore
\begin{equation}
\frac{
\Gamma(\ell+p+1)
}{
\Gamma\!\left(\frac{\ell+p+1}{2}\right)
\Gamma\!\left(\frac{\ell+p+2}{2}\right)
}
=
\frac{2^{\ell+p}}{\sqrt{\pi}}.
\end{equation}
The recurrence relation for the Gamma function also gives
\begin{align}
&\Gamma\!\left(\frac p2+\frac14-\frac\nu2\right)
\Gamma\!\left(\frac p2+\frac14+\frac\nu2\right)
\nonumber\\
&\qquad=
\frac{\Delta_p}{4}
\Gamma\!\left(\frac p2-\frac34-\frac\nu2\right)
\Gamma\!\left(\frac p2-\frac34+\frac\nu2\right),
\end{align}
and therefore, using the definition of $\mathcal C_{p,\nu}$ already introduced in the main text,
\begin{equation}
\frac{
\Gamma\!\left(\frac p2+\frac14-\frac\nu2\right)
\Gamma\!\left(\frac p2+\frac14+\frac\nu2\right)
}{\Delta_p}
=
2^{\frac32-p}\mathcal C_{p,\nu}.
\end{equation}
The first term generated by \cref{eq:folded3F2secondtransformation} consequently reduces to
\begin{align}
&\frac{2^{\ell+\frac32}\mathcal C_{p,\nu}}{\sqrt{\pi}}
\Gamma(-\ell-2)
e^{-\frac{\ii a\pi}{2}(\ell+p+1)}
\nonumber\\
&\qquad\times
\frac{
\Gamma\!\left(\frac\ell2+\frac54+\frac\nu2\right)
\Gamma\!\left(\frac\ell2+\frac74+\frac\nu2\right)
}{
\Gamma\!\left(-\frac\ell2-\frac34+\frac\nu2\right)
\Gamma\!\left(-\frac\ell2-\frac14+\frac\nu2\right)
}.
\label{eq:foldedGammaFrom3F2}
\end{align}
The two homogeneous contributions can be brought to the same normalization by another application of the duplication formula,
\begin{align}
2^{-\nu}\Gamma\!\left(\ell+\frac52+\nu\right)
&=
\frac{2^{\ell+\frac32}}{\sqrt{\pi}}
\Gamma\!\left(\frac\ell2+\frac54+\frac\nu2\right)
\Gamma\!\left(\frac\ell2+\frac74+\frac\nu2\right),
\\
2^{\nu}\Gamma\!\left(\ell+\frac52-\nu\right)
&=
\frac{2^{\ell+\frac32}}{\sqrt{\pi}}
\Gamma\!\left(\frac\ell2+\frac54-\frac\nu2\right)
\Gamma\!\left(\frac\ell2+\frac74-\frac\nu2\right).
\end{align}
At this stage the three purely Gamma-function contributions have the common prefactor
\begin{equation}
\frac{2^{\ell+\frac32}\mathcal C_{p,\nu}}{\sqrt{\pi}}\Gamma(-\ell-2).
\end{equation}
The reflection formula, applied directly to the two denominator pairs that occur in \cref{eq:Rvertexfoldedgeneric,eq:foldedGammaFrom3F2}, gives
\begin{align}
&\frac{1}{
\Gamma\!\left(-\frac\ell2-\frac34+\frac\nu2\right)
\Gamma\!\left(-\frac\ell2-\frac14+\frac\nu2\right)}
\nonumber\\
&\qquad=
\frac{
\Gamma\!\left(\frac\ell2+\frac54-\frac\nu2\right)
\Gamma\!\left(\frac\ell2+\frac74-\frac\nu2\right)
}{2\pi^2}
\cos\!\left[\pi(\ell-\nu)\right],
\label{eq:foldedReflectionPlus}
\\[1mm]
&\frac{1}{
\Gamma\!\left(-\frac\ell2-\frac34-\frac\nu2\right)
\Gamma\!\left(-\frac\ell2-\frac14-\frac\nu2\right)}
\nonumber\\
&\qquad=
\frac{
\Gamma\!\left(\frac\ell2+\frac54+\frac\nu2\right)
\Gamma\!\left(\frac\ell2+\frac74+\frac\nu2\right)
}{2\pi^2}
\cos\!\left[\pi(\ell+\nu)\right].
\label{eq:foldedReflectionMinus}
\end{align}
Hence all three terms contain the same product
\begin{equation}
\Gamma\!\left(\frac\ell2+\frac54+\frac\nu2\right)
\Gamma\!\left(\frac\ell2+\frac74+\frac\nu2\right)
\Gamma\!\left(\frac\ell2+\frac54-\frac\nu2\right)
\Gamma\!\left(\frac\ell2+\frac74-\frac\nu2\right).
\end{equation}
The remaining phase and trigonometric coefficient is
\begin{align}
&\cos\!\left[\pi(\ell-\nu)\right]
 e^{-\frac{\ii a\pi}{2}(\ell+p+1)}
\nonumber\\
&+\cos\!\left[\pi(\ell-\nu)\right]
\left[\sin\theta_{p,\nu}-\cos\theta_{p,\nu}\cot(\pi\nu)\right]
 e^{-\frac{\ii a\pi}{2}(\ell+\frac52+\nu)}
\nonumber\\
&+\cos\!\left[\pi(\ell+\nu)\right]
\cos\theta_{p,\nu}\csc(\pi\nu)
 e^{-\frac{\ii a\pi}{2}(\ell+\frac52-\nu)}.
\end{align}
Using
\begin{equation}
\sin\theta_{p,\nu}-\cos\theta_{p,\nu}\cot(\pi\nu)
=
-\cos(\theta_{p,\nu}+\pi\nu)\csc(\pi\nu),
\end{equation}
together with
\begin{equation}
\theta_{p,\nu}=\frac\pi2\left(p-\frac52-\nu\right),
\end{equation}
this coefficient reduces to
\begin{equation}
-2\cos\theta_{p,\nu}\,
\sin(\pi\ell)\,
e^{-\frac{\ii a\pi}{2}\left(\ell+\frac52-\nu\right)}.
\label{eq:foldedTrigReduction}
\end{equation}
The product of the four Gamma functions above is simplified once more with the duplication formula,
\begin{align}
&\Gamma\!\left(\frac\ell2+\frac54+\frac\nu2\right)
\Gamma\!\left(\frac\ell2+\frac74+\frac\nu2\right)
\Gamma\!\left(\frac\ell2+\frac54-\frac\nu2\right)
\Gamma\!\left(\frac\ell2+\frac74-\frac\nu2\right)
\nonumber\\
&\qquad=
2^{-2\ell-3}\pi\,
\Gamma\!\left(\ell+\frac52+\nu\right)
\Gamma\!\left(\ell+\frac52-\nu\right).
\label{eq:foldedFourGammaDuplication}
\end{align}
Combining \cref{eq:foldedReflectionPlus,eq:foldedReflectionMinus,eq:foldedTrigReduction,eq:foldedFourGammaDuplication}, the only apparently singular factor left is $-\Gamma(-\ell-2)\sin(\pi\ell)$.  The reflection relation for these two Gamma functions is
\begin{equation}
\Gamma(-\ell-2)\Gamma(\ell+3)
=-\frac{\pi}{\sin(\pi\ell)},
\end{equation}
and hence
\begin{equation}
-\Gamma(-\ell-2)\sin(\pi\ell)
=
\frac{\pi}{\Gamma(\ell+3)}.
\label{eq:foldedApparentPoleCancellation}
\end{equation}
Thus the factors $\Gamma(-\ell-2)$ appearing separately in \cref{eq:Rvertexfoldedgeneric} are not poles of the complete folded remainder.  After the three terms are combined, they are replaced by the regular factor $1/\Gamma(\ell+3)$.

Collecting this result with \cref{eq:foldedRegular3F2piece}, the folded finite part takes the manifestly regular form
\begin{equation}
\begin{aligned}
\operatorname{Fin}_{u\to1}\mathcal R^\ell_{a,p}(u)
={}&
\frac{
2^{-\ell-\frac32}
\mathcal C_{p,\nu}
\cos\theta_{p,\nu}
}{\sqrt{\pi}}
\frac{
\Gamma\!\left(\ell+\frac52+\nu\right)
\Gamma\!\left(\ell+\frac52-\nu\right)
}{\Gamma(\ell+3)}
e^{-\frac{\ii a\pi}{2}\left(\ell+\frac52-\nu\right)}
\\[2mm]
&-
\frac{
\Gamma(\ell+p+1)
e^{-\frac{\ii a\pi}{2}(\ell+p+1)}
}{
\left(p-\frac32+\nu\right)
\left(\ell+\frac52+\nu\right)
}
\\
&\qquad\times
{}_3F_2\!\left(
\begin{matrix}
1,
\frac{\ell+p+1}{2},
-\frac\ell2-\frac34+\frac\nu2\\[1mm]
\frac p2+\frac14+\frac\nu2,
\frac\ell2+\frac94+\frac\nu2
\end{matrix};1\right).
\end{aligned}
\label{eq:Rvertexfoldedregular}
\end{equation}

\medskip
\noindent\textbf{8. Integer values of $\ell$.}
The unsimplified finite-part formulas above are written for noninteger balance.  In particular, at $\ell=-2$ or $\ell=0$ the factor $\Gamma(-\ell-2)$ is singular term by term, so the three finite parts must not be evaluated separately at those values.  Following Refs.~\cite{Qin:2023closed,Buhring:1987unit}, the correct prescription is to keep $\ell$ generic, combine the three finite contributions in \cref{eq:Rvertexfoldedgeneric}, and only then take the desired integer limit.  The same ordering applies to the zero-balanced logarithm at $\ell=-2$: it is the degenerate continuation of the nonanalytic sector already shown to cancel in \cref{eq:RFfoldedcancel}.  Equation~\eqref{eq:Rvertexfoldedregular} is precisely the result of carrying out this combination algebraically: the common $\Gamma(-\ell-2)$ has disappeared and the regular factor $1/\Gamma(\ell+3)$ is explicit.  The remaining possible singularities are therefore those associated with the original endpoint structure of the one-vertex integral, rather than with the folded continuation itself.  In particular, for the physical bispectrum value $\ell=-2$ the common balance is zero and the separate hypergeometric blocks develop the corresponding logarithmic continuation.  The cancellation in \cref{eq:RFfoldedcancel} is to be performed before taking this integer limit; the folded logarithm therefore cancels in the complete retarded combination, and the only remaining subtlety is the independent $x\to0$ endpoint discussed in the main text.

\clearpage

\section{Bootstrap equations for the disorder-averaged dS}
\label{app:RFbootstrap}

The direct integration in \cref{subsec:RFclosed} gives the one-vertex transform in closed form as a sum of one ${}_3F_2$ and two ${}_2F_1$ functions.  In this appendix we derive the same structure independently from differential equations.  It is useful to begin from the full disorder-averaged propagator rather than from the random-field correction alone, because this makes explicit the relation to the standard de Sitter bootstrap and separates two physically distinct inhomogeneities: the usual contact source of the time-ordered de Sitter propagator and the separable source induced by the random field.

For momentum magnitude $s$, define the Klein--Gordon operator
\begin{equation}
  \mathcal L_{\eta,s}
  \equiv
  \partial_\eta^2-\frac{2}{\eta}\partial_\eta
  +s^2+\frac{m^2}{H^2\eta^2}.
  \label{eq:bootstrapKGoperator}
\end{equation}
The complete disorder-averaged contour propagator is
\begin{equation}
  G^{\rm dis}_{ab}(s;\eta_1,\eta_2)
  =G^{(0)}_{ab}(s;\eta_1,\eta_2)
  +f(s)R_{s,p}(\eta_1)R_{s,p}(\eta_2),
  \qquad a,b=\pm1.
  \label{eq:bootstrapFullPropagator}
\end{equation}
The clean Wightman functions are homogeneous,
\begin{equation}
  \mathcal L_{\eta_1,s}G^{(0)}_{+-}
  =\mathcal L_{\eta_1,s}G^{(0)}_{-+}=0,
  \label{eq:bootstrapWightmanHomogeneous}
\end{equation}
whereas differentiation of the step functions in the time-ordered propagators gives
\begin{equation}
  \mathcal L_{\eta_1,s}G^{(0)}_{++}
  =-\ii\frac{\delta(\eta_1-\eta_2)}{a^2(\eta_2)},
  \qquad
  \mathcal L_{\eta_1,s}G^{(0)}_{--}
  =+\ii\frac{\delta(\eta_1-\eta_2)}{a^2(\eta_2)}.
  \label{eq:bootstrapCleanContact}
\end{equation}
This contact term is the microscopic origin of the standard non-homogeneous de Sitter bootstrap equation.

The random-field response obeys instead
\begin{equation}
  \mathcal L_{\eta,s}R_{s,p}(\eta)
  =Q_p(\eta),
  \qquad
  Q_p(\eta)
  \equiv\frac{(-H\eta)^p}{H^2\eta^2}.
  \label{eq:bootstrapResponseSource}
\end{equation}
Consequently,
\begin{equation}
  \mathcal L_{\eta_1,s}\Delta G_{{\rm RF},ab}
  =f(s)Q_p(\eta_1)R_{s,p}(\eta_2),
  \label{eq:bootstrapDisorderPropSource}
\end{equation}
for all four contour assignments.  Unlike the clean contact term, the random-field source is present also in the mixed Wightman sectors and is separable in the two time variables.  Equations~\eqref{eq:bootstrapCleanContact} and \eqref{eq:bootstrapDisorderPropSource} are the two independent inhomogeneities of the disorder-averaged problem.

For a generic function $X(\eta)$ introduce the one-vertex transform
\begin{equation}
  \mathcal V_a^{\ell}[X](u)
  \equiv
  \int_{-\infty_a}^{0}\dd\eta\,
  (-\eta)^\ell
  e^{\ii aK\eta}X(\eta),
  \qquad
  u\equiv\frac{s}{K}.
  \label{eq:bootstrapGenericVertex}
\end{equation}
Two integrations by parts, with the Bunch--Davies $\ii\epsilon$ prescription regulating the past endpoint, give
\begin{equation}
  \mathcal D_u^\ell\,\mathcal V_a^{\ell}[X](u)
  =
  \int_{-\infty_a}^{0}\dd\eta\,
  (-\eta)^{\ell+2}
  e^{\ii aK\eta}
  \mathcal L_{\eta,s}X(\eta),
  \label{eq:bootstrapTransformIdentity}
\end{equation}
where
\begin{equation}
  \begin{aligned}
  \mathcal D_u^\ell
  ={}&u^2(1-u^2)\partial_u^2
  -2u\big(u^2+\ell+2\big)\partial_u
  +\left[\left(\ell+\frac52\right)^2-\nu^2\right].
  \end{aligned}
  \label{eq:bootstrapOperator}
\end{equation}
It is often useful to introduce $\vartheta_u\equiv u\partial_u$ and $L_\ell\equiv\ell+5/2$.  The same operator then takes the factorized form
\begin{equation}
  \mathcal D_u^\ell
  =\big(\vartheta_u-L_\ell\big)^2-\nu^2
  -u^2\vartheta_u(\vartheta_u+1).
  \label{eq:bootstrapOperatorFactorized}
\end{equation}
The two forms are exactly equivalent.  In particular, on a monomial,
\begin{equation}
  \mathcal D_u^\ell u^q
  =\left[\left(q-\ell-\frac52\right)^2-\nu^2\right]u^q
  -q(q+1)u^{q+2}.
  \label{eq:bootstrapMonomialAction}
\end{equation}

Apply \cref{eq:bootstrapTransformIdentity} to the first time variable of the clean seed \eqref{eq:cleanseeddef}.  The mixed sectors remain homogeneous,
\begin{equation}
  \mathcal D_u^{\ell_1}
  \widehat{\mathcal I}^{\,\ell_1\ell_2}_{+-}(u,v)
  =
  \mathcal D_u^{\ell_1}
  \widehat{\mathcal I}^{\,\ell_1\ell_2}_{-+}(u,v)
  =0.
  \label{eq:bootstrapCleanMixed}
\end{equation}
For the equal-sign sectors, the delta function in \cref{eq:bootstrapCleanContact} collapses the two time integrations.  Let
\begin{equation}
  N\equiv5+\ell_1+\ell_2.
\end{equation}
For $a=\pm1$ one obtains
\begin{align}
  \mathcal D_u^{\ell_1}
  \widehat{\mathcal I}^{\,\ell_1\ell_2}_{aa}(u,v)
  ={}&
  \ii a\,s^N
  \int_{-\infty_a}^{0}\dd\eta\,
  (-\eta)^{N-1}
  e^{\ii a(k_{12}+k_{34})\eta}
  \nonumber\\
  ={}&
  \ii a\,
  e^{-\ii a\pi N/2}\Gamma(N)
  \frac{s^N}{(k_{12}+k_{34})^N}.
  \label{eq:bootstrapCleanContactIntegral}
\end{align}
Using
\begin{equation}
  k_{12}+k_{34}
  =s\left(\frac1u+\frac1v\right)
  =s\frac{u+v}{uv}
\end{equation}
and
\begin{equation}
  \ii a\,e^{-\ii a\pi N/2}
  =e^{-\ii a\pi(\ell_1+\ell_2)/2},
\end{equation}
we recover the standard de Sitter bootstrap source,
\begin{equation}
  \mathcal D_u^{\ell_1}
  \widehat{\mathcal I}^{\,\ell_1\ell_2}_{\pm\pm}(u,v)
  =e^{\mp\ii\pi(\ell_1+\ell_2)/2}
  \Gamma(5+\ell_1+\ell_2)
  \left(\frac{uv}{u+v}\right)^{5+\ell_1+\ell_2}.
  \label{eq:bootstrapCleanSourceFinal}
\end{equation}
The factor $(uv/(u+v))^N$ is therefore nothing but the momentum-space image of the local time contact $\delta(\eta_1-\eta_2)$.  Notice that the overall $H^2$ has been scaled out by the definition \eqref{eq:cleanseeddef}, which contains an explicit factor $H^{-2}$.  In the alternative convention where the seed is defined without this factor, the right-hand side of \eqref{eq:bootstrapCleanSourceFinal} is multiplied by $H^2$, reproducing the normalization commonly used in the cosmological-bootstrap literature and in our previous bootstrap conventions.

\subsection{Bootstrap equation for the random-field contribution}

For the seed equations in this appendix we now adopt the same white-noise normalization $f(s)=f_0H^3$ used in the main text.  The propagator-level equations above remain valid for a generic $f(s)$.

For the random-field response, \cref{eq:bootstrapTransformIdentity,eq:bootstrapResponseSource} give directly
\begin{equation}
  \mathcal D_u^\ell\mathcal R_{a,p}^{\ell}(u)
  =J_{a,p}^{\ell}(u),
  \label{eq:bootstrapOneVertexEquation}
\end{equation}
with
\begin{equation}
  J_{a,p}^{\ell}(u)
  \equiv
  \Gamma(\ell+p+1)
  e^{-\frac{\ii a\pi}{2}(\ell+p+1)}
  u^{\ell+p+1}.
  \label{eq:bootstrapOneVertexSource}
\end{equation}
Equivalently, acting on the factorized two-time seed gives
\begin{align}
  \mathcal D_u^{\ell_1}
  \Delta\widehat{\mathcal I}^{\,\ell_1\ell_2}_{{\rm RF},ab}(u,v,s,p)
  ={}&
  -ab\,f_0\left(\frac{s}{H}\right)^{3-2p}
  J_{a,p}^{\ell_1}(u)
  \mathcal R_{b,p}^{\ell_2}(v),
  \label{eq:bootstrapRFSingle}\\
  \mathcal D_u^{\ell_1}\mathcal D_v^{\ell_2}
  \Delta\widehat{\mathcal I}^{\,\ell_1\ell_2}_{{\rm RF},ab}(u,v,s,p)
  ={}&
  -ab\,f_0\left(\frac{s}{H}\right)^{3-2p}
  J_{a,p}^{\ell_1}(u)
  J_{b,p}^{\ell_2}(v).
  \label{eq:bootstrapRFDouble}
\end{align}
This should be contrasted with \cref{eq:bootstrapCleanSourceFinal}: the clean source couples $u$ and $v$ because the delta function identifies the two interaction times, whereas the random-field source remains separable because $\Delta G_{ab}=fR_pR_p$ is rank one in time.

The full disorder-averaged seed,
\begin{equation}
  \widehat{\mathcal I}^{\rm dis}_{ab}
  \equiv
  \widehat{\mathcal I}^{(0)}_{ab}
  +\Delta\widehat{\mathcal I}_{{\rm RF},ab},
\end{equation}
therefore obeys
\begin{equation}
  \begin{aligned}
  \mathcal D_u^{\ell_1}
  \widehat{\mathcal I}^{\rm dis}_{ab}(u,v,s,p)
  ={}&
  \delta_{ab}^{\rm K}\,
  e^{-\ii a\pi(\ell_1+\ell_2)/2}
  \Gamma(N)
  \left(\frac{uv}{u+v}\right)^N
  \\
  &-ab\,f_0\left(\frac{s}{H}\right)^{3-2p}
  J_{a,p}^{\ell_1}(u)
  \mathcal R_{b,p}^{\ell_2}(v),
  \end{aligned}
  \label{eq:bootstrapFullSeedEquation}
\end{equation}
where $\delta_{ab}^{\rm K}=1$ for $a=b$ and zero otherwise.  Thus the disorder does not replace the usual de Sitter contact source; it adds a second, statistically generated inhomogeneity.

We now solve \cref{eq:bootstrapOneVertexEquation} without performing the oscillatory integral.  Define
\begin{equation}
  r\equiv\ell+p+1,
  \qquad
  j_{a,p}^{\ell}\equiv
  \Gamma(r)e^{-\ii a\pi r/2}.
\end{equation}
For generic non-resonant parameters, take the particular ansatz
\begin{equation}
  \mathcal R_{a,p,{\rm part}}^{\ell}(u)
  =u^r\sum_{n=0}^{\infty}c_nu^{2n}.
  \label{eq:bootstrapParticularAnsatz}
\end{equation}
Using \cref{eq:bootstrapMonomialAction}, the lowest coefficient satisfies
\begin{equation}
  \Delta_p c_0=j_{a,p}^{\ell},
  \qquad
  \Delta_p=\left(p-\frac32\right)^2-\nu^2,
\end{equation}
and hence
\begin{equation}
  c_0=\frac{\Gamma(r)}{\Delta_p}e^{-\ii a\pi r/2}.
\end{equation}
For $n\geq1$ the coefficients obey
\begin{equation}
  \frac{c_n}{c_{n-1}}
  =
  \frac{
  \left(n-1+\frac r2\right)
  \left(n-1+\frac{r+1}{2}\right)
  }{
  \left(n-1+\frac p2+\frac14-\frac\nu2\right)
  \left(n-1+\frac p2+\frac14+\frac\nu2\right)
  }.
  \label{eq:bootstrapParticularRecurrence}
\end{equation}
The absence of an explicit $n$ in the denominator is precisely compensated by $(1)_n/n!=1$ in the generalized hypergeometric series.  Resumming gives
\begin{equation}
  \begin{aligned}
  \mathcal R_{a,p,{\rm part}}^{\ell}(u)
  ={}&
  \frac{\Gamma(\ell+p+1)}{\Delta_p}
  e^{-\frac{\ii a\pi}{2}(\ell+p+1)}
  u^{\ell+p+1}
  \\
  &\times
  {}_3F_2\!\left(
  \begin{matrix}
  1,\frac{\ell+p+1}{2},\frac{\ell+p+2}{2}\\
  \frac p2+\frac14-\frac\nu2,
  \frac p2+\frac14+\frac\nu2
  \end{matrix};u^2\right).
  \end{aligned}
  \label{eq:bootstrapParticularSolution}
\end{equation}
This reproduces exactly the first line of the direct result \eqref{eq:Rvertexclosed}.

The homogeneous equation is
\begin{equation}
  \mathcal D_u^\ell\mathcal Y(u)=0.
\end{equation}
A Frobenius ansatz
\begin{equation}
  \mathcal Y_\sigma^\ell(u)
  =u^{q_\sigma}\sum_{n=0}^{\infty}d_n^{(\sigma)}u^{2n},
  \qquad \sigma=\pm1,
\end{equation}
first gives the indicial equation
\begin{equation}
  \left(q-\ell-\frac52\right)^2-\nu^2=0,
\end{equation}
so that
\begin{equation}
  q_\sigma=\ell+\frac52+\sigma\nu.
  \label{eq:bootstrapIndicialExponents}
\end{equation}
For $n\geq1$,
\begin{equation}
  \frac{d_n^{(\sigma)}}{d_{n-1}^{(\sigma)}}
  =
  \frac{
  \left(n-1+\frac{q_\sigma}{2}\right)
  \left(n-1+\frac{q_\sigma+1}{2}\right)
  }{n(n+\sigma\nu)}.
  \label{eq:bootstrapHomogeneousRecurrence}
\end{equation}
Therefore a convenient homogeneous basis is
\begin{equation}
  \mathcal Y_\sigma^\ell(u)
  =u^{\ell+\frac52+\sigma\nu}
  {}_2F_1\!\left(
  \frac\ell2+\frac54+\frac{\sigma\nu}{2},
  \frac\ell2+\frac74+\frac{\sigma\nu}{2};
  1+\sigma\nu;u^2\right).
  \label{eq:bootstrapHomogeneousBasis}
\end{equation}
The two characteristic powers $u^{\ell+5/2\pm\nu}$ are independent of $p$.  Thus the temporal disorder profile changes the source and the amplitudes with which the massive branches are excited, but not the collider exponents themselves.

The most general solution of the one-vertex bootstrap equation is consequently
\begin{equation}
  \mathcal R_{a,p}^{\ell}(u)
  =\mathcal R_{a,p,{\rm part}}^{\ell}(u)
  +B_{a,+}^{\ell,p}\mathcal Y_+^\ell(u)
  +B_{a,-}^{\ell,p}\mathcal Y_-^\ell(u).
  \label{eq:bootstrapGeneralSolution}
\end{equation}

The differential equation fixes the functional form but not the two homogeneous coefficients.  They are fixed by the microscopic retarded boundary condition.  From the Lommel connection formula \eqref{eq:SminusS},
\begin{align}
  S_{p-\frac52,\nu}(x)
  ={}&s_{p-\frac52,\nu}(x)
  +\mathcal C_{p,\nu}
  \left[\sin\theta_{p,\nu}-\cos\theta_{p,\nu}\cot(\pi\nu)\right]J_\nu(x)
  \nonumber\\
  &+\mathcal C_{p,\nu}\cos\theta_{p,\nu}\csc(\pi\nu)J_{-\nu}(x).
  \label{eq:bootstrapLommelBoundaryDecomposition}
\end{align}
Using the leading small-$x$ behavior of $J_{\pm\nu}$ and the elementary oscillatory transform gives the retarded soft coefficients
\begin{align}
  B_{a,+}^{\ell,p}
  ={}&
  \mathcal C_{p,\nu}
  \left[\sin\theta_{p,\nu}-\cos\theta_{p,\nu}\cot(\pi\nu)\right]
  \frac{2^{-\nu}\Gamma\!\left(\ell+\frac52+\nu\right)}{\Gamma(1+\nu)}
  e^{-\frac{\ii a\pi}{2}(\ell+\frac52+\nu)},
  \label{eq:bootstrapBplus}\\
  B_{a,-}^{\ell,p}
  ={}&
  \mathcal C_{p,\nu}\cos\theta_{p,\nu}\csc(\pi\nu)
  \frac{2^{\nu}\Gamma\!\left(\ell+\frac52-\nu\right)}{\Gamma(1-\nu)}
  e^{-\frac{\ii a\pi}{2}(\ell+\frac52-\nu)}.
  \label{eq:bootstrapBminus}
\end{align}
Substitution of \cref{eq:bootstrapParticularSolution,eq:bootstrapHomogeneousBasis,eq:bootstrapBplus,eq:bootstrapBminus} into \cref{eq:bootstrapGeneralSolution} reproduces term by term the direct integration result \eqref{eq:Rvertexclosed}.  The two derivations are therefore independent in their treatment of the time integral: direct integration resums the oscillatory transforms explicitly, while the bootstrap reconstructs the same hypergeometric basis from the differential equation and fixes only the two homogeneous constants from retarded boundary data.

For the physical Schwinger--Keldysh sum it is useful to define
\begin{equation}
  \mathcal Q_p^\ell(u)
  \equiv
  \mathcal R_{+,p}^\ell(u)-\mathcal R_{-,p}^\ell(u).
\end{equation}
Equation~\eqref{eq:bootstrapOneVertexEquation} immediately gives
\begin{equation}
  \mathcal D_u^\ell\mathcal Q_p^\ell(u)
  =-2\ii\Gamma(\ell+p+1)
  \sin\!\left[\frac\pi2(\ell+p+1)\right]
  u^{\ell+p+1}.
  \label{eq:bootstrapPhysicalDifference}
\end{equation}
Since
\begin{equation}
  \Delta\widehat{\mathcal I}_{\rm RF}^{\ell_1\ell_2}(u,v,s,p)
  =-f_0\left(\frac{s}{H}\right)^{3-2p}
  \mathcal Q_p^{\ell_1}(u)\mathcal Q_p^{\ell_2}(v),
\end{equation}
we also obtain the compact double-bootstrap equation
\begin{equation}
  \begin{aligned}
  \mathcal D_u^{\ell_1}\mathcal D_v^{\ell_2}
  \Delta\widehat{\mathcal I}_{\rm RF}^{\ell_1\ell_2}(u,v,s,p)
  ={}&4 f_0\left(\frac{s}{H}\right)^{3-2p}
  \Gamma(\ell_1+p+1)\Gamma(\ell_2+p+1)
  \\
  &\times
  \sin\!\left[\frac\pi2(\ell_1+p+1)\right]
  \sin\!\left[\frac\pi2(\ell_2+p+1)\right]
  u^{\ell_1+p+1}v^{\ell_2+p+1}.
  \end{aligned}
  \label{eq:bootstrapPhysicalSeedDouble}
\end{equation}
This equation makes the separation from the clean de Sitter source especially transparent.  The clean contact term is non-separable and proportional to $(uv/(u+v))^{5+\ell_1+\ell_2}$, whereas the disorder source is a product of one-variable powers.  Both are governed by the same second-order de Sitter operator \eqref{eq:bootstrapOperator}; the difference lies in the microscopic origin and boundary data of the corresponding inhomogeneities.  Setting $p=0$ throughout this appendix immediately recovers the persistent-disorder bootstrap source and the corresponding $S_{-5/2,\nu}$ solution used in the minimal model.

\section{Mellin--Barnes check of the retarded response}
\label{app:MBresponse}

In this appendix we verify the exact response \cref{eq:Rlommel} directly from the retarded representation \cref{eq:Rdef} for arbitrary power $p$.  The calculation shows explicitly how the persistent result arises at $p=0$ and isolates the Mellin pole responsible for the homogeneous massive completion.

Set
\begin{equation}
  x=-k\eta,
  \qquad
  z=-k\eta'.
\end{equation}
Using the Bunch--Davies Hankel modes in the retarded Green function, one obtains
\begin{equation}
  R_{k,p}(x)
  =\left(\frac{H}{k}\right)^p
  \frac{\ii\pi}{4H^2}x^{3/2}
  \left[
  H_\nu^{(1)}(x)\mathcal I_{\nu,p}^{(2)}(x)
  -H_\nu^{(2)}(x)\mathcal I_{\nu,p}^{(1)}(x)
  \right],
  \label{eq:RwithI}
\end{equation}
where
\begin{equation}
  \mathcal I_{\nu,p}^{(1,2)}(x)
  \equiv
  \int_x^\infty\dd z\,z^{p-5/2}H_\nu^{(1,2)}(z).
  \label{eq:Idef}
\end{equation}
At $p=0$ these expressions reduce to the persistent formulas.

\subsubsection*{Mellin--Barnes representation}

We use
\begin{equation}
  H_\nu^{(1)}(z)
  =\frac1\pi
  \int_{\mathcal C}\frac{\dd\omega}{2\pi\ii}
  \Gamma\left(\omega+\frac\nu2\right)
  \Gamma\left(\omega-\frac\nu2\right)
  e^{\ii\pi(\omega-\nu/2-1/2)}
  \left(\frac z2\right)^{-2\omega},
  \label{eq:MBH1}
\end{equation}
and
\begin{equation}
  H_\nu^{(2)}(z)
  =\frac1\pi
  \int_{\mathcal C}\frac{\dd\omega}{2\pi\ii}
  \Gamma\left(\omega+\frac\nu2\right)
  \Gamma\left(\omega-\frac\nu2\right)
  e^{-\ii\pi(\omega-\nu/2-1/2)}
  \left(\frac z2\right)^{-2\omega}.
  \label{eq:MBH2}
\end{equation}
The time integral is now
\begin{equation}
  \int_x^\infty\dd z\,z^{p-5/2-2\omega}
  =\frac{x^{p-3/2-2\omega}}{2\omega+3/2-p},
  \qquad
  \Re\omega>\frac p2-\frac34,
  \label{eq:zIntegralMB}
\end{equation}
where the result elsewhere is defined by analytic continuation.  Therefore
\begin{align}
  \mathcal I_{\nu,p}^{(1)}(x)
  ={}&\frac{x^{p-3/2}}{\pi}
  \int_{\mathcal C}\frac{\dd\omega}{2\pi\ii}
  \frac{\Gamma(\omega+\nu/2)\Gamma(\omega-\nu/2)}{2\omega+3/2-p}
  \nonumber\\
  &\times e^{\ii\pi(\omega-\nu/2-1/2)}
  \left(\frac x2\right)^{-2\omega},
  \label{eq:MBI1}
\end{align}
and
\begin{align}
  \mathcal I_{\nu,p}^{(2)}(x)
  ={}&\frac{x^{p-3/2}}{\pi}
  \int_{\mathcal C}\frac{\dd\omega}{2\pi\ii}
  \frac{\Gamma(\omega+\nu/2)\Gamma(\omega-\nu/2)}{2\omega+3/2-p}
  \nonumber\\
  &\times e^{-\ii\pi(\omega-\nu/2-1/2)}
  \left(\frac x2\right)^{-2\omega}.
  \label{eq:MBI2}
\end{align}
The pole families are
\begin{equation}
  \omega=-\frac\nu2-n,
  \qquad
  \omega=+\frac\nu2-n,
  \qquad
  \omega_\star=\frac p2-\frac34,
  \qquad n=0,1,2,\ldots .
  \label{eq:MBpoles}
\end{equation}
The first two families are the Hankel Gamma-function towers; the third is introduced solely by the time integration.  At $p=0$ the additional pole becomes $\omega_\star=-3/4$.  A collision of the additional pole with a Gamma-function pole occurs when
\begin{equation}
  p+2n=\frac32\pm\nu,
\end{equation}
which is precisely the resonance condition \cref{eq:generalResonance}; the corresponding double Mellin poles generate the logarithmic solutions of the degenerate Frobenius sectors.

Closing the contour to the left, the two Gamma towers combine in the antisymmetric Hankel structure of \cref{eq:RwithI}.  Their sum can be organized as
\begin{equation}
  R_{k,p}^{\rm an}(x)
  =\left(\frac{H}{k}\right)^p\frac1{H^2}
  \sum_{n=0}^{\infty}
  \frac{(-1)^n x^{p+2n}}{a_{n+1}(p-5/2,\nu)},
  \label{eq:MBanalyticseries}
\end{equation}
where
\begin{equation}
  a_n(\rho,\nu)
  =4^n
  \left(\frac{\rho-\nu+1}{2}\right)_n
  \left(\frac{\rho+\nu+1}{2}\right)_n.
  \label{eq:anLommel}
\end{equation}
This is precisely
\begin{equation}
  R_{k,p}^{\rm an}(x)
  =\left(\frac{H}{k}\right)^p
  \frac{x^{3/2}}{H^2}s_{p-5/2,\nu}(x).
  \label{eq:MBsmallLommel}
\end{equation}
Since
\begin{equation}
  a_1(p-5/2,\nu)
  =\left(p-\frac32\right)^2-\nu^2
  =\Delta_p,
\end{equation}
the leading term is $(-H\eta)^p/(H^2\Delta_p)$.  Resumming gives
\begin{equation}
  R_{k,p}^{\rm an}(x)
  =\left(\frac{H}{k}\right)^p
  \frac{x^p}{H^2\Delta_p}
  {}_1F_2\left(
  1;
  \frac p2+\frac14-\frac\nu2,
  \frac p2+\frac14+\frac\nu2;
  -\frac{x^2}{4}
  \right).
  \label{eq:MBhyper}
\end{equation}
At $p=0$ this becomes the persistent $m^{-2}{}_1F_2$ solution.

Define
\begin{equation}
  \Phi_{p,\nu}
  \equiv
  \frac{5\pi}{4}+\frac{\pi\nu}{2}-\frac{\pi p}{2}
  =-\theta_{p,\nu}.
\end{equation}
At $\omega_\star=p/2-3/4$ the two Mellin integrals have residues
\begin{align}
  \mathcal I_{\nu,p,\star}^{(1)}
  ={}&
  \frac{2^{p-5/2}}{\pi}
  \Gamma\left(\frac p2-\frac34+\frac\nu2\right)
  \Gamma\left(\frac p2-\frac34-\frac\nu2\right)
  e^{-\ii\Phi_{p,\nu}},
  \label{eq:resI1}\\
  \mathcal I_{\nu,p,\star}^{(2)}
  ={}&
  \frac{2^{p-5/2}}{\pi}
  \Gamma\left(\frac p2-\frac34+\frac\nu2\right)
  \Gamma\left(\frac p2-\frac34-\frac\nu2\right)
  e^{+\ii\Phi_{p,\nu}}.
  \label{eq:resI2}
\end{align}
Substituting into \cref{eq:RwithI} yields
\begin{align}
  R_{k,p}^{(\star)}(x)
  ={}&-\left(\frac{H}{k}\right)^p
  \frac{2^{p-7/2}}{H^2}x^{3/2}
  \Gamma\left(\frac p2-\frac34+\frac\nu2\right)
  \Gamma\left(\frac p2-\frac34-\frac\nu2\right)
  \nonumber\\
  &\times
  \left[
  \sin\Phi_{p,\nu}\,J_\nu(x)
  +\cos\Phi_{p,\nu}\,Y_\nu(x)
  \right].
  \label{eq:MBhomogeneous}
\end{align}
Since $\Phi_{p,\nu}=-\theta_{p,\nu}$, this is exactly
\begin{equation}
  R_{k,p}^{(\star)}(x)
  =\left(\frac{H}{k}\right)^p
  \frac{x^{3/2}}{H^2}
  \left[S_{p-5/2,\nu}(x)-s_{p-5/2,\nu}(x)\right].
  \label{eq:MBdifference}
\end{equation}
Combining the Gamma towers and the additional pole gives
\begin{equation}
  R_{k,p}(x)
  =\left(\frac{H}{k}\right)^p
  \frac{x^{3/2}}{H^2}S_{p-5/2,\nu}(x),
  \label{eq:MBfinal}
\end{equation}
which reproduces \cref{eq:Rlommel}.  Thus changing $p$ shifts the extra time-integration pole but does not alter the Hankel pole towers carrying the heavy-field index $\nu$.  This is the Mellin-space version of the separation between the local forced response and the collider mass signal.

\subsection{Useful properties and limiting forms of the Lommel function}
\label{app:lommelproperties}

For completeness, we summarize the properties of the Lommel function that enter the analysis in the main text.  We use the conventions of Refs.~\cite{NIST:DLMF,Watson:Bessel}.  The functions $s_{\rho,\nu}$ and $S_{\rho,\nu}$ solve the inhomogeneous Bessel equation
\begin{equation}
  w''+\frac{1}{x}w'
  +\left(1-\frac{\nu^2}{x^2}\right)w
  =x^{\rho-1}.
  \label{eq:LommelDefAppendix}
\end{equation}
The particular solution $s_{\rho,\nu}$ is represented by the convergent series
\begin{equation}
  s_{\rho,\nu}(x)
  =x^{\rho+1}
  \sum_{j=0}^{\infty}
  \frac{(-1)^j x^{2j}}{a_{j+1}(\rho,\nu)},
  \qquad
  a_j(\rho,\nu)
  =4^j
  \left(\frac{\rho-\nu+1}{2}\right)_j
  \left(\frac{\rho+\nu+1}{2}\right)_j.
  \label{eq:smallLommelGeneral}
\end{equation}
The second Lommel function differs from it by a homogeneous Bessel combination,
\begin{align}
  S_{\rho,\nu}(x)
  ={}&s_{\rho,\nu}(x)
  +2^{\rho-1}
  \Gamma\left(\frac{\rho+\nu+1}{2}\right)
  \Gamma\left(\frac{\rho-\nu+1}{2}\right)
  \nonumber\\
  &\times
  \left[
  \sin\left(\frac{\pi(\rho-\nu)}{2}\right)J_\nu(x)
  -\cos\left(\frac{\pi(\rho-\nu)}{2}\right)Y_\nu(x)
  \right].
  \label{eq:LommelConnectionGeneral}
\end{align}
This representation is understood by analytic continuation at the exceptional values for which the individual Gamma functions or the separated terms become singular; the function $S_{\rho,\nu}$ itself has a finite limiting continuation in the cases relevant below~\cite{NIST:DLMF,Watson:Bessel}.

For large argument, $S_{\rho,\nu}$ has the algebraic asymptotic expansion
\begin{equation}
  S_{\rho,\nu}(x)
  \sim x^{\rho-1}
  \sum_{j=0}^{\infty}
  (-1)^j a_j(-\rho,\nu)x^{-2j},
  \qquad x\rightarrow\infty,
  \label{eq:LommelLargeGeneral}
\end{equation}
which is the property used to select the retarded solution in \cref{subsec:lommel}.  For the present family, $\rho=p-5/2$, and therefore
\begin{equation}
  S_{p-5/2,\nu}(x)
  \sim x^{p-7/2}
  \left[1+\mathcal O(x^{-2})\right],
\end{equation}
so that
\begin{equation}
  R_{k,p}(x)
  \sim
  \left(\frac Hk\right)^p\frac{x^{p-2}}{H^2}
  \left[1+\mathcal O(x^{-2})\right],
  \qquad x\rightarrow\infty.
  \label{eq:RlargeAppendix}
\end{equation}

The small-$x$ behavior is most transparent for noninteger $\nu$ and away from resonant parameter values.  Using
\begin{equation}
  Y_\nu(x)
  =\frac{\cos(\pi\nu)J_\nu(x)-J_{-\nu}(x)}{\sin(\pi\nu)}
\end{equation}
and the power series for $J_{\pm\nu}$, one obtains the three-sector decomposition
\begin{align}
  S_{\rho,\nu}(x)
  ={}&\mathcal B_-(\rho,\nu)x^{-\nu}
  \sum_{j=0}^{\infty}
  \frac{(-1)^j}{j!(1-\nu)_j}
  \left(\frac{x^2}{4}\right)^j
  \nonumber\\
  &+\mathcal B_+(\rho,\nu)x^{\nu}
  \sum_{j=0}^{\infty}
  \frac{(-1)^j}{j!(1+\nu)_j}
  \left(\frac{x^2}{4}\right)^j
  \nonumber\\
  &+x^{\rho+1}
  \sum_{j=0}^{\infty}
  \frac{(-1)^j x^{2j}}{a_{j+1}(\rho,\nu)},
  \label{eq:LommelSmallGeneral}
\end{align}
where
\begin{align}
  \mathcal B_-(\rho,\nu)
  ={}&\frac{2^{\rho+\nu-1}}{\pi}
  \Gamma(\nu)
  \Gamma\left(\frac{\rho+\nu+1}{2}\right)
  \Gamma\left(\frac{\rho-\nu+1}{2}\right)
  \cos\left(\frac{\pi(\rho-\nu)}{2}\right),
  \label{eq:BminusGeneral}\\
  \mathcal B_+(\rho,\nu)
  ={}&\frac{2^{\rho-\nu-1}}{\pi}
  \Gamma(-\nu)
  \Gamma\left(\frac{\rho+\nu+1}{2}\right)
  \Gamma\left(\frac{\rho-\nu+1}{2}\right)
  \cos\left(\frac{\pi(\rho+\nu)}{2}\right).
  \label{eq:BplusGeneral}
\end{align}
Thus $S_{\rho,\nu}$ contains two homogeneous Frobenius sectors, $x^{-\nu}$ and $x^{+\nu}$, together with the particular series $x^{\rho+1+2j}$.  For the response problem, $\rho=p-5/2$, and \cref{eq:LommelSmallGeneral} becomes
\begin{align}
  S_{p-5/2,\nu}(x)
  ={}&\mathcal B_-(p-5/2,\nu)x^{-\nu}\left[1+\mathcal O(x^2)\right]
  +\mathcal B_+(p-5/2,\nu)x^{\nu}\left[1+\mathcal O(x^2)\right]
  \nonumber\\
  &+\frac{x^{p-3/2}}{\Delta_p}
  \left[
  1-\frac{x^2}{M^2+(p+2)(p-1)}+\mathcal O(x^4)
  \right].
  \label{eq:LommelSmallOurCase}
\end{align}
Multiplication by $(H/k)^p x^{3/2}/H^2$ reproduces \cref{eq:Rlate}: the particular sector gives the powers $x^{p+2n}$, whereas the homogeneous sectors give $x^{3/2\pm\nu}$.  Degeneracies occur whenever
\begin{equation}
  p+2n=\frac32\pm\nu,
\end{equation}
in agreement with \cref{eq:generalResonance}.  The persistent examples $p=0$, $\nu=3/2$ and $p=0$, $\nu=1/2$ yield respectively the logarithms $\log x$ and $x^2\log x$ displayed in \cref{eq:RmasslessLate,eq:RconformalLate}.

A further degeneracy occurs at $\nu=0$ ($m^2=9H^2/4$), where the two homogeneous exponents coincide; the second independent homogeneous solution then contains the familiar logarithmic behavior associated with the $J_\nu$--$Y_\nu$ basis.  These logarithms are therefore resonance effects of the local Frobenius expansion, not singularities of the exact retarded response.

\bibliographystyle{JHEP}
\bibliography{biblio}

\end{document}